\documentclass[prl,twocolumn,superscriptaddress]{revtex4-1}
\usepackage{amsmath}
\usepackage{graphicx}

\newcommand{\ii}{\text{i}}

\newcommand{\ket}[1]{|#1\rangle}

\usepackage{color}

\begin{document}
\title{Long-lived circulating currents in strongly correlated nanorings}
\author{B. M. Schoenauer}
\affiliation{Institute for Theoretical Physics, Center for Extreme Matter and Emergent Phenomena, Utrecht University, Princetonplein 5, 3584 CE Utrecht, The Netherlands}
\author{N. M. Gergs}
\affiliation{Institute for Theoretical Physics, Center for Extreme Matter and Emergent Phenomena, Utrecht University, Princetonplein 5, 3584 CE Utrecht, The Netherlands}
\author{P. Schmitteckert}
\affiliation{Institute for Theoretical Physics and Astrophysics, Julius-Maximilians University of W\"urzburg, Am Hubland, 97074 W\"urzburg, Germany}
\affiliation{HQS Quantum Simulations GmbH, 76131 Karlsruhe, Germany}
\author{F. Evers}
\affiliation{Institute of Theoretical Physics, University of Regensburg, 93040 Regensburg, Germany}
\author{D. Schuricht}
\affiliation{Institute for Theoretical Physics, Center for Extreme Matter and Emergent Phenomena, Utrecht University, Princetonplein 5, 3584 CE Utrecht, The Netherlands}

\begin{abstract}
We study the time evolving currents flowing in an interacting, ring-shaped nanostructure after a bias voltage has been switched on. The source-to-drain current exhibits the expected relaxation towards its quasi-static equilibrium value at a rate $\Gamma_0$ reflecting the lead-induced broadening of the ring states. In contrast, the current circulating within the ring decays with a different rate $\Gamma$, which is a rapidly decaying function of the interaction strength and thus can take values orders of magnitude below $\Gamma_0$. This implies the existence of a regime in which the nanostructure is far from equilibrium even though the transmitted current is already stationary. We discuss experimental setups to observe the long-lived ring transients. 
\end{abstract}
\date{3 July 2019}
\maketitle

\emph{Introduction.}---Isolated quantum systems, such as small molecules, feature a discrete set of energy levels. When brought to contact with two electrodes, a nano-junction can form and a current begins to flow. At weak coupling, the associated level broadening, $\Gamma_0$, is still small as compared to the typical energy spacing, $\Delta E$, of the isolated system. One might perhaps suspect that these energies by themselves set the only relevant time scales. But in fact a prominent exception is known, the Kondo phenomenon~\cite{Hewson93}, which occurs in a situation where $\Delta E$ is dominated by strong on-site repulsion between the charge carriers. This suppresses charge fluctuations but allows for spin fluctuations, leading to an emergent energy scale, the Kondo temperature $T_\text{K}$, which is parametrically small compared to the native scales $\Gamma_0$ and $\Delta E$. 

In this work, we report another example of an emergent energy scale, $\Gamma$; it manifests in the relaxation of circulating currents in mesoscopic nanostructures. Like the Kondo temperature, the new scale is a many-body phenomenon, originating from interactions between particles on the nanostructure. However, the manifestation of the new relaxation rate $\Gamma$ requires the nanostructure to be brought out of equilibrium.

\begin{figure}[b]
\includegraphics[width=0.4\textwidth]{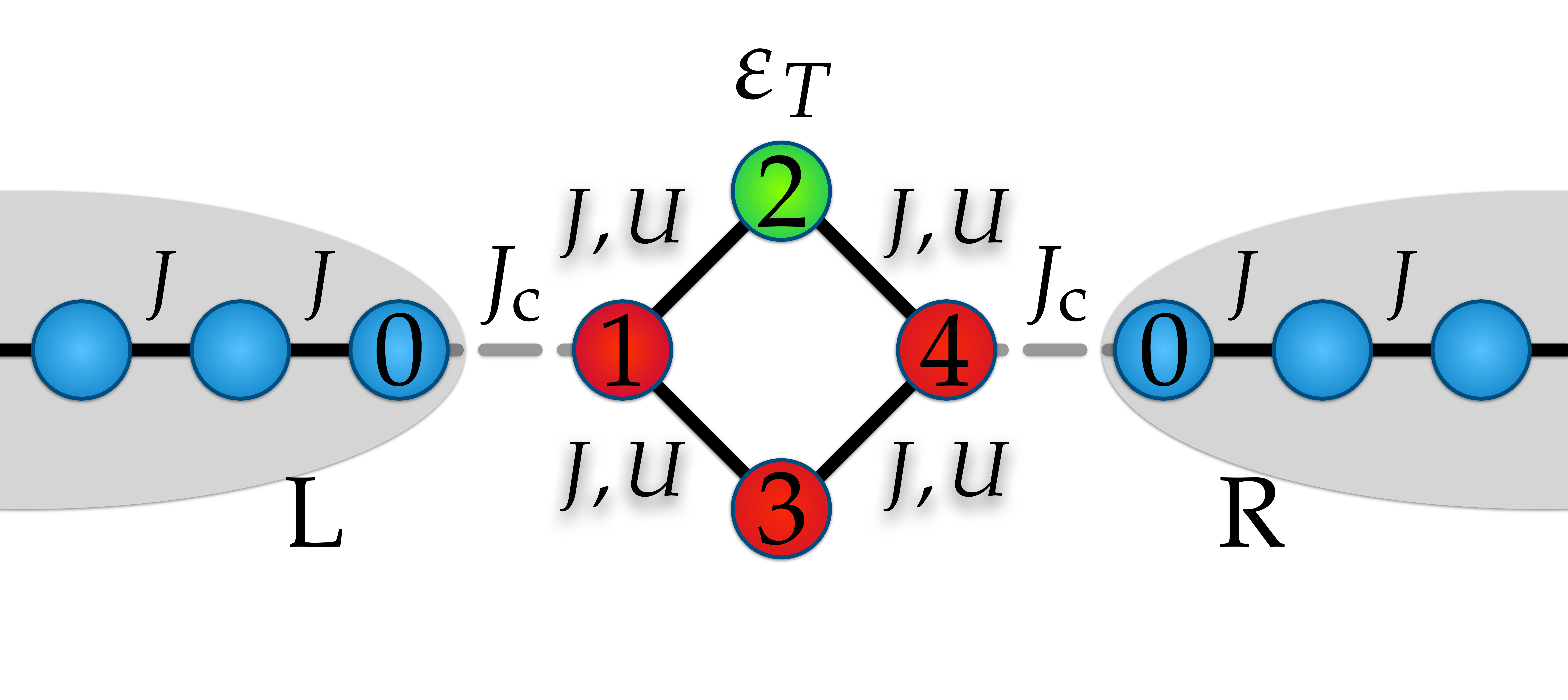}
\caption{Schematic representation of the nanostructure. The ring (red and green dots) is coupled by $J_\text{c}$ to left and right leads (blue dots). Spinless fermions can hop within the ring and leads with amplitude $J$, the top site (site 2) on the ring is subject to the potential $\varepsilon_T$, and inside the ring a nearest-neighbour interaction $U$ is present.}
\label{fig:f1}
\end{figure}

A sketch of a minimal model system that exhibits the novel scale $\Gamma$ is displayed in Fig.~\ref{fig:f1}. Originally, similar ring-shaped devices served as a toy-models to study the interplay of interaction and interference~\cite{BohrSchmitteckert12,Schmitteckert13} and to explain quantum-interference effects in transport through functionalised graphene ribbons~\cite{Walz-14}. The ring geometry supports stationary circulating (``orbital'') currents that can exceed the source-drain (``transport'') current by orders of magnitude at Fermi-energies situated close to a Fano-resonance. 

Strong circulating currents in ring-shaped devices, Fig.~\ref{fig:f1}, generically arise as transients after a voltage quench. They then carry an oscillating amplitude with a  frequency resembling the lowest lying excitation gap of the nanostructure. We here report results from time-dependent density matrix renomalisation group (tdDMRG)~\cite{Vidal04,WhiteFeiguin04,Daley-04,Schmitteckert04} simulations showing that in situations where the interaction $U$ is the dominating native scale of the nanoring, these oscillations can be very pronounced and very long-lived. They exhibit a lifetime $\Gamma^{-1}$ that exceeds the transients in transport currents, $\Gamma_0^{-1}$, by orders of magnitude if the repulsive interaction $U$ becomes strong. The computational finding is complemented with perturbative arguments that explain this effect and clarify the relevant physical processes. In essence, the strong suppression of $\Gamma$ originates in a large energy gap between the two low-lying states and the rest of the spectrum on the nanoring (see Fig.~\ref{fig:spectrum}), with ring current connecting the low-lying states. Possible experimental signatures of the effect proposed here are discussed. We note that in contrast to previously discussed~\cite{MedenSchollwoeck03,Molina-03,RejekRamsak03} persistent ring currents driven by magnetic fields, the oscillating ring current we observe is a pure non-equilibrium effect.

\begin{figure}[t] 
\begin{center}
\includegraphics[width=0.4\textwidth]{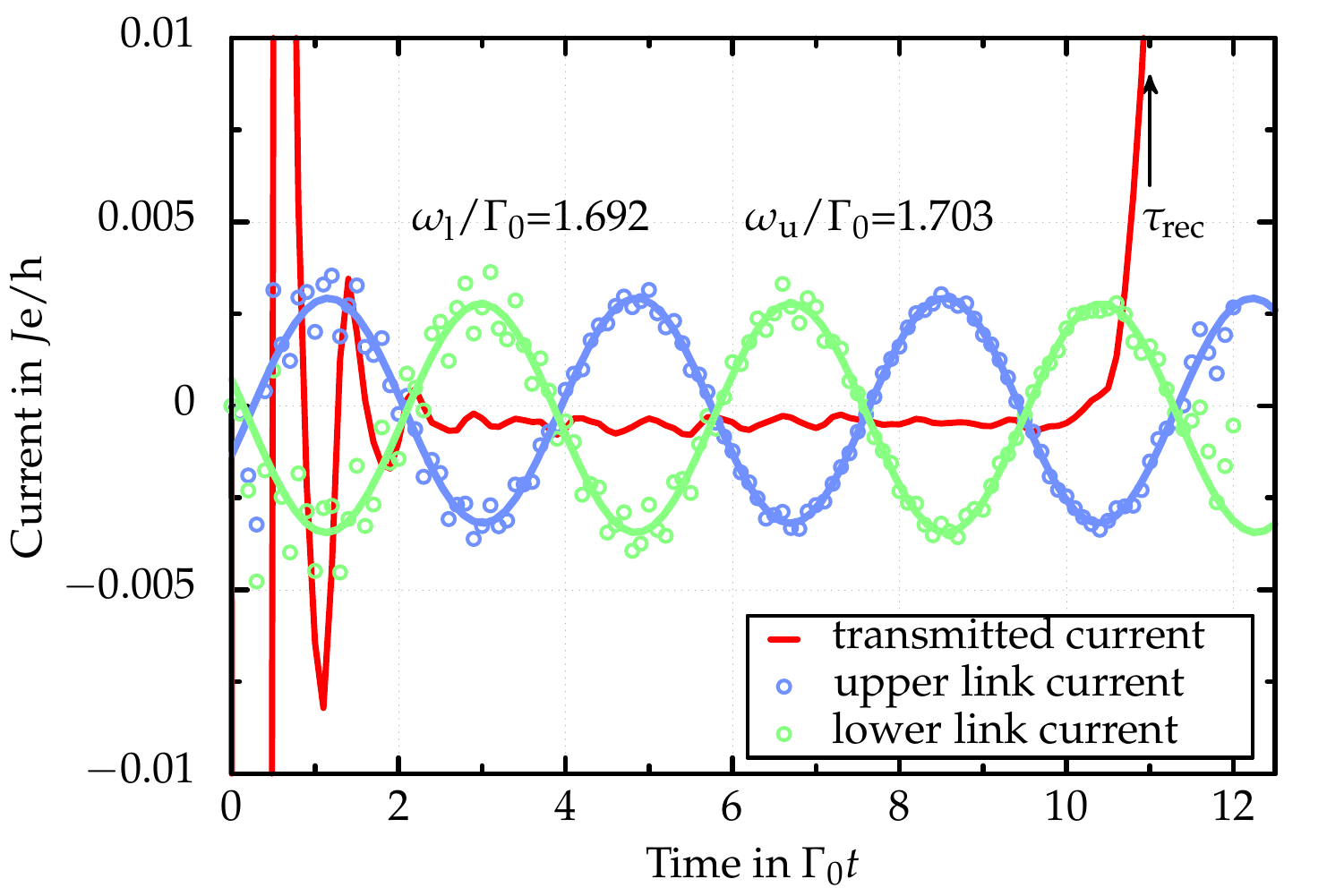}
\caption{Time evolution of the transmitted and ring currents, $\langle I_\text{t} \rangle(t)$ and $\langle I_\text{r} \rangle (t)$, evaluated using tdDMRG on the links $1\to 2$ and $1\to 3$ in Fig.~\ref{fig:f1}. The recurrence time $\tau_{\text{rec}}=L/(2v_\text{F})\simeq 44$ is indicated by the black arrow. While the transmitted current quickly relaxes to a stationary value, the ring currents show persistent oscillations with frequencies $\omega_\text{l,u}$ over the accessible times. The simulation parameters are $L=96$, $U=4J$, $\varepsilon_T=J/2$, $J_\text{c}=J/2$ and $\text{e}V=0.4J$.}
\label{fig:ltime}
\end{center}
\end{figure}
\emph{Nanostructure.}---The model associated with Fig.~\ref{fig:f1} is represented by the Hamiltonian $H=H_\text{r}+H_\text{l}+H_\text{c}$ describing the ring, the leads and their mutual coupling, respectively. The ring Hamiltonian is given by 
\begin{align} 
    H_\text{r}=&-J\sum_{\left\langle i,j\right\rangle}
    \left(d^{\dagger}_{i}d_{j} +d_j^\dagger d_i\right)\nonumber\\
    &+U\sum_{\left\langle i,j\right\rangle}\left(n_{i} n_{j}-\frac{n_i+n_j}{2}\right)
      +\varepsilon_T n_2,
\end{align}
with operators $d^{\dagger}_j$ and $d_j$ creating/annihilating spinless fermions at site $j$ and $n_j=d^{\dagger}_j d_j$ denoting the corresponding density. The first term describes hopping of the fermions between nearest neighbours, while the second represents the repulsive nearest-neighbour interaction. The last term is an external potential at the top site which breaks the symmetry between the upper and lower path through the ring. The lead Hamiltonian reads 
\begin{equation}
   H_\text{l}=-J\sum_{\alpha= \text{L,R}} \sum_{n\ge 0}\bigl(c^{\dagger}_{\alpha,n+1} c_{\alpha,n} + c^{\dagger}_{\alpha,n} c_{\alpha,n+1} \bigr),
   \label{e1} 
\end{equation} 
where $c^{\dagger}_{n,\alpha}$ and $c_{n,\alpha}$ create and annihilate a spinless fermion at site $n$ in the lead $\alpha{=}\text{L,R}$. For simplicity we assume the hopping parameter $J$ in the ring and lead to be equal. Finally, the coupling between both subsystems is facilitated by 
\begin{equation}
  H_\text{c}=-J_\text{c} \left(d^{\dagger}_{1} c_{\text{L},0} + c_{\text{L},0}^\dagger d_1+d^{\dagger}_{4}c_{\text{R},0}+c_{\text{R},0}^\dagger d_4\right),
\end{equation}
coupling the outer sites on the ring to 
the leads.

In the following we analyse the non-equilibrium currents in the nanostructure by three different methods: (i) tdDMRG simulations, (ii) a reduced density-operator transport theory (RDTT)~\cite{Schoeller09,SaptsovWegewijs12}, and (iii) mapping to an effective two-state nanostructure~\cite{Bravyi-11}. 

\begin{figure}[t] 
\begin{center}
\includegraphics[width=0.48\textwidth]{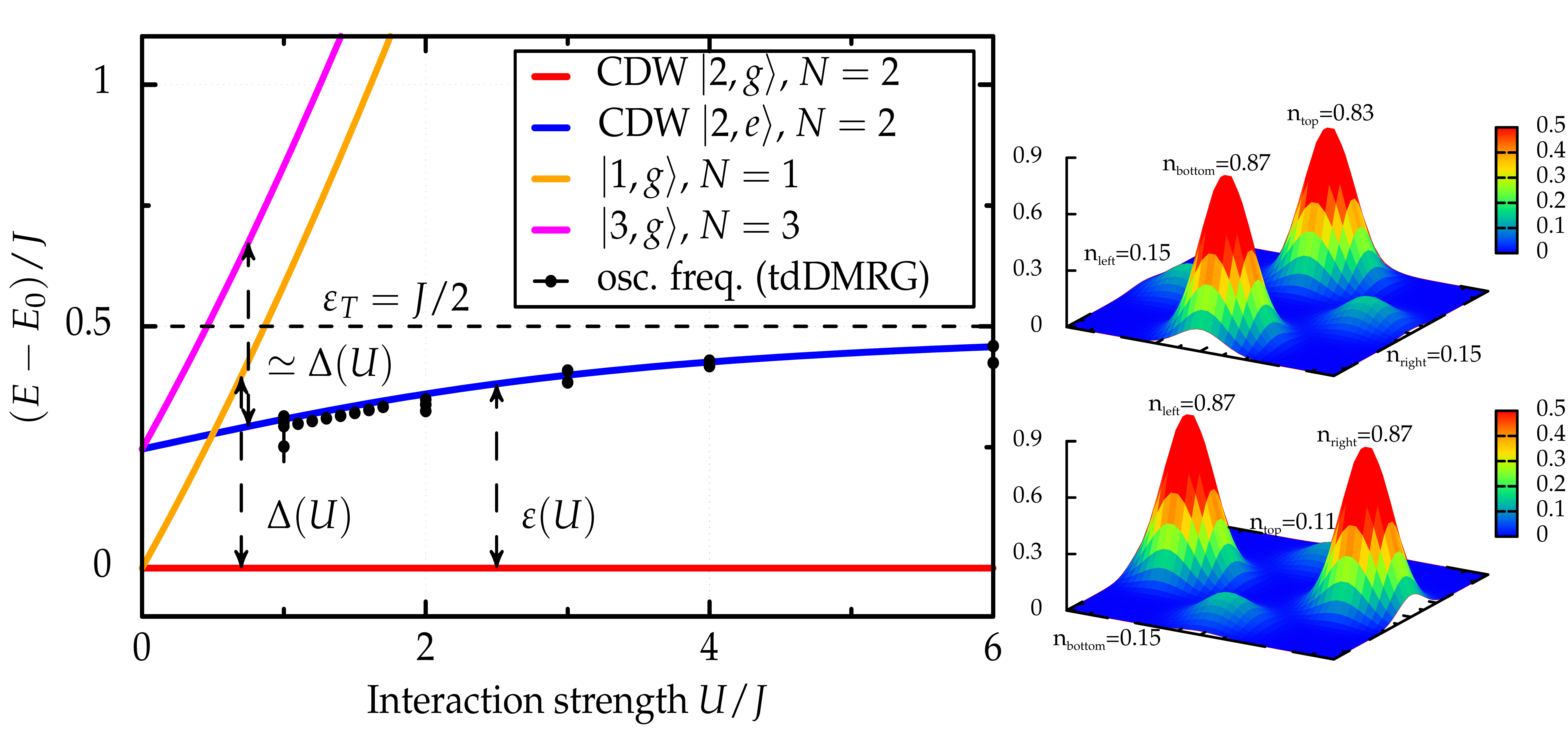}
\caption{Spectrum of the uncoupled ring $H_\text{r}$ relative to the ground-state energy $E_0$. The ground state $\ket{2,g}$ is a CDW state with $N=2$ particles, for $U>J$ the first excited state $\ket{2,e}$ is also a CDW state with two particles. The corresponding particle densities are shown for $U=2J$. The observed oscillation frequencies of the ring currents match the energy difference $\varepsilon(U)$ between these two states. The higher excited states are obtained by adding or removing particles, with $\Delta(U)$ denoting the corresponding energies.}
\label{fig:spectrum}
\end{center}
\end{figure}
\emph{tdDMRG simulations.}---First, we study the time evolution after a voltage quench using the tdDMRG algorithm~\cite{BohrSchmitteckert07,Boulat-08,Kirino-08,DiasdaSilva-08,Heidrich-Meisner-09,Branschadel-10,Schwarz-18}. Specically we use the time evolution scheme outlined in Refs.~\cite{Schmitteckert04,Branschadel-10,supplement} performing the evaluation of the time evolution via matrix exponentials within the framework of Krylov spaces. At times $t<0$ the system is prepared in the ground state of the model with an additional charge excess induced by a stationary gating with $V/2\left(\sum_i n_{\text{L},i} -\sum_i n_{\text{R},i}\right)$. At $t=0$ the gate is switched off, so the electrodes begin to discharge and currents start to flow through the system. We simulate the time evolution with finite leads which are long enough to be able to study the transient regime all the way into the quasi-stationary, non-equilibrium limit. Finite-size effects will interfere only at times exceeding the recurrence time $\tau_\text{rec}{=}L/(2v_\text{F})$, at which the electrons reach the boundary of the leads. (For details of the quenching protocol see Ref.~\cite{Branschadel-10}.) Here $L$ denotes the total number of sites, ie, the length of the leads is given by $(L-4)/2\approx L/2$, and $v_\text{F}=2J$ is the Fermi velocity of the lead electrons.

During the time evolution we determine the expectation values of the local currents $I_\text{t} \propto \text{Im} ( c^{\dagger}_i  c_{i-1})$ and $I_\text{r} \propto \text{Im} ( d^{\dagger}_k d_l)$ flowing in the leads and the impurity, respectively, where $l$ and $k$ are neighbouring sites. The local current densities after quenching are displayed in Fig.~\ref{fig:ltime}. The transport (``transmitted'') current $I_\text{t}$ initially fluctuates in response to the quench for times $\Gamma_0t\leq 3$, where we use $\Gamma_0=2\pi\rho_0J_\text{c}^2$ with the density of states in the leads $\rho_0=1/(2\pi J)$ as our time unit. After this transient the transmitted current appears to have reached a largely time-independent steady state in line with predictions from non-equilibrium Green function formalism~\cite{Jauho-94,Tuovinen-13}.

In contrast, for the local currents in the ring $I_\text{r}$  we observe a drastically different behaviour. Although some transient features decay quickly, the ring currents oscillate with a distinct frequency $\omega$ for long times. In fact, for sufficiently strong Coulomb repulsions $U$ we do not observe a significant reduction of the oscillation amplitude within the observation times accessible to our simulations. Qualitatively similar results were obtained for a ring structure with eight sites~\cite{supplement}.

The frequency of the oscillations can be understood based on the spectrum~\cite{supplement} of the uncoupled ring $H_\text{r}$ shown in Fig.~\ref{fig:spectrum}. We find that the frequency $\omega$ extracted from the tdDMRG simulations matches the energy gap between the two lowest-lying states on the ring. These two states can be identified as charge-density wave (CDW) states with $N=2$ particles on the ring, one being the ground state $\ket{2,g}$ and the other the first excited state $\ket{2,e}$. Thus we confirm that the ring current originates from the mixing of these two states by the time evolution, which is driven by the coupling of the ring to the leads as exemplified by the proportionality of the ring current to the coupling $\Gamma_0$ shown in the inset of Fig.~\ref{fig:comparison}.

\begin{figure}[t]
\begin{center}
\includegraphics[width=0.4\textwidth]{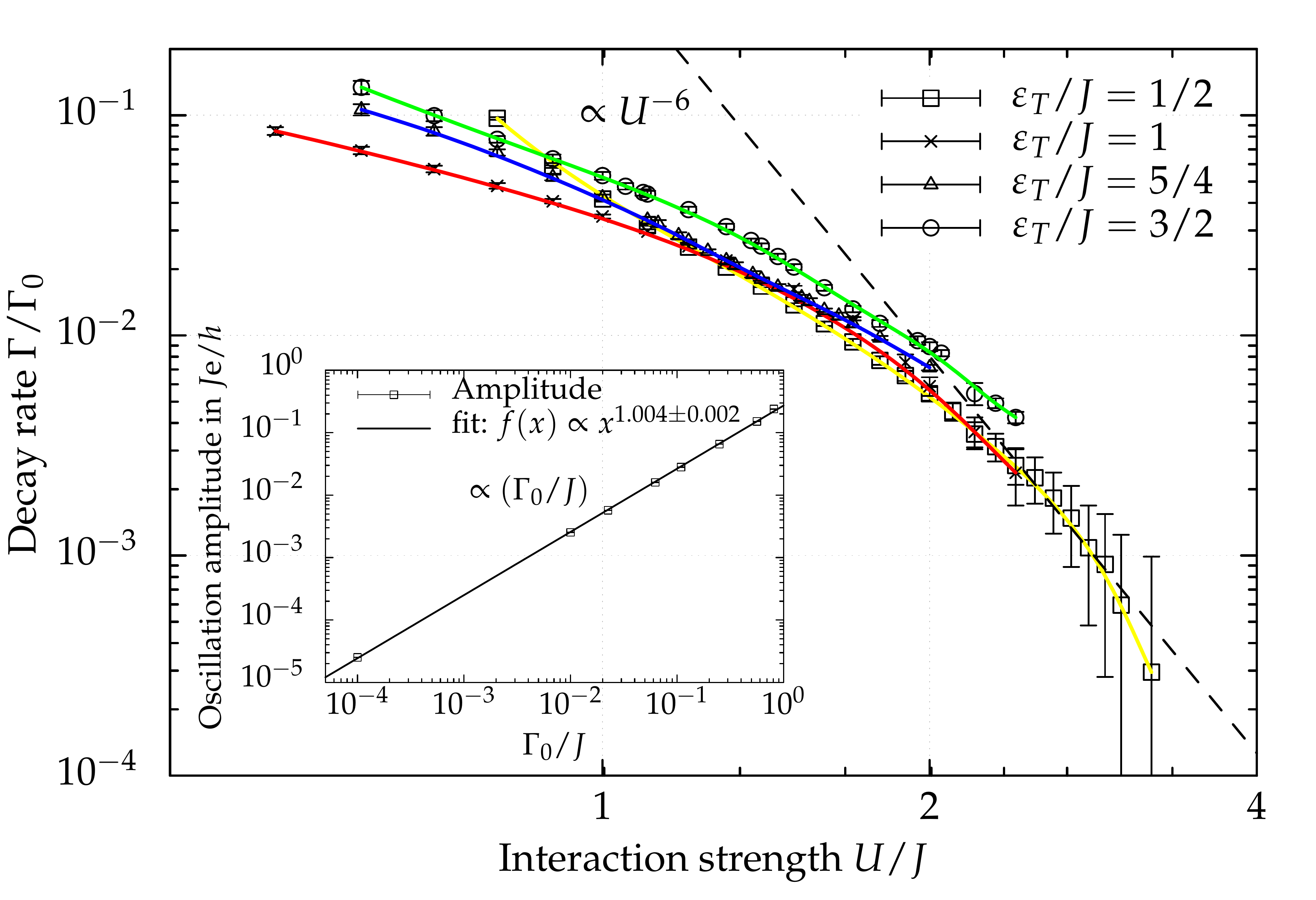}
\caption{Decay rate $\Gamma$ of the ring current extracted from tdDMRG simulations. For $U/\varepsilon_T\simeq 1$ the decay rate appears to be exponentially suppressed in $U$.
For $U\gg\varepsilon_T$ the decay is consistent with $\Gamma\sim U^{-6}$ predicted using an effective two-level system \eqref{eq:SWrate}, as is indicated by the dashed line. All other parameters as in Fig.~\ref{fig:ltime}. Inset: Dependence of the amplitude of the ring current on the coupling $\Gamma_0$ to the leads.}
\label{fig:comparison}
\end{center}
\end{figure}
The decay rate $\Gamma$ of the ring currents is rapidly decreasing with the interaction strength $U$, see Fig.~\ref{fig:comparison}, exhibiting a wide regime with $\Gamma\ll\Gamma_0$. To understand the origin of this regime, we proceed with the RDTT analysis. 

\begin{figure*}
\includegraphics[width=0.98\textwidth]{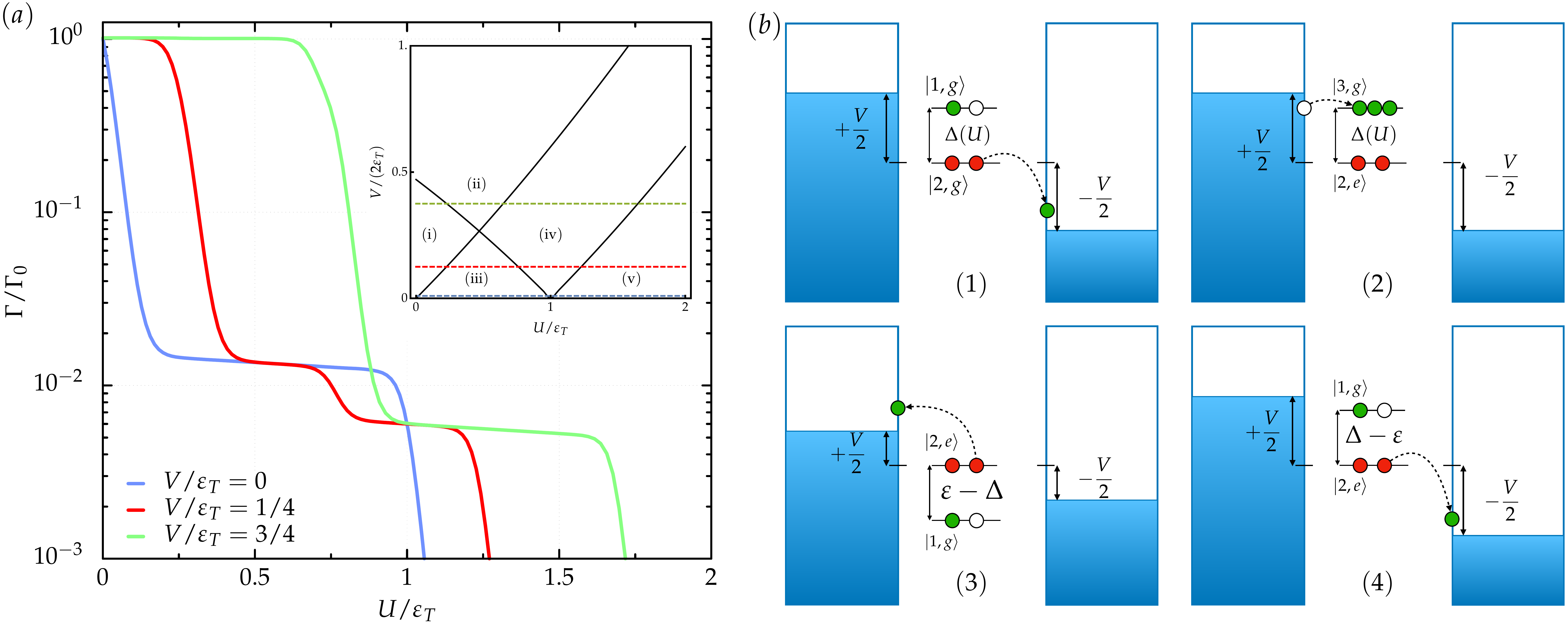}
\caption{(a) Decay rate $\Gamma$ obtained from RDTT for the temperature $T=10\,\Gamma_0$. Inset: In $U$-$V$-parameter space we identify five distinct regions labeled (i) to (v), in which $\Gamma$ takes strongly different values. The dashed lines indicate cuts shown in the main figure. (b) Relaxation processes contributing to the rate $\Gamma$, which result in the distinct regions (i)--(v). Red and green dots represent initial and final configurations, respectively, $\Delta=\Delta(U)$ denotes the energy required to add or remove a particle (see Fig.~\ref{fig:spectrum}), while $\varepsilon\approx\varepsilon_T$ is the energy gap between the two CDW states.}
\label{fig:rtptres}
\end{figure*}
\emph{RDTT analysis.}---The RDTT~\cite{Schoeller09,SaptsovWegewijs12} method aims at determining the time evolution of the reduced density matrix of the nanostructure, $\rho_\text{ns}(t)=\text{tr}_\text{l}\rho(t)$, where the trace is taken over the lead degrees of freedom in the density matrix $\rho(t)$ of the full system. The time evolution of $\rho_\text{ns}(t)$ can be cast in the form $\dot{\rho}_\text{ns}(t)=-\ii L_\text{ns}\rho_\text{ns}(t)$, with the effective Liouvillian $L_\text{ns}$ governing the relaxation of the nanostructure. Since the ring current originates from the mixing of the two CDW states $\ket{2,g}$ and $\ket{2,e}$, its decay is related to the decay of the off-diagonal elements $\rho_{ge}$ and $\rho_{eg}$ of $\rho_\text{ns}$. We have determined the corresponding decay rate from the Liouvillian $L_\text{ns}$ calculated~\cite{supplement} to first order in the bare coupling rate $\Gamma_0$, with the perturbative regime set by $\Gamma_0\ll T$ with the temperature $T$. 

The obtained results for the decay rate $\Gamma$ of the ring current are shown in Fig.~\ref{fig:rtptres}(a). The results are qualitatively similar to the ones obtained via tdDMRG shown in Fig.~\ref{fig:comparison} in the sense that the rate is strongly suppressed at large $U$. The quantitative  differences between the RDTT and tdDMRG results reflect the fact that both methods operate in different parameter regimes. 

Furthermore, the RDTT allows us to identify~\cite{supplement} the relaxation processes contributing to the decay rate, which are visualised in Fig.~\ref{fig:rtptres}(b). The dominant processes are shown in sketches (1) and (2), which involve the tunneling of a particle off or onto the ring, while the sub-leading processes are shown in sketches (3) and (4). All processes are constraint by energetics: (1) and (2) only contribute in the regions (i) and (ii) in Fig.~\ref{fig:rtptres}(a), (3) only in regions (i) and (iii), and (4) is relevant in the regions (i)--(iv). We stress that in region (v) no relaxation processes in order $\Gamma_0$ exist. Thus at sufficiently large interaction strengths $U$ the rate $\Gamma$ essentially drops to zero (to order $\Gamma_0^2$), explaining the very slow decay of the ring current. 

\emph{Schrieffer--Wolff transformation.}---Finally we focus on the regime of strong interactions, $U/\text{max}(\varepsilon_T,J) \rightarrow \infty$, where we can derive the analytic dependence $\Gamma \sim U^{-6}$ consistent with our computational results, Fig.~\ref{fig:comparison}. As can be seen from the spectrum of the bare ring (Fig.~\ref{fig:spectrum}), in this limit the two CDW states $\ket{2,g}$ and $\ket{2,e}$ will be well separated by an energy splitting $\Delta(U)\sim U$ from the higher excited states. It is thus instructive to construct an effective two-level system containing only these states, where the couplings to the higher excited states are treated using a Schrieffer--Wolff transformation~\cite{Bravyi-11} in fourth order in the couplings $J$ and $J_\text{c}$. Going to this order in the expansion is necessary since all off-diagonal matrix elements exactly cancel in second order due to the mirror symmetry of the isolated ring structure~\cite{supplement}.

The resulting two-level system can be written in the form of an electronic Kondo model, with the localised spin identified with the CDW states as $|\!\downarrow\rangle=\ket{2,g}$ and ${|\!\uparrow\rangle}=\ket{2,e}$ and the corresponding spin operator denoted by $\vec{S}$. An effective reservoir electronic degree of freedom can be formed via $c_{\text{res},\uparrow\downarrow}=(c_\text{L}\pm c_\text{R})/\sqrt{2}$ from the leads \eqref{e1} of the original model; the effective spin operator formed from the first sites ($n=0$) is denoted by $\vec{S}_\text{res}$. With this notation the effective model reads~\cite{supplement}
\begin{equation}
\begin{split}
H_\text{SW}=&\sum_{k,\sigma}\epsilon_k c_{\text{res},k\sigma}^\dagger c_{\text{res},k\sigma}+hS^z+\tilde{h}S_\text{res}^z\\
&+J_\perp\bigl(S^xS_\text{res}^x+S^yS_\text{res}^y\bigr)+J_zS^zS_\text{res}^z,
\end{split}
\label{eq:Kondo}
\end{equation}
where the first term is the energy of the electronic reservoir, the second and third are effective magnetic fields $h\approx\varepsilon_T$ and $\tilde{h}=\mathcal{O}(U^{-4})\ll h$ acting on the two-level system and spin of the electron reservoir, and the fourth and fifth term represent a Kondo coupling between the two, with the coupling being strongly anisotropic with $J_z\simeq 10 J^2J_\text{c}^2/U^3$ and $J_\perp=\mathcal{O}(U^{-5})$, and thus $\vert J_{\perp}\vert\ll \vert J_z\vert \ll J_\text{c}, J$.

Due to the formation of the effective reservoir electron spin from the leads $\text{L,R}$ the bias voltage $V$ enters the effective Kondo model in the form of a transverse field in the reservoir, ie, as $V/2 \sum_{k\sigma\sigma'}c_{\text{res},k\sigma}\tau^x_{\sigma\sigma'}c_{\text{res},k\sigma'}$ with $\tau^x$ being the x-component of the Pauli matrices. Finally, the ring current corresponds to oscillations between the two CDW states and thus is related to the localised spin via $I_\text{r}\sim S^y$. Performing a suitable spin rotation in the electronic reservoir we calculated~\cite{supplement} the corresponding relaxation rate using standard perturbation theory in the Kondo system~\cite{Schoeller09,Rosch-03prl,SchoellerReininghaus09} with the result 
\begin{equation}
\Gamma=\frac{\pi\rho_0^2 J_\perp^2}{16}\bigl(|\varepsilon_T+V|+|\varepsilon_T-V|+2|\varepsilon_T|\bigr)+\frac{\pi\rho_0^2 J_z^2}{8}V.
\label{eq:SWrate}
\end{equation}
We stress that in the considered regime of strong interactions this rate is vanishingly small, $\Gamma\sim \rho_0^2J_z^2V\sim \rho_0^2J^4J_\text{c}^4V/U^6$, in accordance with our finding of long-lived oscillations in the ring current. In particular, the predicted behaviour $\Gamma\sim U^{-6}$ is consistent with our tdDMRG simulations shown in Fig.~\ref{fig:comparison}. We note that the result \eqref{eq:SWrate} is applicable deep in region (v) of Fig.~\ref{fig:rtptres}(a), where we found that processes of order $\Gamma_0$ vanish. Furthermore, the effective model \eqref{eq:Kondo} will show the Kondo effect, however, the relevant energy scale $T_\text{K}$ will be much smaller than the energy scales we consider here, in particular $T_\text{K}\ll\varepsilon_T$. Thus the equilibrium Kondo effect is not observable in our setup. 

Finally we note that nanostructures with two energetically well separated low-lying states can generically be approximated by an effective Kondo model using a Schrieffer--Wolff transformation. In the absence of the above mentioned mirror symmetry the exchange couplings will be of the order $J_z,J_\perp\sim J_\text{c}^2/U\ll J_\text{c}$, resulting in a relaxation rate $\Gamma\sim \rho_0^2J_{z}^2V\sim U^{-2}\ll \Gamma_0$. Thus ring currents that couple to these low-lying states are still expected to decay very slowly.

\emph{Experimental verification.}---We see a possible experimental realisation of the ring-shaped model system, Fig.~\ref{fig:f1}, in molecules such as porphyrines or phthalocyanines. Single molecule conductance measurements have indeed been performed at these systems~\cite{Sedghi-11,Schmaus-11,Bagrets-12} so the possibility for bias-ramping has also been demonstrated already. As an observable indicating the slow decay of the ring currents we propose to measure the photons that are emitted when these currents decay via coupling to the radiation field. In this context we note that single-molecule electroluminscence measurements have been performed~\cite{Marquardt-10,Reecht-14} already and thus are indeed experimentally feasible. An alternative realisation of our ring-shaped model may be provided by quantum dot arrays~\cite{Mukhopadhyay-18}, which in particular offer a high level of control of the couplings and allow to enter the regime of strong interactions essential for the long-lived ring currents.

\emph{Conclusion.}---We have studied the relaxation of transport processes in an interacting ring-shaped nanostructure.  Owing to a mirror symmetry of the Hamiltonian, the system supports oscillating ring currents long after the transmitted current has died out, with the ratio $\Gamma/\Gamma_0$ of the respective relaxation rates being strongly suppressed by the interactions. Our work provides a striking example for an untypical situation in thermodynamic relaxation processes: Two observable currents approach their equilibrium values on timescales that are parametrically separated with rates differing by orders of magnitude. In addition, our system provides new insight into the field of quantum devices as we show that internal oscillations can be longer-lived than observed in currents through the system. While we have focused on a ring-shaped nanostructure, the appearance of the suppressed relaxation rate $\Gamma$ is generally expected in systems that can be effectively described by a two-level model with the ring current connecting the low-lying states.

We thank Theo Costi, Mikhail Pletyukhov and Peter W\"olfle for useful discussions. This work is part of the D-ITP consortium, a program of the Netherlands Organisation for Scientific Research (NWO) that is funded by the Dutch Ministry of Education, Culture and Science (OCW). BMS and PS thank the HPC project QWHISTLE at the Steinbuch Centre of Computing at Karls\-ruhe Institute of Technology (KIT). PS was supported by ERC-StG-Thomale-TOPOLECTRICS-336012. FE thanks the DFG for support under grant  EV30/08-1 and SFB 1277 project A03. BMS and DS were supported by the Netherlands Organisation for Scientific Research (NWO) under FOM 14PR3168.


\newpage
\phantom{O}
\newpage
\onecolumngrid
\setcounter{figure}{0}
\setcounter{page}{0}

\thispagestyle{empty}
\begin{center}
{\large\bf Supplementary material for\\[5mm]
Long-lived circulating currents in strongly correlated nanorings}\\[15mm]
{B. M. Schoenauer, N. M. Gergs, P. Schmitteckert, F. Evers, D. Schuricht}
\end{center}

\newpage
\section{Exact diagonalization of the decoupled ring impurity}

\begin{figure}
  \centering
  \includegraphics[width=0.68\textwidth]{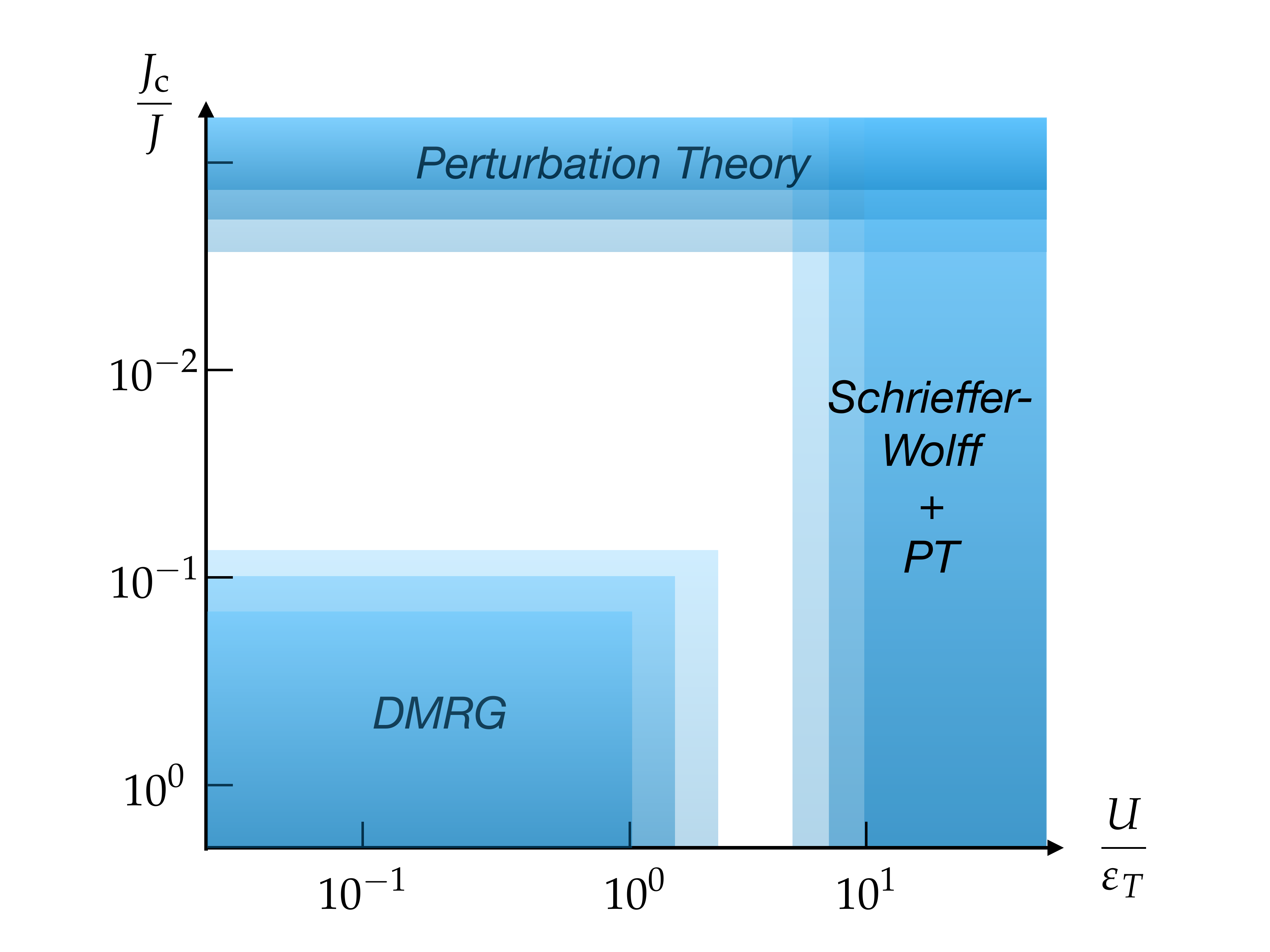}
  \caption{Parameter ranges $U/\varepsilon_T$ and $J_{\text{c}}/J$ in which our
    employed methods are applicable. For
  the DMRG time evolutions we require a coupling $J_{\text{c}}/J$ between leads and impurity
which is large enough to allow relaxation to the nonequilibrium steady
state within the maximum simulation time $L/(2v_F)$. The coupling $J_{\text{c}}/J$
also needs to be
larger than the typical level splitting $2\pi J/L$. The range of
interaction strengths for our DMRG method is restricted by the limitations
of our fitting procedure. For large enough interaction strength the
fitting error exceeds the value of the fitted decay rate. The
perturbation theory (more precisely reduced density-operator transport theory) is perturbative in $\rho_0 J_{\text{c}}^2/T$ and therefore
requires small $J_{\text{c}}/J \ll 1$ to be valid. The Schrieffer-Wolff transformation is
perturbative in $J^2 J_{\text{c}}^2 / U^3$. It is thus only valid in the regime
$U\gg J \simeq \varepsilon_T$.}
\end{figure}

\paragraph{Spectrum and particle densities}

\begin{figure}
\begin{center}
\includegraphics[width=0.68\textwidth]{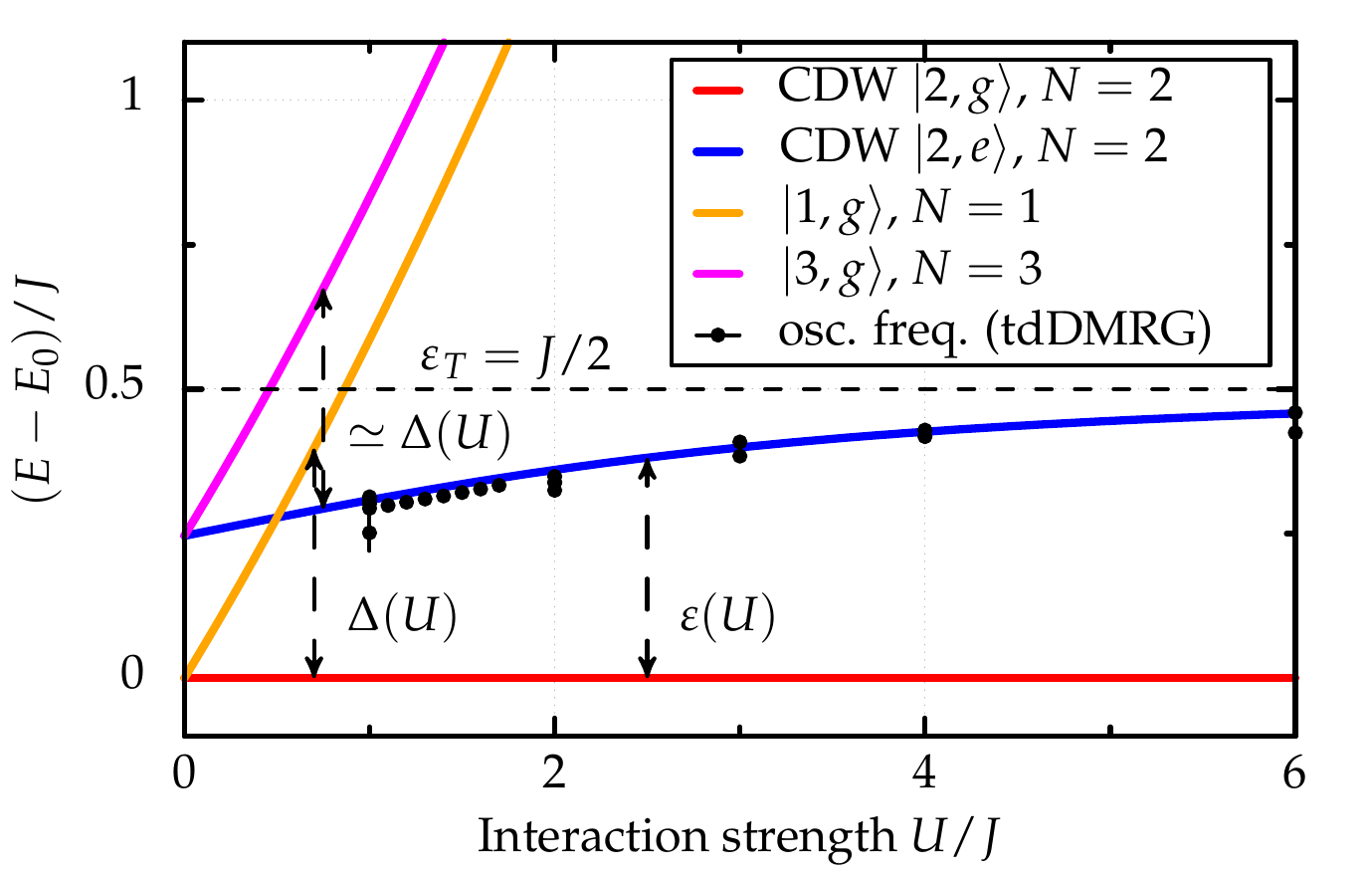}
\caption{Relative, low-energy spectrum of the bare ring impurity as a
  function of the interaction strength $U/J$ in the repulsive regime $U>0$. The 
red line indicates the ground state energy $E_0$. The blue line shows the
energy of the excited charge density wave (CDW) state. $\varepsilon_T=J/2$ denotes 
the applied gate potential. The points indicate the values obtained
within
DMRG calculations for the oscillation frequency of the local currents inside the ring 
impurity.}
\end{center}
\label{fig:spectrum}
\end{figure}

\begin{figure}
  \centering
  \small{(a)}
  \includegraphics[width=0.46\textwidth]{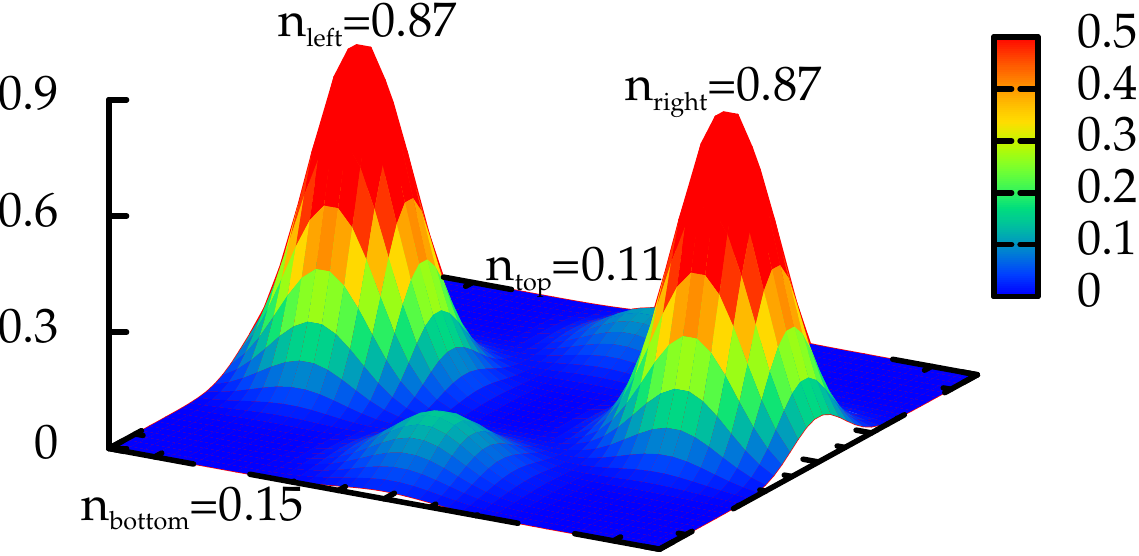}
  \small{(b)}
  \includegraphics[width=0.46\textwidth]{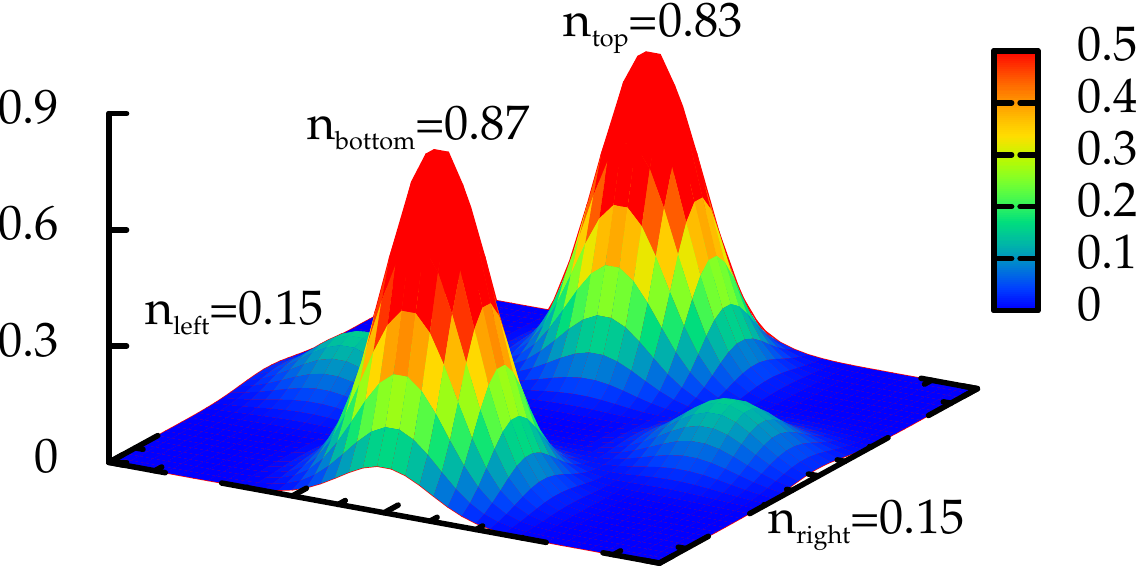}\\
  \small{(c)}
  \includegraphics[width=0.46\textwidth]{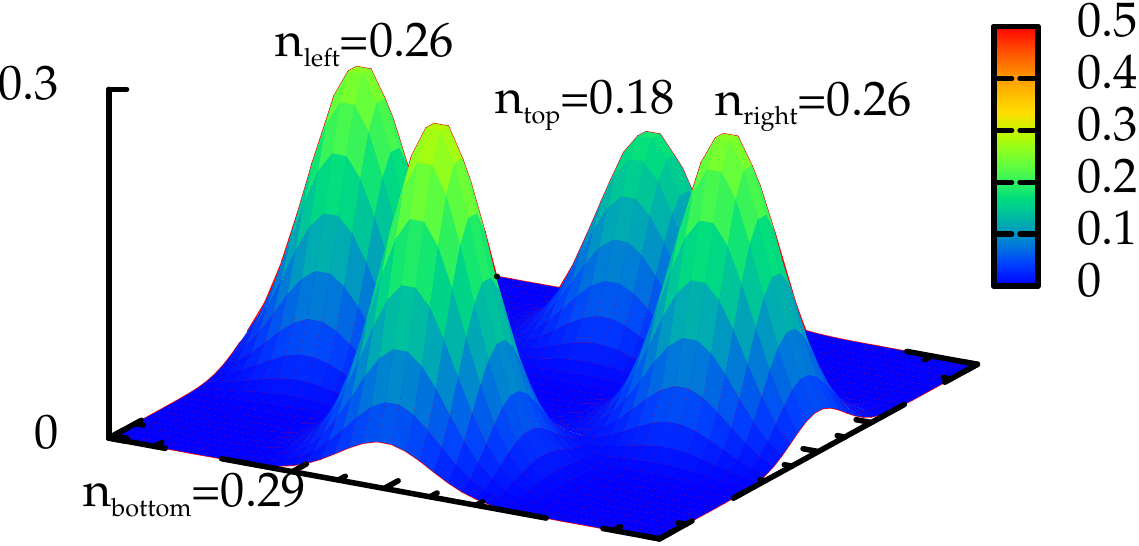}
  \small{(d)}
  \includegraphics[width=0.46\textwidth]{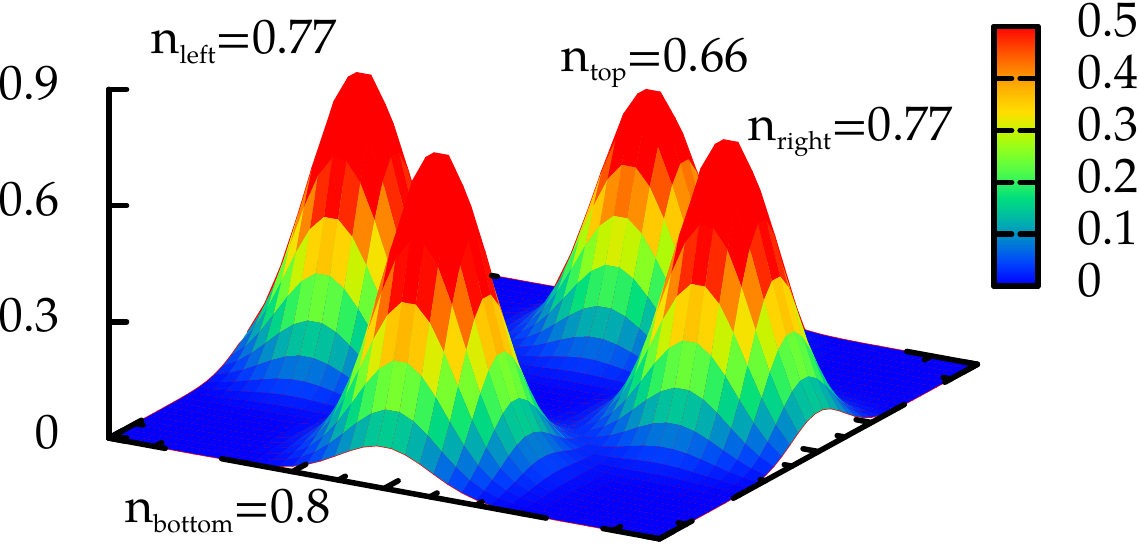}
  \caption{Local electron density on the four lattice sites in the ring for
  $U/J=2$ and $\varepsilon_T/J=0.5$. (a) Density for $\vert
2,g\rangle$. (b) Density for $\vert 2,e\rangle$. We find the
characteristics of charge density waves for (a) and (b). (c) Density for
$\vert 1,g\rangle$. (d) Density for $\vert 3,g\rangle$.}
  \label{fig:particledensity}
\end{figure}
We have performed an exact numerical diagonalization of the Hamiltonian
matrix $H_{\text{r}} (U,\varepsilon_T,J)$ of the ring impurity in the absence of the leads. In figure 2 we plot the relative spectrum $(E-E_0)$ for the for
eigenstates with the lowest energy. The energy of these states is shown as a
function of the interaction strength $U$ and a gate potential
$\varepsilon_T = J/2$. The ground state features half-filling of the
ring ($n=2$) and is indicated by the red line. The other eigenstate in
the spectrum with half-filling is shown as the blue line. The
state marked by the orange line features only a single electron in the
ring while the state indicated by the magenta line has three electrons
in the ring. For interaction $U/\varepsilon_T \geq 1$ we observe an
increasing energy separation between the two 
eigenstates at half-filling and the rest of the spectrum. When comparing
the frequency of the observed oscillations of the local currents in the
ring with the relative spectrum of the ring, we find an excellent
agreement of the frequencies with the energy gap between the ground
state $\vert 2,g\rangle$ and the second
eigenstate at half-filling $\vert 2,e\rangle$. The frequencies that we
have obtained from the fit of a cosine function to the data of the ring
current are displayed as black dots in figure 2. We show the local electron density on the ring sites for the
four low energy eigenstates in figure 2. We find that the two eigenstates at half-filling exhibit
characteristics of charge density waves. The ground state has a
significantly increased electron density on site $1$ and $4$ of the
ring, while the excited state features an increased density on sites $2$
and $3$. The other two states have a more evenly distributed electron
density. We will therefore refor to the states $\vert 2,g\rangle$ and
$\vert 2,e\rangle$ as charge density wave (CDW) states from now on.
\paragraph{Time evolution of an initial superposition}
We have performed DMRG calculations of the time-dependent reduced
density matrix of the ring impurity. We find finite occupation
probabilities for both CDW states at time $t=0$. With increasing bias
voltage, the occupation probability of the excited CDW state tends to
grow as well. We have used these occupation probabilities from the DMRG
to construct an initial pure state 
\begin{align}
  \vert \psi_0\rangle = \sqrt{N}\left(\sqrt{\rho_{gg}}\vert 2,g \rangle \pm
 \sqrt{\rho_{ee}} \vert 2,e\rangle\right)\,,
 \label{eq:psi0}
\end{align}
where $\sqrt{N}$ is a normalization factor, $\rho_{gg}$ refers to the ground state occupation probability and
$\rho_{ee}$ to the occupation probability of the excited CDW state.
Using exact diagonalization we
then perform the time evolution of this initial state in the bare ring
impurity as
\begin{align}
  \vert \psi (t) \rangle = \exp(-i H_{\text{r}} t) \vert \psi_0 \rangle\,,
  \label{}
\end{align}
and calculate the expectation values $\langle I_{\text{u}} \rangle$ and $\langle
I_{\text{l}} \rangle$ of the local currents in the ring.
The results of this calculation are in good agreement with our DMRG
results in both amplitude and frequency.

\section{DMRG}
\paragraph{DMRG implementation}
For our numerical calculation of the time evolution of the complete
system including ring impurity and leads we have employed a typical
finite lattice Density
Matrix Renormalization Group (DMRG) algorithm. We keep a maximum of
$N_{\text{cut}}=2800$
states per block and set the maximum amount of discarded entanglement entropy to
$\delta S_{\text{max}}=10^{-7}$ in each DMRG step. We use a Krylov subspace method to
calculate the matrix exponential, allowing us to chose larger time steps
up to $\Delta t$ of order one. Each state that is reached through
application of the matrix exponential onto the initial state $\vert
\psi_0 \rangle$ is included
into the density matrix from which we determine the subspace of the
Hilbert that we project onto in each DMRG step. At each time step we
measure the observables of interest as $\langle \psi(t) \vert
\mathcal{O} \vert \psi(t)\rangle$ where the operator $\mathcal{O}$ has
also been projected onto the retained subspace of the Hilbert space. 
\paragraph{Quench protocol}
At time $t=0$ we prepare the system in the ground state of
\begin{align}
H(t=0)=H+\frac{V}{2}\left(\sum_i n_{\text{L},i} -\sum_i n_{\text{R},i}\right)\,,
\end{align}
and perform the time evolution using $H(t>0)=H$. We simulate time
evolution up
$t \leq L/2 v_F$, where $L$ is the 
length of the chain (usually $L\geq 72$) and $v_F=2 J$ is the Fermi
velocity of the fermions in the leads. During the time evolution we
measure the expectation value of the local currents in the leads as
\begin{align}
  I_{\text{t}} &= -2 e J \left( c^{\dagger}_i c_{i-1}-\text{h.c.}\right)\,,
\end{align}
and on specific bonds $1\to 2$ and $1\to 3$ in the ring (see Fig. 1 in the main paper) as
\begin{align}
I_{\text{u}} &= - e J \left( d^{\dagger}_2 d_1-\text{h.c.}\right),\\
I_{\text{l}} &= - e J \left( d^{\dagger}_3 d_1-\text{h.c.}\right)\,.
\end{align}
For the majority of our calculations we have used a set of default parameters, namely
$L=72$, $\varepsilon_T=J/2$, $J_{\text{c}}=J/2$ and $V=0.4\,J/e$.
\paragraph{Detailed discussion of the DMRG time evolution results}
\begin{figure}[]
  \centering
  \small{(a)}
  \includegraphics[width=0.46\textwidth]{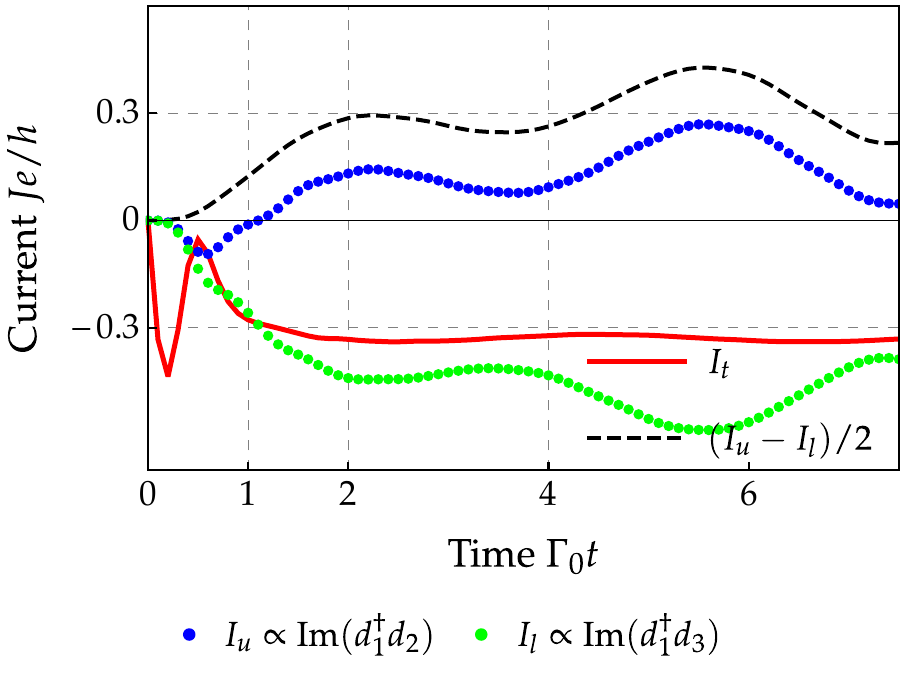}
  \small{(b)}
  \includegraphics[width=0.46\textwidth]{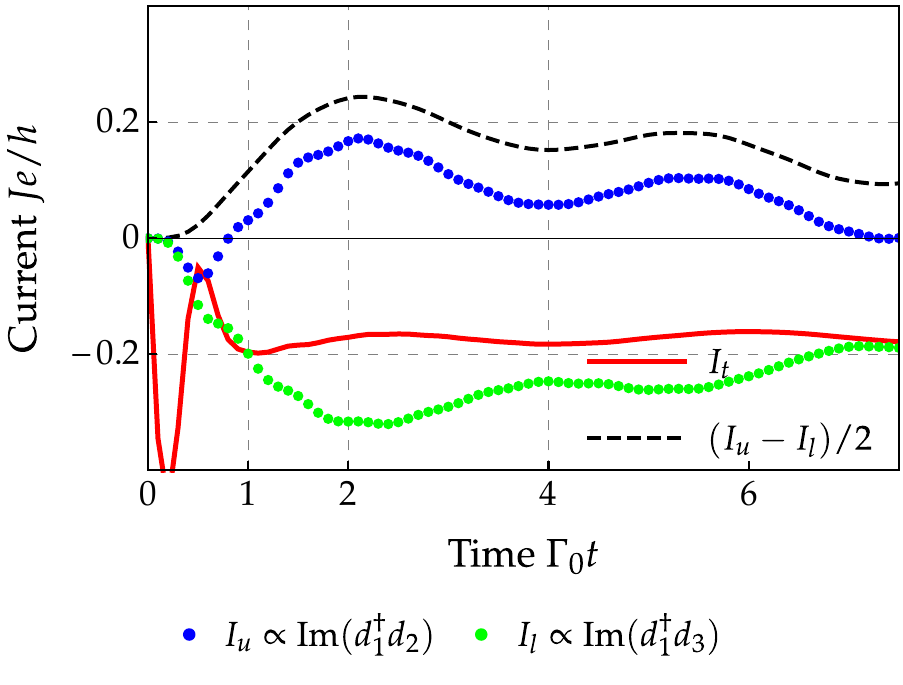}
  \small{(c)}
  \includegraphics[width=0.46\textwidth]{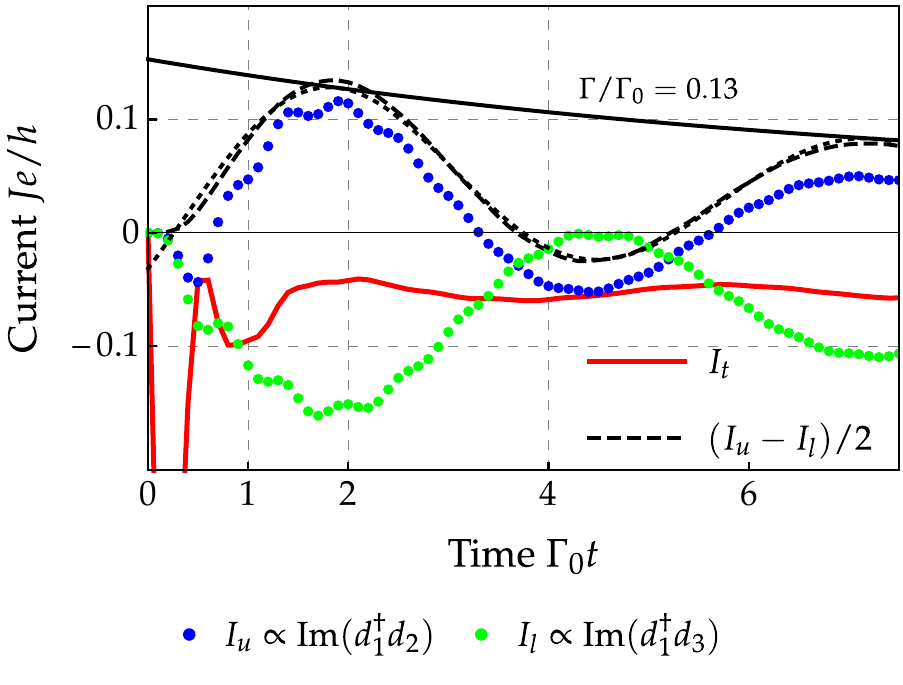}
  \small{(d)}
  \includegraphics[width=0.46\textwidth]{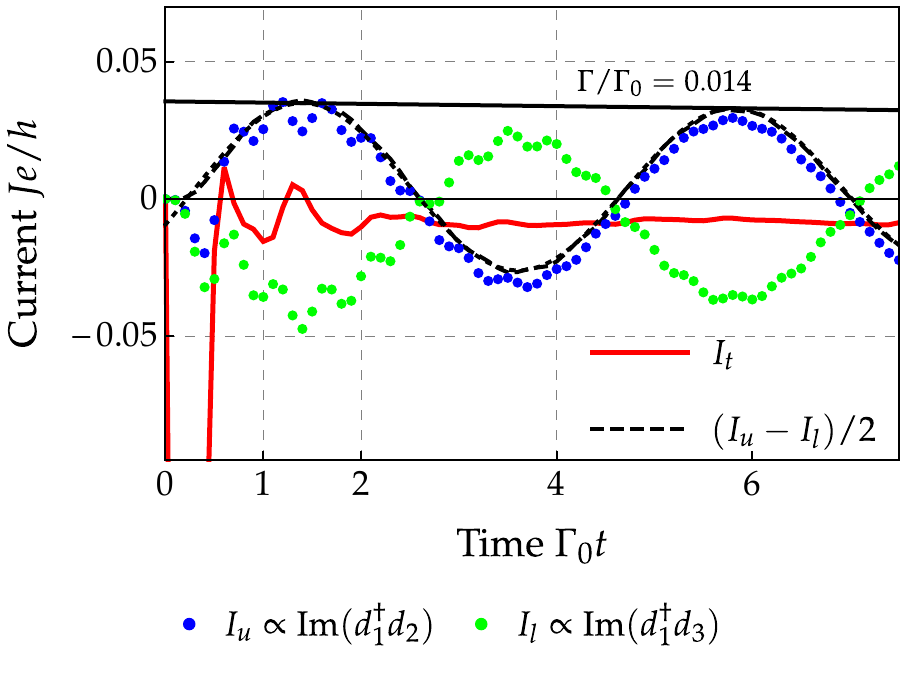}
  \caption{Time dependent currents calculated within DMRG. The red line
  denotes the transport (``transmitted'') current $I_{\text{t}}$. Blue dots mark the upper link
current $I_{\text{u}}$ and green dots the lower link current $I_{\text{l}}$. (a):
$U/J=0.1$, $\varepsilon_T/J=0.5$, $eV/J=0.4$, (b): $U/J=0.5$,
$\varepsilon_T/J=0.5$, $eV/J=0.4$, (c): $U/J=1.0$, $\varepsilon_T/J=0.5$,
$eV/J=0.4$, (d): $U/J=2.0$, $\varepsilon_T/J=0.5$, $eV/J=0.4$. The solid
black lines indicate a fit function $f(\Gamma_0 t) \propto \exp(\Gamma t)$.}
  \label{fig:tdcurr}
\end{figure}
In figure \ref{fig:tdcurr} we plot the time-dependent expectation values of the
operators $I_{\text{t}}$, $I_{\text{u}}$ and $I_{\text{l}}$ using our default parameters and
interaction strengths $U/J\in \left\lbrace
0.1,0.5,1.0,2.0\right\rbrace$. We begin by discussing the results for
weak interaction $U/J=0.1$ shown in figure \ref{fig:tdcurr} (a). For the transmitted current $\langle I_{\text{t}} \rangle (t)$ we observe
significant initial oscillations inside the typical transient regime
$\Gamma_0 t\leq 1$ that appear to have decayed for $\Gamma_0 t >1$
while a weak periodic oscillation remains even for large times. This
periodic oscillation is not physical but a known finite size effect with a frequency
$\omega\equiv V$. For the local currents in the ring we first verify that
$I_{\text{u}}+I_{\text{l}} = I_{\text{t}}$ as a consistency check of our results. For times
$\Gamma_0 t \leq 1$ we find the oscillations of $\langle I_{\text{u}} \rangle
(t)$ and $\langle I_{\text{l}} \rangle (t)$ small when compared to the
oscillations of $\langle I_{\text{t}} \rangle (t)$. The finite size effect with
$\omega = V$ for
the the local currents in the ring on the other hand is large when
compared to the transmitted current. We also indicate $(\langle I_{\text{u}}
\rangle -\langle I_{\text{l}}\rangle)/2$ as a
dashed black line in fig. \ref{fig:tdcurr}. This observable corresponds to a ring current in
clockwise direction. For interaction strength $U/J=0.5$, shown in fig.
\ref{fig:tdcurr} (b), we solely
observe quantitative differences for $\langle I_{\text{t}} \rangle (t)$. While
the initial transient features remain largely unchanged, the steady
state current for $\Gamma_0 t >1$ is reduced. For $(\langle I_{\text{u}} \rangle
- \langle I_{\text{l}} \rangle)/2$ we observe what seems to be an initial 
oscillatory feature that is not due to finite size effect for $\Gamma_0
t < 4$. Due to the small window $1 \leq \Gamma_0 t \leq 4$ a fit
does not yield reliable results for frequency and decay rate. For
$U/J=1$ the steady state value of the transmitted current experiences
yet another significant reduction, whereas the transient features
remain of similar size as for $U/J=0.1$. We stil observe that the
transient features of the transmitted current have largely decayed by
$\Gamma_0 t =1$. For the ring currents we find a qualitatively different
behavior. The ring current exhibits periodic oscillations with a distinct
frequency and a visible decay rate $\Gamma$ which is an order of magnitude smaller than
$\Gamma_0$. For the directional ring current $(\langle I_{\text{u}} \rangle -
\langle I_{\text{l}} \rangle)/2$ there is even a window in which the direction
of the current has changed. By increasing the interaction strength to
$U/J=2$ we find yet another decrease of the steady state trnamsitted
current. In the transient regime $\Gamma_0 t \leq 1$ we now also observe
an additional sign change of the transmitted current. We also no longer
see the oscillations due to the finite system size. The oscillations of
the local currents in the ring $I_{\text{u}}$ and $I_{\text{l}}$ become even more
pronounced and feature a periodic change of direction. Through a fit we find that the decay rate of these
oscillations is an order of magnitude smaller than in the case $U/J=1$
and now amounts to $\Gamma/\Gamma_0 \approx 1/100$. There is
a clear separation of scales between the typical decay rate
$\Gamma_0$ which holds for the transmitted current and the decay rate
$\Gamma$ of the local currents in the ring impurity.
Calculations for stronger interaction $U/J>2$ show a continuation of
this trend.

\paragraph{Fitting procedure for the computation of
$\Gamma$}
To determine the oscillation frequency $\varepsilon$ and decay rate
$\Gamma$ we fit a function
\begin{align}
  f(t)=a\exp(-\Gamma t)\cos(\varepsilon t + b)+c\,,
  \label{}
\end{align}
to our DMRG data for the local currents where $\Gamma$, $\varepsilon$, $a$, $b$
and $c$ are fitting parameters. The fit is performed for $\Gamma_0 < t <
L/2 v_F$. This fitting procedure only yields reliable results for $0.5 <
U/J \leq 5$. For weak interaction $U/J \leq 0.5$ the decay time is too
short to observe the amount of sine waves necessary to reliably
determine the decay rate. For very strong interaction the decay rate
becomes so small that it does not lead to a visible reduction in
oscillation amplitude for $t < L/2 v_F$. As a result, the fitting error
associated with decay rate becomes larger than the decay rate itself.
These limitation of the fitting procedure limit the application of our
DMRG method as a tool to determine the decay rate $\Gamma$
to a parameter range $0.5 < U/J \leq 5$ as indicated in figure
\ref{fig:comparison}.

\subsection{DMRG calculations for the decay rate $\Gamma$}
We have performed a set of DMRG calculations to study
the behavior of the decay rate $\Gamma$ as a function 
of $U/J$ for a range of specifically chosen parameters $U$,
$\varepsilon_T$ and $V$. The results of these calculations are shown in
figures \ref{fig:comparison} (a) and (b). Due to 
the aforementioned limitations of our fitting procedure
it is not possible to quantify
$\Gamma$ for 
$0.5< U/J \leq 5$. In the vicinity of $U/\varepsilon_T\simeq 1$ a
comparison of the log-linear and log-log plots indicates a small region of
exponential suppression. 
For stronger interactions $U/\varepsilon_T > 1$ we observe a power law
behavior of the 
decay rates as a function of $U/J$. The fit of a power law to the data
indicates a smaller exponent for smaller values of $\varepsilon_T$.
In the case of $\epsilon_T=0.5$ we are safely in the regime
$U/\epsilon_T \gg 1$ for $U/J \geq 4$. In this regime one could consider the data
comparable to results obtained in the limit $U/\varepsilon_T \rightarrow
\infty$. The fit of a power law finds an exponent $\alpha=6.0\pm 0.4$ in
this case.
\begin{figure}
\begin{center}
  \small{(a)}
\includegraphics[width=0.46\textwidth]{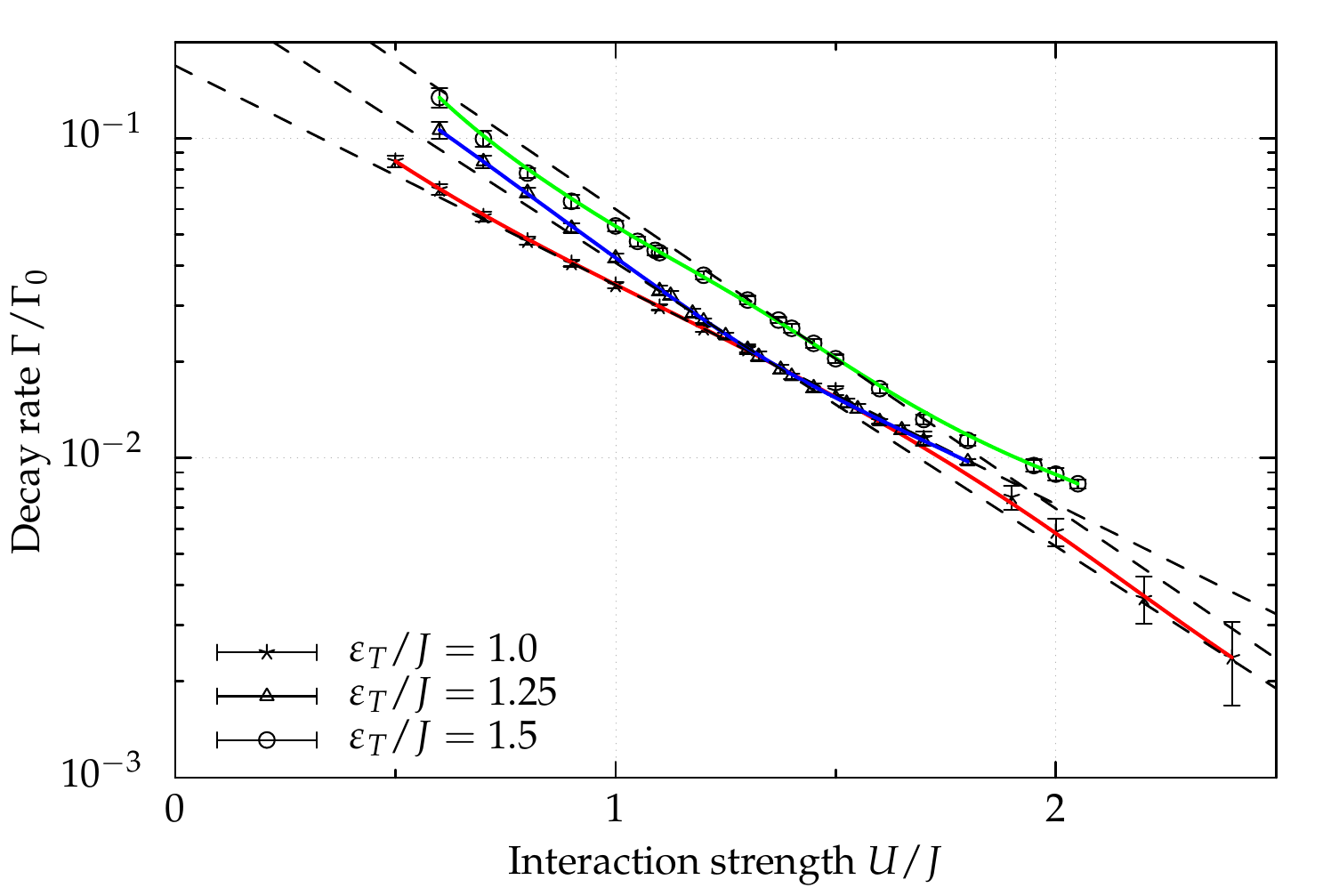}
\small{(b)}
\includegraphics[width=0.46\textwidth]{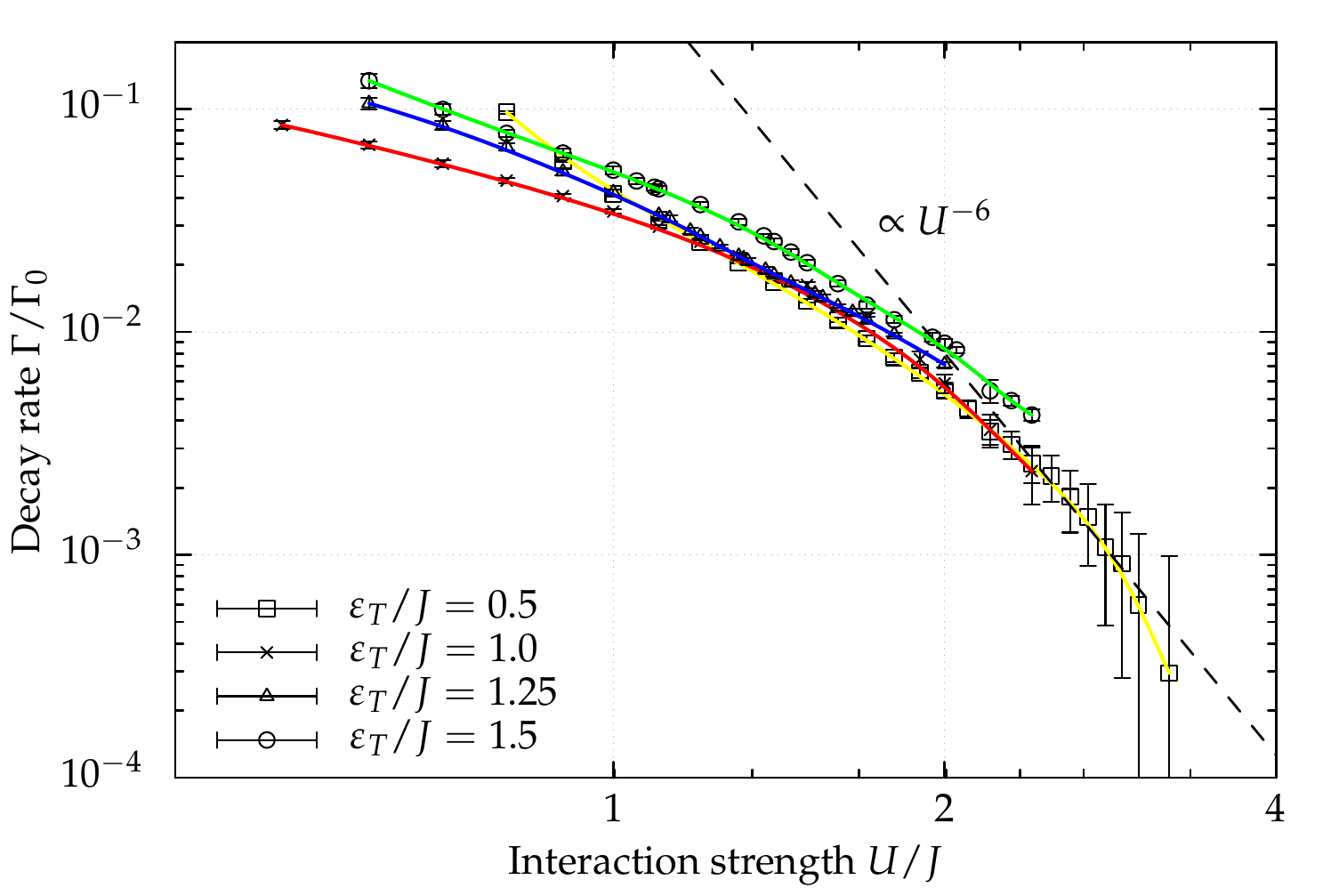}
\caption{Decay rate of the oscillating ring current obtained within DMRG calculations for several values
of $\varepsilon_T$ on a log-linear (a) and a log-log scale (b). We find that for
$U\simeq\varepsilon$ the decay rate appears to be exponential in $U$
whereas
for $U\gg \varepsilon_T$ the decay rate exhibits an algebraic behavior.
We plot a power law $f(U) \propto U^{-6}$ in (b) for comparison.}
\label{fig:comparison}
\end{center}
\end{figure}

\subsection{DMRG calculations for an eight-site ring}
\begin{figure}
\begin{center}
  \small{(a)}
\includegraphics[width=0.46\textwidth]{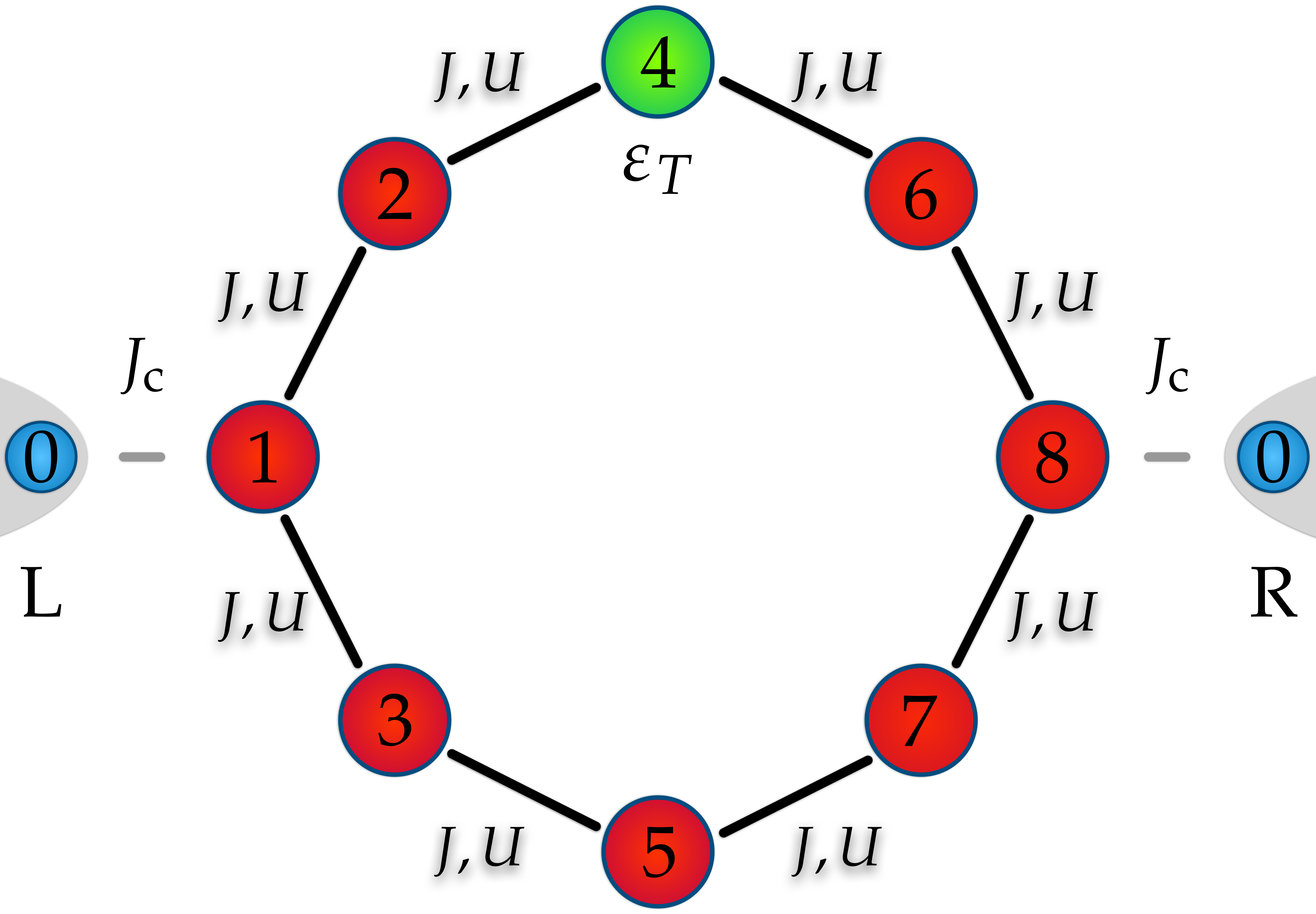}
\small{(b)}
\includegraphics[width=0.46\textwidth]{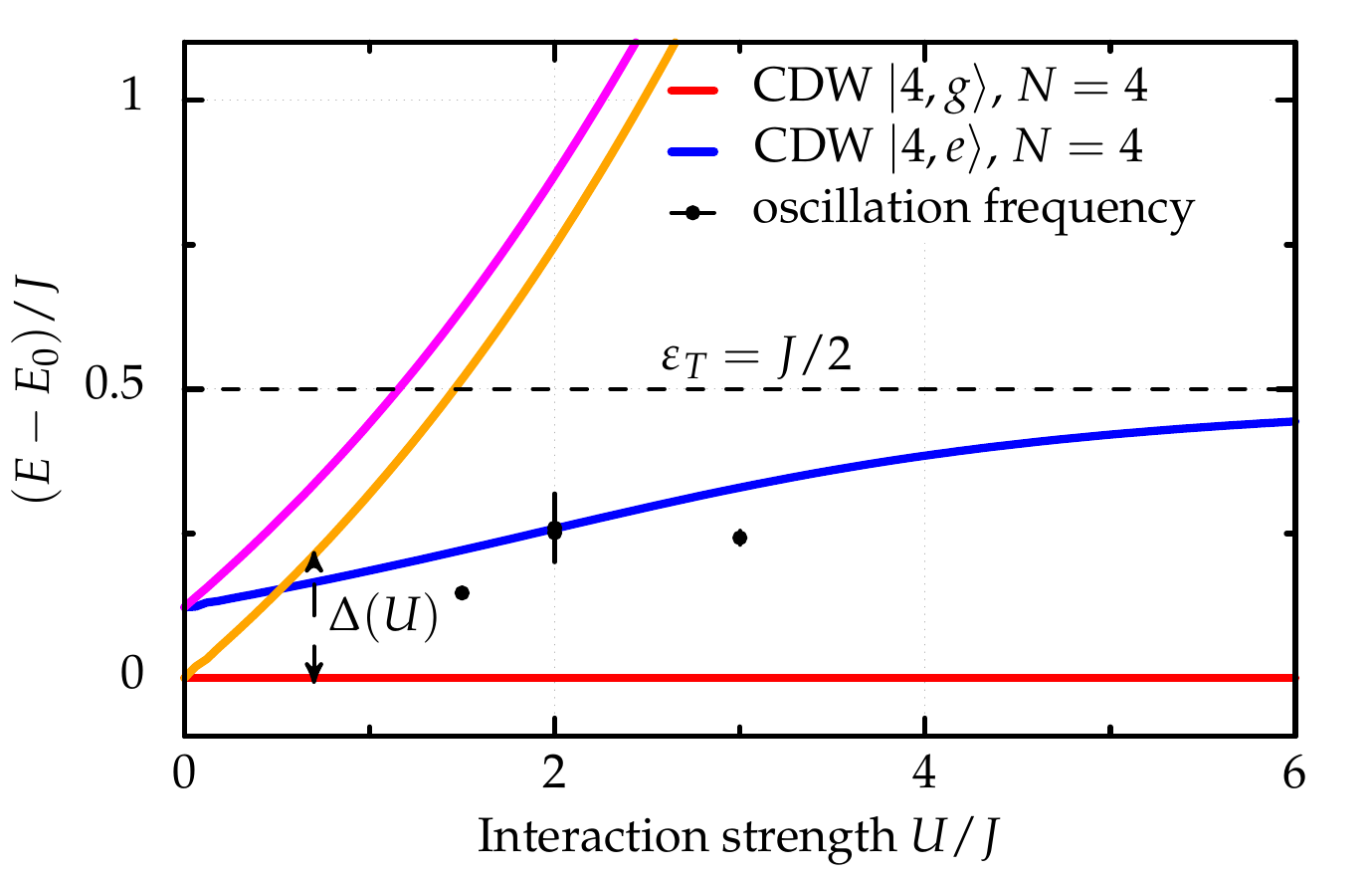}
\caption{(a) Schematic representation of the ring consisting of eight sites.
A gate potential $\varepsilon_T$ is applied to the site in green $j=4$.
Between neighboring sites inside the ring there is a hopping amplitude
$J$ and a repulsive fermion-fermion interaction $U$. The ring is
connected to two tight-binding leads with an amplitude $J_{\text{c}}$.
(b) Low-energy spectrum of the eight site ring disconnected from the
leads as a function of the interaction strength $U$. The black circles
indicate the fitted values for the oscillation frequency for the current
data obtained with td-DMRG. As can be seen in figure~\ref{fig:8site-current}, 
the fitting is accurate only for intermediate values $U=2J$ of the interaction strength, since at weak interactions $U=3J/2$ only one oscillation cycle has finished in the available time window, while for stronger interactions $U=3J$ additional fast modulations appear.}
\label{fig:8site}
\end{center}
\end{figure}
As a test of generality of the ring current oscillations, we have
performed additional calculations for an asymmetric ring consisting
of eight lattice sites.
The corresponding Hamiltonian reads
\begin{equation}
  H_{\text{r},8}= -J\sum_{\langle i,j \rangle} \left(d^{\dagger}_i d_j +
  \text{h.c.}\right) + U \sum_{\langle i,j \rangle}\left(n_i n_j -
  \frac{n_i + n_j}{2}\right) + \varepsilon_T n_4\,,
\end{equation}
and
\begin{align}
  H_{\text{c}} = J_{\text{c}} \left( d^{\dagger}_1 c_{\text{L},0} +
  d^{\dagger}_8 c_{\text{R},0} + \text{h.c.} \right)\,,
\end{align}
where $\langle i,j \rangle$ again denotes neighboring sites, and the
gate potential $\varepsilon_T$ is now applied to the site with index $j=4$. A sketch of
the ring is shown in figure~\ref{fig:8site} (a). We plot the low-energy spectrum of
the uncoupled ring, which was obtained by means of exact
diagonalization, as a function of the interaction strength $U/J$ in
figure.~\ref{fig:8site} (b).
Once again, we find a large separation in energy between the two lowest
eigenstates ($\vert 4,g \rangle$ and $\vert 4,e\rangle$) and the
remainder of the spectrum for $U/\varepsilon_T \gg 1$. The particular eigenstates
again correspond to CDWs at half-filling, namely $N=4$. 
We have performed several td-DMRG calculations in the same fashion as
for the four site ring. We have chosen parameters for interaction
strength $U$ and coupling $J_{\text{c}}$ as well as lead sizes
$\ell_{\text{L},\text{R}}=(L-8)/2$, for which we have previously
observed slowly decaying ring current oscillations in the four site
ring.
\begin{figure}
\begin{center}
  \small{(a)}
\includegraphics[width=0.45\textwidth]{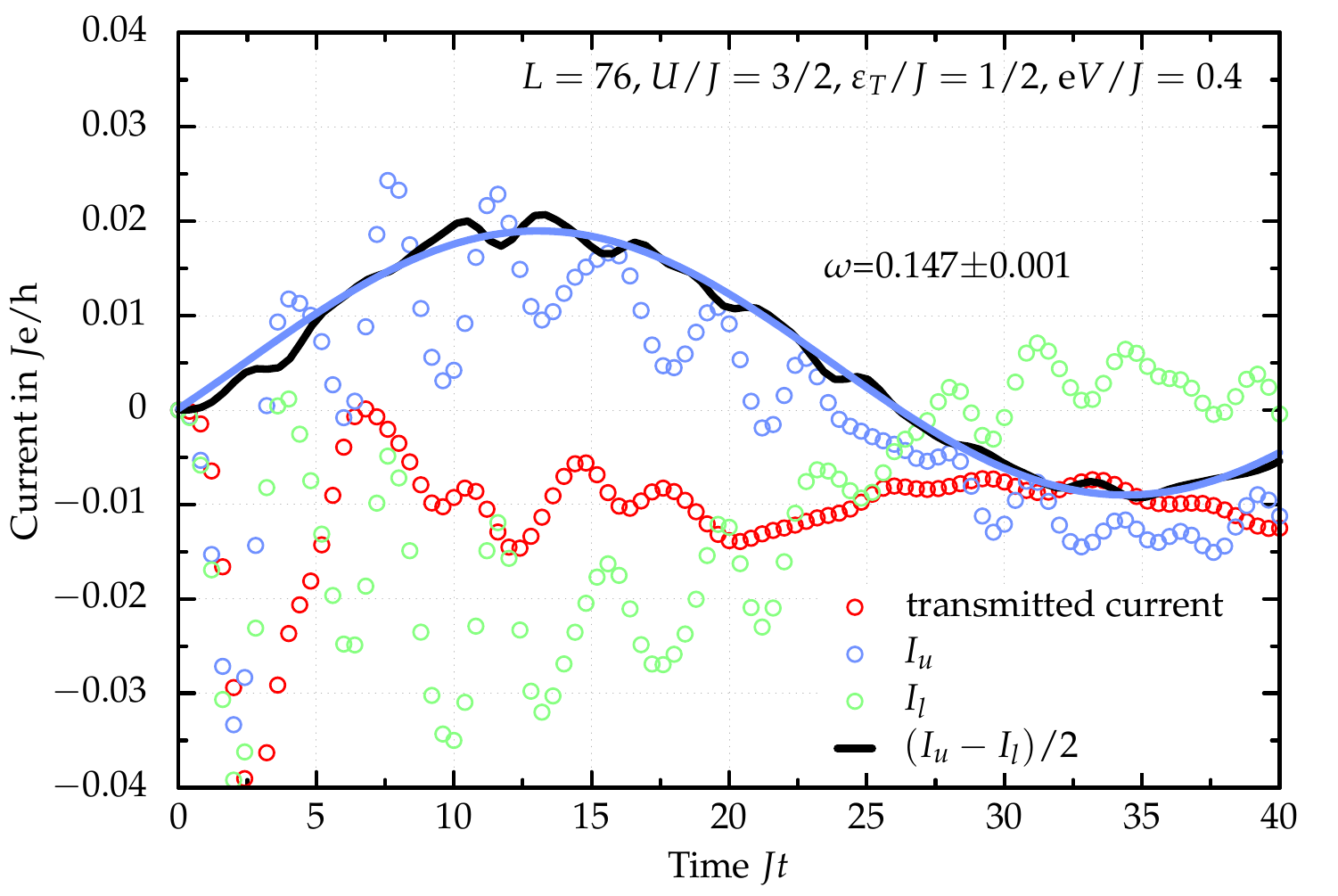}\vspace{5mm}
\small{(b)}
\includegraphics[width=0.45\textwidth]{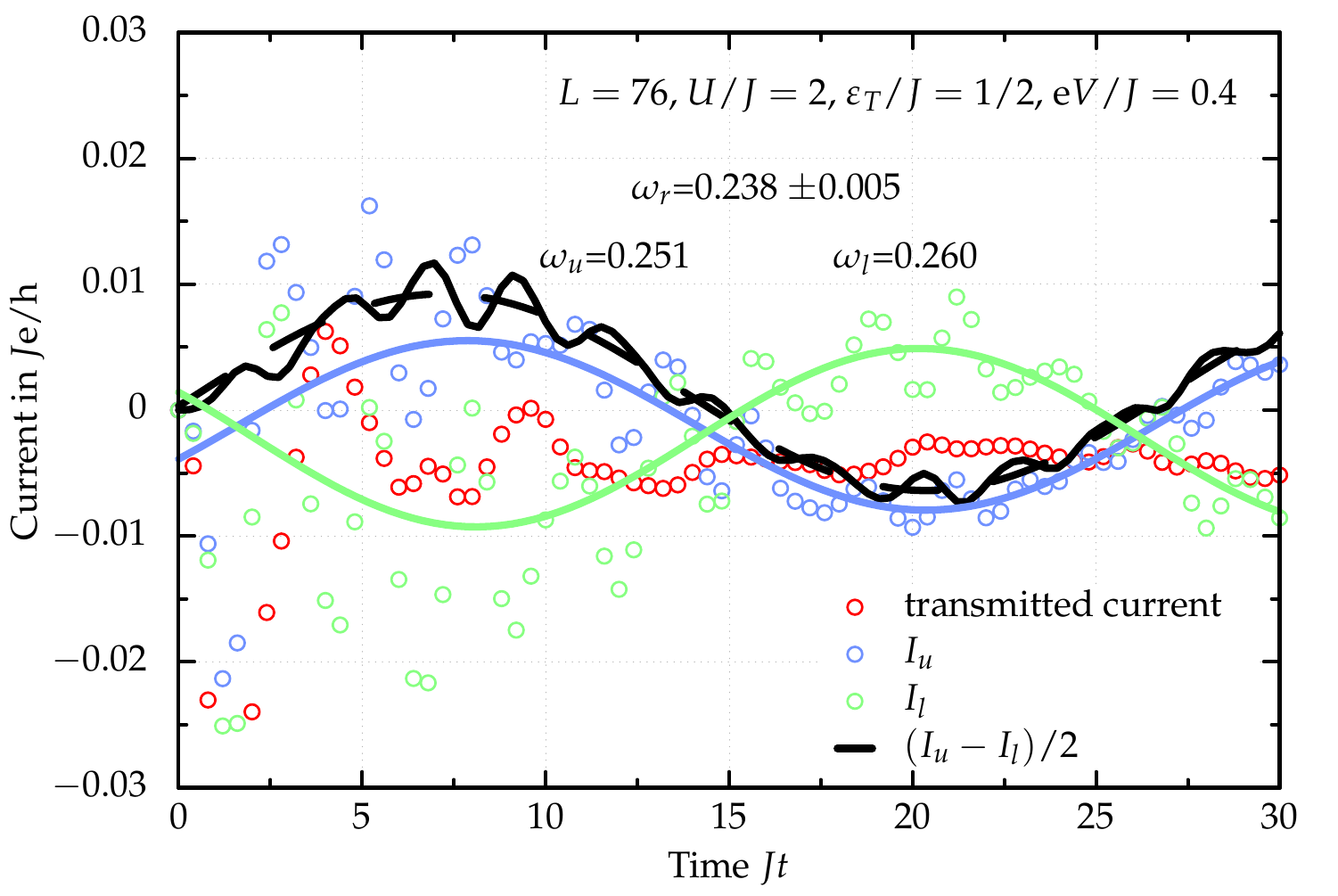}
\small{(c)}
\includegraphics[width=0.45\textwidth]{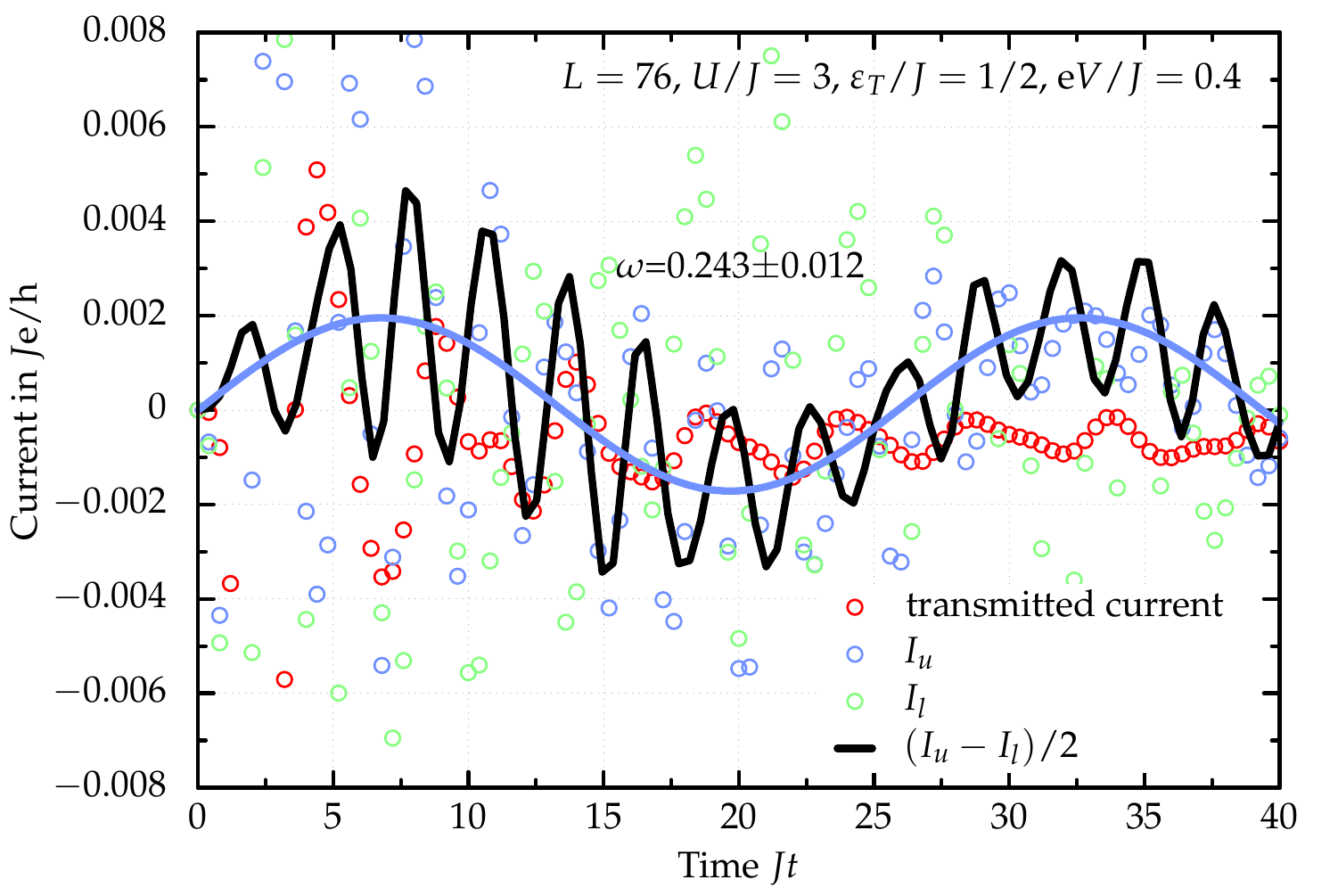}
\caption{td-DMRG data for the transmitted and the local currents in the
ring for a system with $L=76$ lattice sites, a ring-lead coupling
$J_{\text{c}}/J=0.5$, a gate potential $\varepsilon_T/J=0.5$. A bias
voltage $\text{e}V/J=0.4$ was applied to the leads at $t=0$. The
repulsive nearest neighbor interaction is (a) $U/J=3/2$, (b) $U/J=2$ and
(c) $U/J=3$. For all values of the interaction strength $U$, we observe
a slowly decaying oscillation of the ring current $I_r = (I_u -I_l)/2$ with
a frequency of the order of the energy gap between ground state $\vert
4,g \rangle$ and first excited state $\vert 4,e\rangle$ at half-filling
$N=4$.}
\label{fig:8site-current}
\end{center}
\end{figure}
We show the results for the time-dependent currents that have been measured in
the td-DMRG calculations in figure~\ref{fig:8site-current}. In each calculation we observe an
overall increase in transient features, both for the transmitted current and the ring
current $I_{\text{r}}=(I_u -I_l)/2$, where $I_u =
2\text{e}\,\text{Im}(d^{\dagger}_4 d_2)$ and $I_l =
2\text{e}\,\text{Im}(d^{\dagger}_5 d_3)$. In case of the ring current, these transient features are modulated on top of a single
dominant oscillation. These modulations increase the difficulty of
fitting an exponentially decaying cosine function to the data, such that
the obtained values are less reliable than in the case of the four site
ring. In 
figure~\ref{fig:8site-current} (a) we find a noticable decay $\Gamma/J\simeq 1/100$ of the ring current oscillation
within the simulation time for $U/J=3/2$. For interaction strength
$U/J=2$ (see figure~\ref{fig:8site-current} (b)), the decay rate
diminishes further to $\Gamma/J\simeq 1/200$.
In the case $U/J=3$, shown in figure~\ref{fig:8site-current} (c), other transient features of higher frequency have
become significantly more pronounced. One can nevertheless still observe the
underlying oscillation of frequency $\omega \simeq E_{4,e}-E_{4,g}$.
A fit of an exponential decay to the data is however no longer feasible due to
the other transient features. We show the fitted values for the
oscillation frequencies
in figure~\ref{fig:8site} (b). For $U/J=2$ we find very good agreement.
The deviation for the other values of the interaction strength, can be explained
with the deficiencies of the fitting procedure.

\section{Perturbation theory in the limit of small hybridization}
For our computation of the decay rate $\Gamma$, which is
associated with the oscillation of the local currents in the ring, we
make use of the reduced density-operator transport theory (RDTT). We mainly follow Schoeller, 
Eur. Phys.~J. Special Topics \textbf{168}, 179 (2009).
The RDTT approach is exact in the
Hilbert space $\mathcal{H}_{\text{r}}$ of the impurity and perturbative in the hybridization
between impurity and reservoirs. It may be applied if the
associated perturbative scale $\Gamma_0 = 2\pi \rho_0 J_{\text{c}}^2$ satisfies
$\Gamma_0 \ll T$, where $T$ denotes the system temperature.
The RDTT
determines the time-dependent reduced density-matrix $\rho_{\text{ns}}(t)$
of an impurity by calculating corrections to the Liouvillian $L$ of the
impurity caused by tunneling processes between impurity and leads.
The Liouvillian can be understood as a \textit{superoperator} that
corresponds to the action of the commutator
between the Hamiltonian $H$ and a second operator $A\in \mathcal{H}$,
\begin{align}
  LA\equiv \left[H,A\right]_{-}\,.
  \label{def_liou}
\end{align}
The von Neumann equation, which governs the time-evolution
of the density matrix $\rho$, can be written in terms of the
Liouvillian as
\begin{align}
  \dot{\rho}(t)=-i\left[ H,\rho (t)\right]_{-}=-iL \rho (t)\,,
  \label{vonNeumann}
\end{align}
and is in turn solved by
\begin{align}
  \rho (t) = \exp\left[-i L(t-t_0)\right]\rho (t_0)\,.
  \label{eq:solLio}
\end{align}
A Laplace transform and a subsequent trace over the reservoir degrees of
freedom of solution (\ref{eq:solLio}) yields the expression
\begin{align}
  \tilde{\rho}_{\text{ns}}(E)=& \text{tr}_{\text{l}}\int_{t_0}^{\infty}
  dt\,\exp\left[i(E-L)(t-t_0)\right]\rho(t_0)\nonumber\\
  =&
  \text{tr}_{\text{l}}\frac{i}{E-L_{\text{l}}-L_{\text{ns}}-L_V}\rho_{\text{l}}
  \rho_{\text{ns}} (t_0)\,,  \label{eq:def_resolv}
\end{align}
where $L_{\text{ns}}$ denotes the original Liouvillian of the impurity
and $L_{\text{l}}$ the Liouvillian of
the reservoirs, i.e., the total Liouvillian is decomposed as $L=L_\text{l}+L_\text{ns}+L_V$. Expression (\ref{eq:def_resolv}) can conveniently be expanded in powers of $L_V$, the contribution
to the Liouvillian
containing the coupling between the impurity and the reservoirs. The
resulting series expansion for $\tilde{\rho}_{\text{ns}}$ reads 
\begin{align}
  \tilde{\rho}_{\text{ns}}(E)=i\sum\text{tr}_{\text{l}}
  \frac{1}{E-L_{\text{l}}-L_{\text{ns}}}L_V\dots
  L_V\frac{1}{E-L_{\text{l}}-L_{\text{ns}}}\rho_{\text{l}}
  \rho_{\text{ns}}(t_0)\,.
\end{align}
In the limit $J_{\text{c}}^2\ll T$ we can set up a perturbation theory in
$L_V$. The effective Liouvillian $L_{\text{eff}}$ of the impurity then obtains
perturbative corrections $\Sigma(E)$ that are functions of the Laplace variable $E$.
It reads
\begin{align}
  L_{\text{eff}}(E)=L_{\text{ns}}+\Sigma(E)\,.
\end{align}
The transient features of the reduced density matrix $\rho_{\text{ns}}(t)$ are encoded in non-zero poles of
\begin{align}
  \frac{i}{E-L_{\text{eff}}(E)}\,.
  \label{}
\end{align}
To obtain these poles we solve for the complex roots of
\begin{align}
 z-L_{\text{eff}}(z)\,.
 \label{eq:Leff}
\end{align}
The Laplace variables $z^{*}_{\pm}$, that are roots of equation
(\ref{eq:Leff}), have a real part that
corresponds to an oscillation frequency $\varepsilon$ of the
associated transient feature and an imaginary 
part denoting its decay rate $\Gamma$. Our aim is to compute
the particular $\Gamma$ of the transient features whose frequency
coincide with the frequency $\varepsilon$ of the oscillation of the currents in the
ring shaped impurity.
\paragraph{Properties of the Liouville space}
In order to represent the Liouvillians $L_{\text{ns}}$ and $L_V$ as well as other
superoperators $G$ as matrices we introduce a new vector space $\mathcal{L}$
that we refer to as Liouville space. Objects that act as matrices in the
Hilbert space $\mathcal{H}_{\text{r}}$ of the impurity can be thought of as
vectors in this Liouville space $\mathcal{L}$. The most relevant example
of such an object is the reduced density matrix $\rho_{\text{ns}}$ of the
impurity. Each matrix element $(\rho_{\text{ns}})_{i,j}=\vert i \rangle \langle j \vert$ of
$\rho_{\text{ns}}$ corresponds to a basis vector $\vert m )$ of the Liouville
space $\mathcal{L}$. We
will subsequently denote vectors in $\mathcal{H}_{\text{r}}$ as $\vert i \rangle$
and vectors in $\mathcal{L}$ as $\vert j )$. To represent each element
of an operator $O\in \mathcal{H}_{\text{r}}$ as a basis vector of $\mathcal{L}$,
the size of the vector space $\mathcal{L}$ has to be chosen such that
$\text{dim}(\mathcal{L})=\text{dim}(\mathcal{H}_{\text{r}})^2$.
\paragraph{Definition of the superoperators}
The coupling Liovilliain $L_V$, which can be interpreted as the
interaction vertex of the perturbation theory, induces charge
fluctuations on the impurity. It has the form
\begin{align}
  L_V = G_{1}^{p_1}:J_1^{p_1}:\,,
  \label{}
\end{align}
where $G_1^{p_1}$ denotes the superoperator acting on the impurity and
$:J_1^{p_1}:$ the normal ordered field superoperator acting on the reservoirs.
The reservoir field superoperator is defined by its action on operators
$A$
acting in the reservoir Hilbert space and reads
\begin{align}
  J^p_1 A = \left\lbrace \begin{array}{ll}
    c_1 A &p=+ \\
    A c_1 &p=- 
  \end{array}\right.\,,
  \label{}
\end{align}
where $1\equiv\eta,\nu,\omega$ is a collection of indices classifying
the field operator $c_1$ such that
\begin{align}
  c_1 = \left\lbrace \begin{array}{ll}
    c^{\dagger}_{\nu,\omega} & \eta=+\\
    c_{\nu,\omega} & \eta=-
  \end{array}\right.\,.
  \label{}
\end{align}
Similarly we define $\bar{1}\equiv-\eta,\nu,\omega$.
The action of the impurity vertex superoperator on this specific
eigenvector is given by
\begin{align}
  G^p_1 A = \left\lbrace \begin{array}{ll}
    d_1 A & p=+ \\
    -\sigma^{p} A d_1 & p=- 
  \end{array}\right.\,.
  \label{}
\end{align}
The index $p$, that appears in the definition of both superoperators,
determines whether the respective field operator acts on the second
operator $A$ from the left ($p=+$) or from the right ($p=-$). It can be
interpreted as indicating the position of the field operator on the Keldysh
contour and is thus sometimes referred to as Keldysh index.
The operator $\sigma^{p}$ accounts for fermionic sign factors. It
returns a negative sign if
\begin{align}
  \vert l ) = G^{-}_1 \vert m ) = \vert i \rangle \langle j \vert\,,
  \label{}
\end{align}
such that 
\begin{align}
  \text{mod} \left[ \left(\sum_i d^{\dagger}_i d_i \vert i \rangle - \sum_i d^{\dagger}_i
  d_i \vert j \rangle \right) ,2\right] = 1\,.
  \label{}
\end{align}
\paragraph{Reservoir contractions} 
We perform the trace $\text{tr}_{\text{l}}$ over the lead degrees of freedom
by contracting pairs of reservoir field superoperators in our series
expansion of $\tilde{\rho}_{\text{ns}}(E)$. We denote these contractions 
\begin{align}
  \gamma^{pp'}_{11'} = \langle J^p_1 J^{p'}_{1'} \rangle_{\text{eq.}}\,,
  \label{}
\end{align}
where $\langle \dots \rangle_{\text{eq.}}$ indicates that we assume the
semi-infinite reservoirs to be in thermal equilibrium. The contractions
are thus proportional to the equilibrium distribution function
$f(\omega)$ at temperature $T$. We can simplify the subsequent
calculations by separating the distribution function $f(\omega)$ into a
symmetric and an antisymmetric contribution. The reservoir contraction
then reads 
\begin{align}
\gamma^{pp'}_{11'}=\delta_{1\bar{1}'} p'\gamma^s_1+\delta_{1\bar{1}'}\gamma^a_1\,,
\label{eq:contr}
\end{align}
with the symmetric contribution
\begin{align}
\gamma^{s}_1=\frac{1}{2}\rho_0\,,
\end{align}
and the antisymmetric contribution
\begin{align}
\gamma^{a}_1=\rho_0\left(f(\omega)-\frac{1}{2}\right)\,,
\end{align}
where $\rho_0$ is the density of states in the reservoir. It is possible
to absorb
the Keldysh index appearing in the contraction (\ref{eq:contr})
by introducing the vertices
\begin{align}
  \bar{G}_1 &=\sum_{p=\pm} G^p_1\,,\\
    \tilde{G}_1 &=\sum_{p=\pm} p G^p_1\,.
  \label{}
\end{align}
\paragraph{Definition of the perturbative corrections}
The leading order correction $\Sigma^{(1)} (E)$ to the effective impurity
Liouvillian $L_{\text{eff}}$, which derives from charge fluctuations, has
the form
\begin{align}
  \Sigma^{(1)}(E)=\int_{-D}^{D} d\omega_1\,\sum_{p,p'=\pm}\sum_{1,1'} G^{p}_1
  \frac{1}{\omega_1+E+\eta_1\mu_1-L_{\text{ns}}}G^{p'}_{1'} \gamma^{pp'}_{11'}\,.
  \label{eq:sig1}
\end{align}
As with the reservoir contractions we can separate $\Sigma^{(1)}(E)$ in
a symmetric and an antisymmetric term,
\begin{align}
  \Sigma^{(1)}(E)=\Sigma_s + \Sigma_a (E)\,,
  \label{}
\end{align}
where $\Sigma_s$ does not depend on the Laplace variable $E$. When using the
redefined vertices $\bar{G}_1$ and $\tilde{G}_1$ we can write $\Sigma_s$
as
\begin{align}
  \Sigma_s =& \frac{1}{2}\rho_0 \sum_{\nu_1,\eta_1}\bar{G}_1
  \int_{-D}^{D}d\omega_1 
  \frac{1}{\omega_1+E+\eta_1\mu_1-L_{\text{ns}}}\tilde{G}_{\bar{1}}\\\nonumber
  =&-i\frac{\pi}{2}\rho_0
  \sum_{\nu_1,\eta_1}\bar{G}_1\tilde{G}_{\bar{1}}\,,
  \label{}
\end{align}
where we have integrated over all reservoir frequencies $\omega_1$
ranging from the
lower to the upper reservoir band edge $D$. The symmetric contribution
$\Sigma_s$ turns out
to be entirely imaginary. It thus adds only to the decay rate of
transient features but not to their oscillation frequency.
The antisymmetric contribution $\Sigma_a(E)$ is a function of the
Laplace variable. It reads
\begin{align}
  \Sigma_a(E) =&-\frac{\rho_0}{2}
  \sum_{j=1}^{d(\mathcal{L})}\sum_{\nu_1,\eta_1} \int_{-D}^{D} d\omega_1
  \frac{\tanh\left(\frac{\omega_1}{2T}\right)}{\omega_1+E+\eta_1\mu_1-\lambda_j}
  \bar{G}_{1} \vert v_j)( v_j \vert \bar{G}_{\bar{1}}\\\nonumber
  =& \rho_0 \sum_{j=1}^{d(\mathcal{L})} \sum_{\nu_1,\eta_1}\left[\psi
    \left(\frac{1}{2}-i\frac{E+\eta_1\mu_1-\lambda_j}{2\pi
  T}\right)-\log\left(\frac{D}{2\pi T}\right)\right] \bar{G}_{1} \vert
  v_j ) (v_j\vert 
  \bar{G}_{\bar{1}}\,,
\end{align}
where $\psi(x)=\partial_x\log(\Gamma(x))$ is the Digamma function and $\vert v_j )$ are the
eigenvectors of the initial impurity Liouvillian $L_{\text{ns}}$ associated
with the eigenvalues $\lambda_j$ of $L_{\text{ns}}$. The imaginary part of
$\Sigma_a(E)$, which is the part contributing to the decay rate, takes a
more simple, intuitive form. It reads
\begin{align}
  \text{Im}\left(\Sigma_a(E)\right)=
  -\frac{\pi}{2}\rho_0 \sum_{j=1}^{d(\mathcal{L})}\sum_{\nu_1,\eta_1}
  \tanh\left(\frac{E+\eta_1
  \mu_1-\lambda_j}{2 T}\right) \bar{G}_{1} \vert v_j) (v_j \vert
  \bar{G}_{\bar{1}}\,.
  \label{}
\end{align}
In the basis spanned by the eigenvectors $\vert l \rangle$ of the
impurity Hamiltonian $H_{\text{r}}$ the initial impurity Liouvillian
$L_{\text{ns}}$ is
diagonal as well and one can easily establish a one-to-one
correspondence between an eigenvector $\vert v_j )$ of $L_{\text{ns}}$ and a matrix element of
$\rho_{\text{ns}}$ in this eigenbasis through 
\begin{align}
  \vert v_j ) =& \vert l \rangle \langle m \vert\,,
  \label{}
\end{align}
with the associated eigenvalue
\begin{align}
  \lambda_j =& E_l - E_m\,,
  \label{}
\end{align}
where $\lambda_j$ is the energy difference between the
two eigenstates $\vert l \rangle$ and $\vert m \rangle$ of the
Hamiltonian $H_{\text{r}}$. There are two eigenvalues
$\lambda_{\varepsilon,\pm}$ of
the impurity Liovilliain $L_{\text{ns}}$ that correspond to the energy difference
between the two charge density wave eigenstates $\vert g \rangle$ and
$\vert e \rangle$. We denote the
eigenvector that corresponds to the positive eigenvalue
$\lambda_{\varepsilon,+}$ as
\begin{align}
  \vert v_\varepsilon ) = \vert 2,e \rangle \langle 2,g \vert \,.
  \label{}
\end{align}
The action of the impurity vertex superoperators on this eigenvector is
given by
\begin{align}
  G^{+}_1 \vert v_\varepsilon ) =& d_1 \vert 2,e \rangle \langle 2,g
  \vert\,, \\
  G^{-}_1 \vert v_\varepsilon ) =& -(-1) \vert 2,e \rangle \langle 2,g
  \vert d_1\,,
  \label{}
\end{align}
where $d_1$ creates or annihilates a particle on lattice sites
$x=1$ or $x=4$ of the impurity. 
\paragraph{Perturbative diagonalization of $L_{\text{eff}}(E)$}
While $L_{\text{ns}}$ is diagonal in the eigenbasis of $H_{\text{r}}$, the corrections
$\Sigma_s$ and $\Sigma_a(E)$ are not. Due to the large size of the
Liouville space, $\text{dim}(\mathcal{L})=256$, an analytical
diagonalization of the effective Liouvillian
$L_{\text{eff}}(E)=L_{\text{ns}}+\Sigma_s+\Sigma_a(E)$ is not feasible. 
To determine the eigenvalues of $L_{\text{eff}}(E)$ we therefore
treat $\Sigma_s+\Sigma_a(E)$ as perturbations to the initial
Liouvillian $L_{\text{ns}}$ and calculate the leading order corrections to its
eigenvalues $\lambda_j$. This approximation is reasonable because
$\Vert \Sigma_s+\Sigma_a(E)\Vert \leq J_{\text{c}}^2\ll\varepsilon_T \approx
\lambda_j$. The eigenvalue corresponding to $\vert v_\varepsilon )$ is
then given by
\begin{align}
  \lambda_{\varepsilon}(E)=( v_\varepsilon \vert L_0 \vert v_\varepsilon
  )+\left[(v_\varepsilon\vert \Sigma_s \vert v_\varepsilon)+(v_\varepsilon
  \vert \Sigma_a (E) \vert v_\varepsilon)\right]\,.
  \label{}
\end{align}
The particle number symmetry of the impurity Hamiltonian
($[H_{\text{r}},\sum_j
n_j]=0$) guarantees that
$\langle 2,e \vert d_1 \vert 2,e \rangle \equiv 0 \equiv \langle 2,g \vert d_1
\vert 2,g \rangle$. We therefore find
\begin{align}
  ( v_\varepsilon \vert G^{+}_1 G^{-}_{\bar{1}} \vert v_\varepsilon )
  &\equiv 0 
  \label{eq:mp}\,,\\
  ( v_\varepsilon \vert G^{-}_1 G^{+}_{\bar{1}} \vert v_\varepsilon
  )&\equiv 0\,.
  \label{eq:pm}
\end{align}
Using (\ref{eq:mp}) and (\ref{eq:pm}) the perturbation theory corrections from the symmetric
contribution $\Sigma_s$ reduce to
\begin{align}
  (v_\varepsilon\vert \Sigma_s \vert v_\varepsilon)
  =&\sum_{\nu_1,\eta_1} -i\frac{\pi}{2}
  \rho_0 (v_\varepsilon\vert [G^{+}_1
  + G^{-}_1] [G^{+}_{\bar{1}}-G^{-}_{\bar{1}}]\vert
  v_\varepsilon)\\\nonumber
  =& \sum_{\nu_1,\eta_1} -i\frac{\pi}{2}\rho_0 (v_\varepsilon \vert G^{+}_1 G^{+}_{\bar{1}} \vert
  v_\varepsilon)
  +i\frac{\pi}{2}\rho_0 (v_\varepsilon\vert G^{-}_1 G^{-}_{\bar{1}} \vert
  v_\varepsilon )\\\nonumber
  =& -2\pi i \rho_0 J_{\text{c}}^2=-i\Gamma_0\,.
  \label{}
\end{align}
We see that the symmetric contribution from the leading order tunneling
processes between reservoirs and impurity causes
a decay rate $\Gamma$ equal to the typical decay rate
$\Gamma_0$ of transient features. However, this contribution does not
yet
factor in the fermion distribution function $f(\omega)$ in the reservoirs,
meaning that
each tunneling process is treated equally. The information about the distribution
function is encoded in the antisymmetric correction $\Sigma_a(E)$.
\paragraph{Matrix elements of the antisymmetric contribution
$\Sigma_a(E)$}
The evaluation of the antisymmetric corrections is more involved as the
contribution from each intermediate eigenstate $\vert v_j )$ of the
Liouvillian $L_{\text{ns}}$ is
individually 
weighted by $\psi [1/2-i(E\pm V-\lambda_j)/(2\pi T)]$. First, we identify 
the intermediate states $\vert v_j )$ that feature in the finite
contributions
\begin{align}
  ( v_\varepsilon \vert \bar{G}_{1} \vert v_j )( v_j \vert
  \bar{G}_{\bar{1}} \vert v_\varepsilon)=&( v_\varepsilon\vert
  \left[G^+_1 +
  G^-_1\right]\vert v_j )(v_j \vert \left[G^+_{\bar{1}}+G^-_{\bar{1}}\right]\vert
  v_\varepsilon)\\\nonumber
  =&( v_\varepsilon\vert G^{+}_1
  \vert v_j )(v_j \vert G^{+}_{\bar{1}} \vert
  v_\varepsilon)+(v_\varepsilon\vert G^{-}_{1} \vert v_j )( v_j \vert
  G^{-}_{\bar{1}} \vert v_\varepsilon)\neq 0\,.
  \label{}
\end{align}
The impurity vertex superoperator $G^{\pm}_1$ either creates or annihilates a
fermion on the impurity.
Finite contributions thus only involve eigenstates $\vert v_j)$
which
satisfy $\vert v_j ) = \vert m \rangle \langle 2,g \vert$ or $\vert v_j ) =
\vert 2,e \rangle \langle m \vert$ such that $\sum_{i=1}^{4} n_i
\vert m \rangle \in \lbrace 1,3 \rbrace$. One finds in total $N=8$
eigenstates in the Hilbert space $\mathcal{H}_{\text{r}}$ with particle number $n=1$ or $n=3$,
implying $N_L\leq 16$ finite matrix elements. A quantitative study of
the matrix elements $(G^{\pm}_1)_{j\varepsilon}$ reveals large contributions $\vert (v_j\vert
G^{\pm}_1 \vert v_\varepsilon)\vert^2 \simeq
J_{\text{c}}^2$ for two eigenstates $\vert v_j )\in \mathcal{L}$. The
two particular eigenstates are
\begin{align}
  \vert v^{-}_1 ) =& \vert 2,e \rangle \langle 1,g \vert\,,\\
  \vert v^{+}_3 ) =& \vert 3,g \rangle \langle 2,g \vert\,,
  \label{}
\end{align}
where $\vert 1,g \rangle$ and $\vert 3,g \rangle$
 are the two
low-energy eigenstates of the impurity Hamiltonian that do not exhibit a
charge density wave character and for
which the energy, $E_1=\langle 1,g\vert H_{\text{r}} \vert 1,g\rangle < \langle
3,g \vert H_{\text{r}} \vert 3,g\rangle=E_3$, is plotted in
figure 2. The matrix elements read  
\begin{align}
  (v_\varepsilon \vert G^{-}_{1}\vert v^{-}_1 )(v^{-}_1 \vert
  G^{-}_{\bar{1}} \vert v_\varepsilon) &\simeq -J^2_{\text{c}}\,,\\
  (v_\varepsilon \vert G^{+}_{1}\vert v^{+}_3 )(v^{+}_3 \vert G^{+}_1
  \vert v_\varepsilon) &\simeq +J^2_{\text{c}}\,,
  \label{}
\end{align}
where we note that the difference in Keldysh index $p=\pm$ of the vertex
superoperators $G^{p}_1$ causes the opposite sign of the matrix
elements. An analysis of the remaining matrix elements reveals a third matrix
element that gives a sizable contribution to the self energy. Here, the
intermediate state is
\begin{align}
  \vert v^{+}_1 ) = \vert 1,g \rangle \langle g \vert\,,
  \label{}
\end{align}
and the matrix element evaluates to
\begin{align}
  ( v_\varepsilon \vert G^{+}_1 \vert v^{+}_1 )( v^{+}_1 \vert
  G^{+}_{\bar{1}} \vert v_\varepsilon) \simeq +\frac{J^2_{\text{c}}}{80} \,.
  \label{}
\end{align}
The contribution from this matrix element becomes particularly relevant
in the vicinity of $U=\varepsilon_T$ due to the analytic structure of its
associated weight function.
\paragraph{Weight function}
\begin{figure}[]
  \centering
  \small{(a)}
  \includegraphics[width=0.46\textwidth]{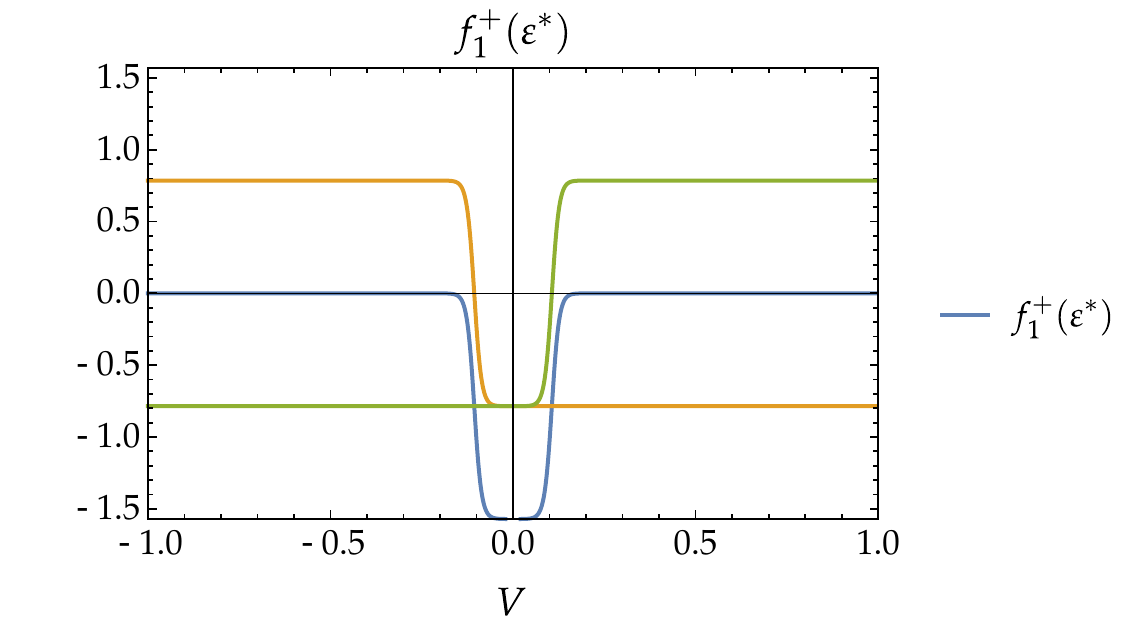}
  \small{(b)}
  \includegraphics[width=0.46\textwidth]{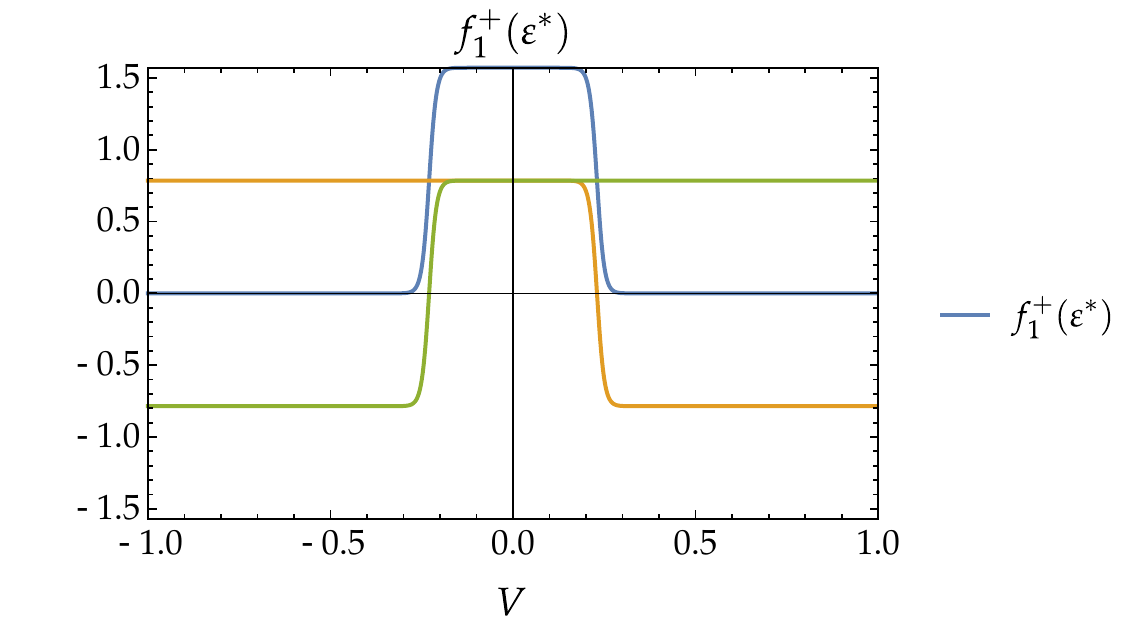}
  \caption{Weight function $f^{+}_1(\varepsilon^{*},V)$ for
two values of the interaction strength, (a) $U/\varepsilon_T=0.9$
and (b) $U/\varepsilon_T=1.2$. The orange line indicates $f(v)=-\pi\tanh
[\varepsilon-\Delta+v/(2T)]/4$ and the green line $f(v)=-\pi\tanh
[\varepsilon-\Delta-v/(2T)]/4$. When 
$U/\varepsilon_T=1$ and $v=0$ the weight function changes sign
and leads to complete a cancellation of the symmetric and antisymmetric
corrections. The weight function $f^+_1(\varepsilon^*,V)$ becomes finite
if $v \leq \vert \varepsilon-\Delta \vert$. (a): $(\varepsilon-\Delta)=0.11$,
(b): $(\varepsilon-\Delta)=-0.23$.}
  \label{fig:vcone}
\end{figure}
The weight function $f(\omega)$ contains the information about the fermionic
distribution function $n(\omega)$ in the leads, which details the probability for an
eigenstate of the lead Hmailtonian $H_{\text{l}}$ with energy $\omega$ to be occupied by a fermion.
Our aim is to determine the decay rate $\Gamma$ which
directly
corresponds to the imaginary part of the the eigenvalue
$\lambda_\varepsilon$ of
the effective Liouvillian that satisfies
\begin{align}
  \varepsilon^*-\lambda_{\varepsilon}(\varepsilon^{*})=0\,.
  \label{}
\end{align}
The imaginary part of $\lambda_\varepsilon$ originates entirely from the
imaginary part of the self-energy $\Sigma(\omega)$
correction, which for the asymmetric correction $\Sigma_a(\omega)$ stems from the weight
function $f_j^{\pm}(\omega)$. The imaginary part of the weight function evaluated at the
eigenvalue $\varepsilon^{*}$ has the simple
form
\begin{align}
  f_j(\varepsilon^{*})=-i\frac{\pi}{2}\sum_{\eta_1=\pm}\tanh\left(\frac{\varepsilon^{*}+\eta_1\frac{V}{2}-\lambda_j}{2T}\right)\,,
  \label{}
\end{align}
We know that $\varepsilon^{*}\simeq E_e -E_g$. We can thus also express the
weigth function as
\begin{align}
  f_j(\varepsilon^{*}) = -i \frac{\pi}{2}\sum_{\eta_1=\pm}\tanh\left(
  \frac{E_e-E_g-\lambda_j+\eta_1 \frac{V}{2}}{2T}\right)\,.
  \label{}
\end{align}
The eigenvalues $\lambda_j$ that correspond to the three largest matrix
elements read
\begin{align}
  \lambda_1^{+}=& E_{1}-E_g\equiv\Delta\,,\\
  \lambda_3^{+}=& E_{3} - E_g=E_3 -E_e +E_e -E_g
  \simeq\varepsilon+\Delta\,,\\
  \lambda_1^{-}=& E_e - E_{1}=E_e -E_g +E_g -E_1 \simeq\varepsilon-\Delta\,,
  \label{}
\end{align}
It is easy to see that, depending on the eigenvalue $\lambda_j$, either the
dependence on $E_g$ or $E_e$ is removed from the argument of the weight
function. To simplify the expression we introduce $v \equiv \vert V/2 \vert$. The weight
function then reads 
\begin{align}
  f_{j}(\varepsilon^{*})=
  -i\frac{\pi}{2}\rho_0\left[\tanh\left(\frac{\varepsilon-\lambda_j+
  v}{2T}\right)+\tanh\left(\frac{\varepsilon-\lambda_j-v}{2T}\right)\right]\,.
  \label{}
\end{align}
In the limit $T\ll\lbrace J,U,\varepsilon,D \rbrace$ we can approximate
the weight function as
\begin{align}
  f_j(\varepsilon^{*})=-i\frac{\pi}{2}\rho_0\left[\text{sign}(\varepsilon-\lambda_j+v)+\text{sign}(\varepsilon-\lambda_j-v)\right]\,.
  \label{}
\end{align}
When evaluating this approximation for the three relevant eigenvalues
$\lambda_j^{\pm}$ one finds
\begin{align}
  f^{+}_1 (\varepsilon^{*})&=
  -i\frac{\pi}{2}\rho_0\left[\text{sign}(\varepsilon-\Delta+v)+\text{sign}(\varepsilon-\Delta-v)\right]\\\nonumber
  &=-i\frac{\pi}{2}\rho_0
  \left[\theta(\varepsilon_T-U)\left(1+\text{sign}(\varepsilon-\Delta-v)\right)+\theta(U-\varepsilon_T)\left(\text{sign}(\varepsilon-\Delta+v)-1\right)\right]\,,\\
  f^{+}_3(\varepsilon^{*})&=-i\frac{\pi}{2}\rho_0\left[\text{sign}(-\Delta+v)+\text{sign}(-\Delta-v)\right]\\\nonumber
  &=-i\frac{\pi}{2}\rho_0 \left[\text{sign}(v-\Delta)-1\right]\,,\\
  f_{1}^{-}(\varepsilon^{*})&=-i\frac{\pi}{2}\rho_0
  \left[\text{sign}(\Delta+v)+\text{sign}(\Delta-v)\right]\\\nonumber
  &=-i\frac{\pi}{2}\rho_0\left[1+\text{sign}(\Delta-v)\right]\,,
  \label{}
\end{align}
where we note that $\varepsilon,\Delta\geq 0$.
\paragraph{Decay channels}
The three matrix elements $(v_\varepsilon\vert G^-_1 \vert
v^-_1)(v^-_1\vert G^-_{\bar{1}} \vert v_\varepsilon)$,
$(v_\varepsilon\vert G^{+}_1 \vert v_3^+ )( v_3^+ \vert G^+_{\bar{1}}
\vert v_\varepsilon )$ and $( v_\varepsilon \vert G^+_1 \vert v^+_1 )(
v^+_1 \vert G^+_{\bar{1}}\vert v_\varepsilon)$ correspond to four
different decay channels that cause the decoherence of a state of the
form $\vert \psi
\rangle = \alpha \vert 2,g \rangle + \beta \vert 2,e \rangle$. A schematic
of these decay channels is shown in figure \ref{fig:dcmodes}. We now turn to the
discussion of the decay channels and why they become
suppressed for specific sets of parameters $U$, $\varepsilon_T$ and $V$.
\paragraph{$(v_\varepsilon \vert G^-_1 \vert v^-_1)(v^-_1 \vert
G^-_{\bar{1}} \vert v_\varepsilon)$:}
An electron is ejected from the ring impurity, which has initially been in the ground state
$\vert 2,g\rangle$. Due to the particle hole symmetry of the repulsive nearest neighbor
interaction $U$, this requires the energy $\Delta(U)$. Said energy
needs to be supplied by the increase of chemical potential energy
$\mu$, which the
electron gains by entering the metallic lead. The process is thus only
possible if $\mu=-V/2< -\Delta(U)$. Here, we have assumed that the
electron can only enter the lead to which a negative chemical
potential $-V/2 \sum_k n_k$ was applied. The condition is reflected by
the weigth function $f^-_1 (\varepsilon^*)$, which evaluates to zero if $v$
surpasses $\Delta$. Then the imaginary part of the asymmetric correction
$\Sigma_a(\omega)$ does not
compensate the constant imaginary part of the symmetric correction
$\Sigma_s$ for this decay channel.
A sketch of the decay process is shown
figure \ref{fig:dcmodes} (1), where red in indicates the initial and green the
final configuration of the decay process.
\paragraph{$( v_\varepsilon \vert G^{+}_1 \vert v^+_3 )( v^+_3 \vert
G^+_{\bar{1}} \vert v_\varepsilon)$:}
An electron tunnels onto the ring impurity, which has initially been in
the excited charge density wave state $\vert 2,e\rangle$. The
additional electron increases the interaction energy on the ring
impurity by $\Delta(U)$. This energy has to be supplied by the
additional electron. The process is thus only possible if the chemical
potential in the lead that the electron originates from satisfies $\mu = V/2
> \Delta(U)$. As with the previous matrix element,
$f^{+}_3(\varepsilon^*)$ vanishes once $v>\Delta$ such that the constant
negative imaginary part of the symmetric correction $\Sigma_s$ is not
compensated. We display a sketch of this decay channel in figure
\ref{fig:dcmodes} (2).\\
\newline
The decay channels $( v_\varepsilon \vert G^{-}_1 \vert v^-_1 )( v^-_1 \vert
G^+_{\bar{1}} \vert v_\varepsilon)$ and $( v_\varepsilon \vert G^{+}_1 \vert v^+_3 )( v^+_3 \vert
G^+_{\bar{1}} \vert v_\varepsilon)$ are closely related - one involves the
ground state while the other one involves the excited state - and are thus
respectively allowed
or suppressed in the same parameter regimes. 
\paragraph{$( v_\varepsilon \vert G^{+}_1 \vert v^+_1 )( v^+_1 \vert
G^{+} \vert v_\varepsilon)$:} 
An electron tunnels out of the ring impurity, which has initially been
in the excited charge density wave state $\vert 2,e\rangle$. Depending
on the interaction strength $U$, this tunneling process is energetically
favorable or unfavorable. For $U<\varepsilon_T$ the one particle state
$\vert 1,g\rangle$ is lower in energy than $\vert 2,e\rangle$. Since
the electron can carry this excess energy it can tunnel into either lead
as long as $\mu = V/2 < \varepsilon-\Delta$. Having two effective decay
channels, one for
each lead, increases the decay rate as can be seen in sector (i) and
(iii) of
figure \ref{fig:rtptres}. The weight function reflects this as
$f^+_1(\varepsilon^*)=-i\pi\rho_0$, which adds to the imaginary part of
the symmetric
correction $\Sigma_s\propto -i\pi\rho_0$ instead of compensating for it.
A schematic of this process is shown in figure \ref{fig:dcmodes} (3) and
(4).
For $U>\varepsilon_T$ the state $\vert 1,g\rangle$ becomes higher in
energy than $\vert 2,e\rangle$. For an electron to tunnel out of the
ring additional energy is now required. This energy needs to be
provided by the increase in chemical potential energy $\mu$ that the
electron gains by entering the lead. The tunneling process is thus only
possible if $\mu=-V/2 < -(\Delta-\varepsilon)$. We sketch this process
in figure \ref{fig:dcmodes} (4).\\
\newline
For $U>\varepsilon_T$ and $v<\Delta-\varepsilon$ each decay channel
becomes suppressed and we find
\begin{align}
  \text{Im}\left[\Sigma_a(\varepsilon^{*})\right]\equiv -\text{Im}\left[\Sigma_s\right]\,.
  \label{}
\end{align}
The first order corrections to the imaginary part of the transient feature with
oscillation frequency $\varepsilon^{*}$ therefore vanish entirely.
\paragraph{Discussion of the phase diagram}
\begin{figure}
\begin{center}
  \small{(a)}
\includegraphics[width=0.44\textwidth]{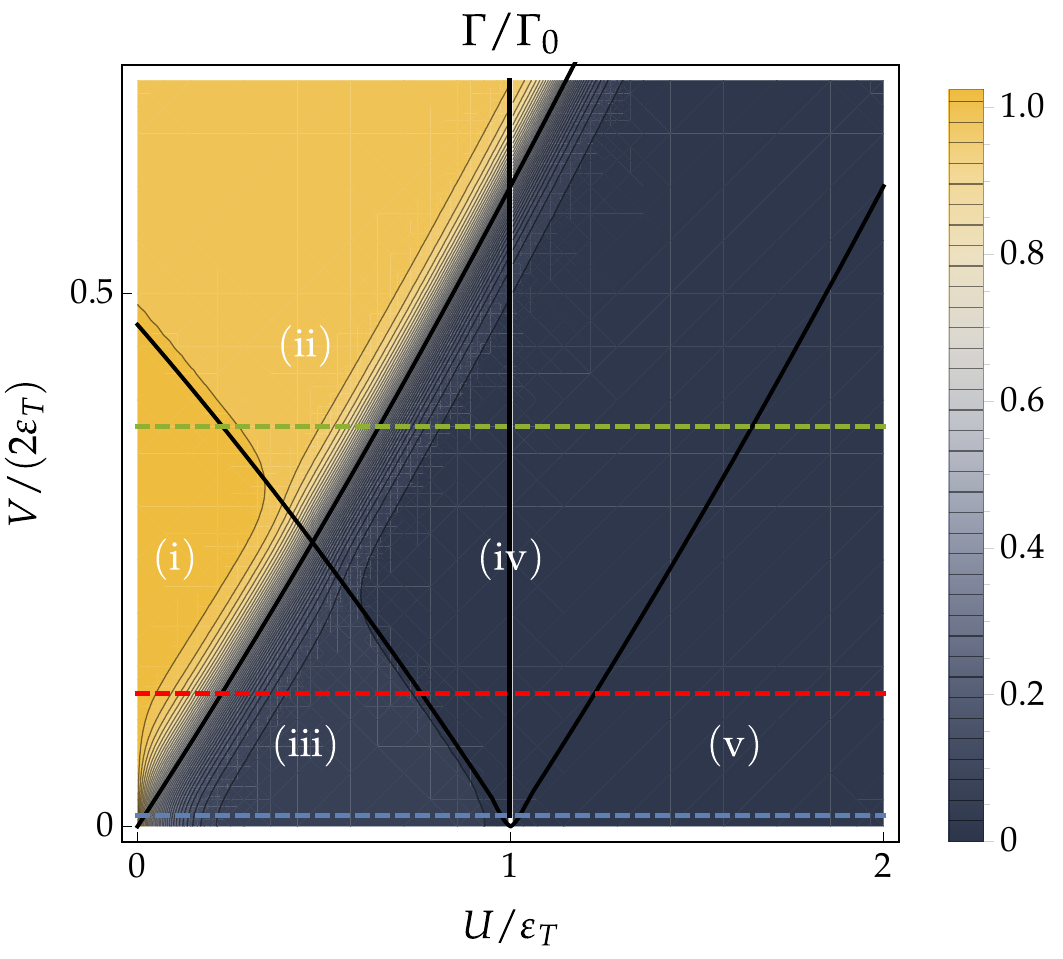}
\small{(b)}
\includegraphics[width=0.47\textwidth]{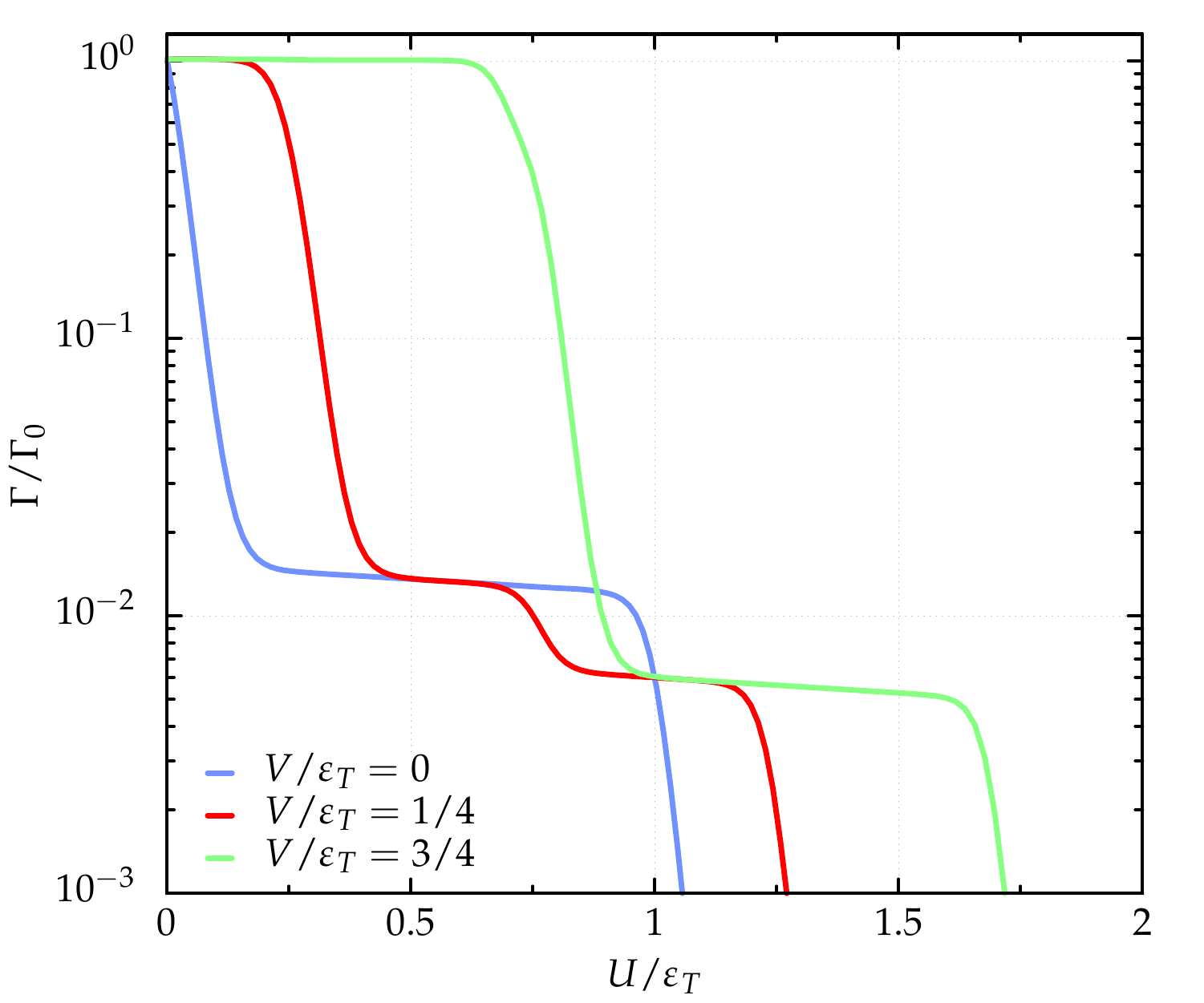}
\caption{(a) Perturbation theory results for the decay rate
  $\Gamma=\text{Im}[\lambda_\varepsilon(\varepsilon^*)]$
  of the transient feature with oscillation frequency $\varepsilon$ as a
  function of the interaction strength $U/\varepsilon_T$ and bias
  voltage $V/(2\varepsilon_T)$ in units of $\Gamma_0$. We observe five
  distinct sectors $\left[ (\rm{i}) - (\rm{v}) \right]$ in which the decay rate
  assumes different values. These sectors are characterized by their available
  decay channels. In sector (i) the decay rate
  $\Gamma$ exceeds $\Gamma_0$ due to the presence of an
  unconventional decay channel, see fig. \ref{fig:dcmodes} (3). For $V/(2\varepsilon_T) < \Delta$
  we find the dominant decay channels suppressed, leading to a decrease
  of the decay rate $\Gamma$ by an order of magnitude
  compared to $\Gamma_0$. In sector (v) each decay channel is suppressed
  leading to an effective decay rate $\Gamma/\Gamma_0
  \rightarrow 0$. (b) shows the decay rate
  $\Gamma/\Gamma_0$ as a function of $U/\varepsilon_T$ for
  three distinct values of the bias voltage $V$. The chosen parameters
are indicated by the blue, red and green dashed lines in (a).}
\label{fig:rtptres}
\end{center}
\end{figure}
\begin{figure}[]
  \centering
  \includegraphics[width=0.8\textwidth]{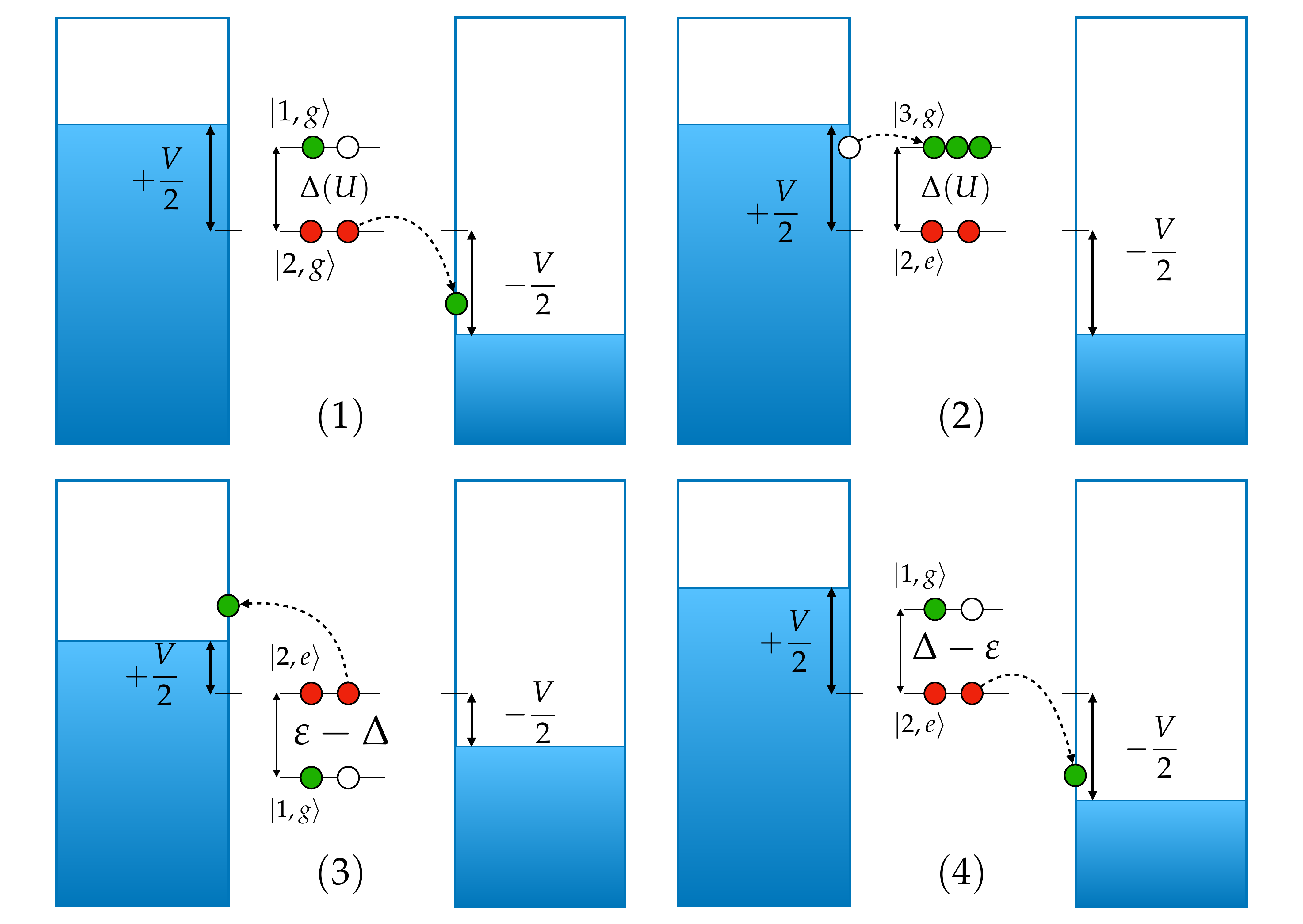}
  \caption{Dominant decay channels in the first order perturbation
  theory. The red circles indicate the initial configuration and the
green circles mark the final configuration of the process. (1): Decay
channel associated with the matrix element $(v_\varepsilon \vert G^{-}_1
\vert v^-_1 )( v^-_1 \vert G^-_{\bar{1}} \vert v_\varepsilon)$. An
electron tunnels from the impurity into a lead, causing a transition
from the ground state $\vert 2,g\rangle$ to the excited state $\vert
1,g\rangle$. (2): Decay
channel associated with the matrix element $(v_\varepsilon \vert G^{+}_1
\vert v^+_3 )( v^+_3 \vert G^+_{\bar{1}} \vert v_\varepsilon)$. An
electron tunnels from a lead onto the impurity, causing a transition
from the excited state $\vert 2,e\rangle$ into the excited state $\vert
3,g\rangle$. (3): First decay
channel associated with the matrix element $(v_\varepsilon \vert G^{+}_1
\vert v^+_1 )( v^+_1 \vert G^+_{\bar{1}} \vert v_\varepsilon)$ for
$U/\varepsilon_T <1$. An
electron tunnels from the impurity into the lead with positive
chemical potential $\mu=+V/2$, causing a transition
from the excited state $\vert 2,e\rangle$ to the excited state $\vert
1,g\rangle$ which releases the energy $\varepsilon-\Delta$. (4): Second decay
channel associated with the matrix element $(v_\varepsilon \vert G^{+}_1
\vert v^+_1 )( v^+_1 \vert G^+_{\bar{1}} \vert v_\varepsilon)$. An
electron tunnels from the impurity into the lead with negative chemical
potential $\mu=-V/2$,
causing a transition
from the excited state $\vert 2,e\rangle$ to the excited state $\vert
1,g\rangle$ which releases the energy $\varepsilon-\Delta$ for
$U/\varepsilon_T\leq 1$ and requires the energy $\Delta-\varepsilon$ for
$U/\varepsilon_T>1$.}
  \label{fig:dcmodes}
\end{figure}
In figure \ref{fig:rtptres} we plot the decay rate $\Gamma/\Gamma_0$ of the eigenvalue
$\lambda_\varepsilon(\varepsilon^*)$ as a function of the ratios
$U/\varepsilon_T$ and $V/(2 \varepsilon_T)$. We identify five different
sectors of these ratios in which the decay rate $\Gamma$
take different values due to the presence or absence of the previously
outlined decay channels respectively. In sector (i) we find the
presence of the decay channels (1), (2), (3) and (4). The decay
channel (3) does not exist for many of the typical quantum dot systems.
Its presence leads to a decay rate $\Gamma$ that exceeds the
typical level broadening $\Gamma_0$. By increasing the bias voltage
$V$ one crosses from sector (i) into sector (ii) where the decay channel
(3) becomes suppressed as there is no remaining unoccupied state with energy
$\omega=\varepsilon_F + (\varepsilon-\Delta)$ available in the left lead
. In sector (ii) we find $\Gamma = \Gamma_0$. By
increasing the interaction strength $U/\varepsilon_T$ sufficiently one
passes from sector (ii) into sector (iv). The increase in interaction
strength causes an increased energy gap $\Delta(U)$. As soon as
$\Delta(U) > V/2$ both the decay channels (1) and (2) simultaneously
become suppressed. This leads to a significant reduction of the decay
rate by almost two orders of magnitude such that
$\Gamma \leq \Gamma_0/80$. For $U/\varepsilon_T<1$ the reduction of the bias voltage facilitates a
crossover from sector (iv) into sector (iii). In this sector, the decay
channel (3) is no longer suppressed leading to small increase of $\Delta
\Gamma \simeq \Gamma_0/80$. For $U/\varepsilon_T > 1$ and
$V/(2\varepsilon_T) < (\Delta-\varepsilon)$ every decay channel is
suppressed as is shown in sector (v).  The corresponding decay rate becomes
$\Gamma/\Gamma_0 \rightarrow 0$. For a finite decay rate, higher order
perturbation theory corrections would be required. However, these
corrections can
induce no more than a decay rate $\Gamma \propto \Gamma^2_0\ll T$.
\paragraph{Eigenvalue spectrum of the effective Liouvillian}
The disappearance of the decay rate $\Gamma$ for an 
eigenvalue $\lambda(z)$ of the effective Liouvillian with finite real
part is unique to the eigenvalues $\pm \varepsilon^*-i\Gamma$.
In figure \ref{fig:poles} we display the real and
imaginary part of each root of 
\begin{align}
  z-\lambda_j(z)=0\,,
  \label{eq:roots}
\end{align}
where $\lambda_j(z)$ are the eigenvalues of the effective Liouvillian.
\begin{figure}[]
  \centering
  \small{(a)}
  \includegraphics[width=0.435\textwidth]{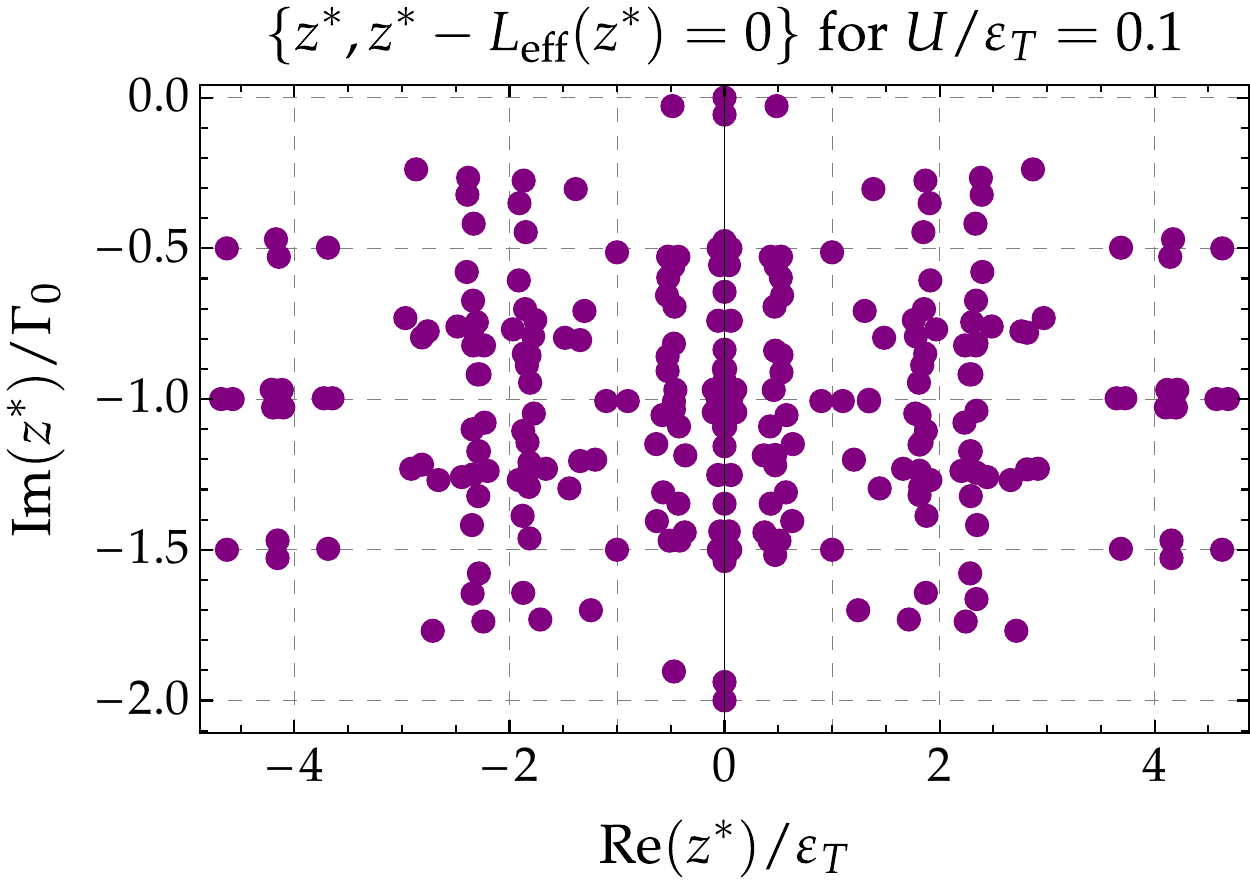}
  \small{(b)}
  \includegraphics[width=0.48\textwidth]{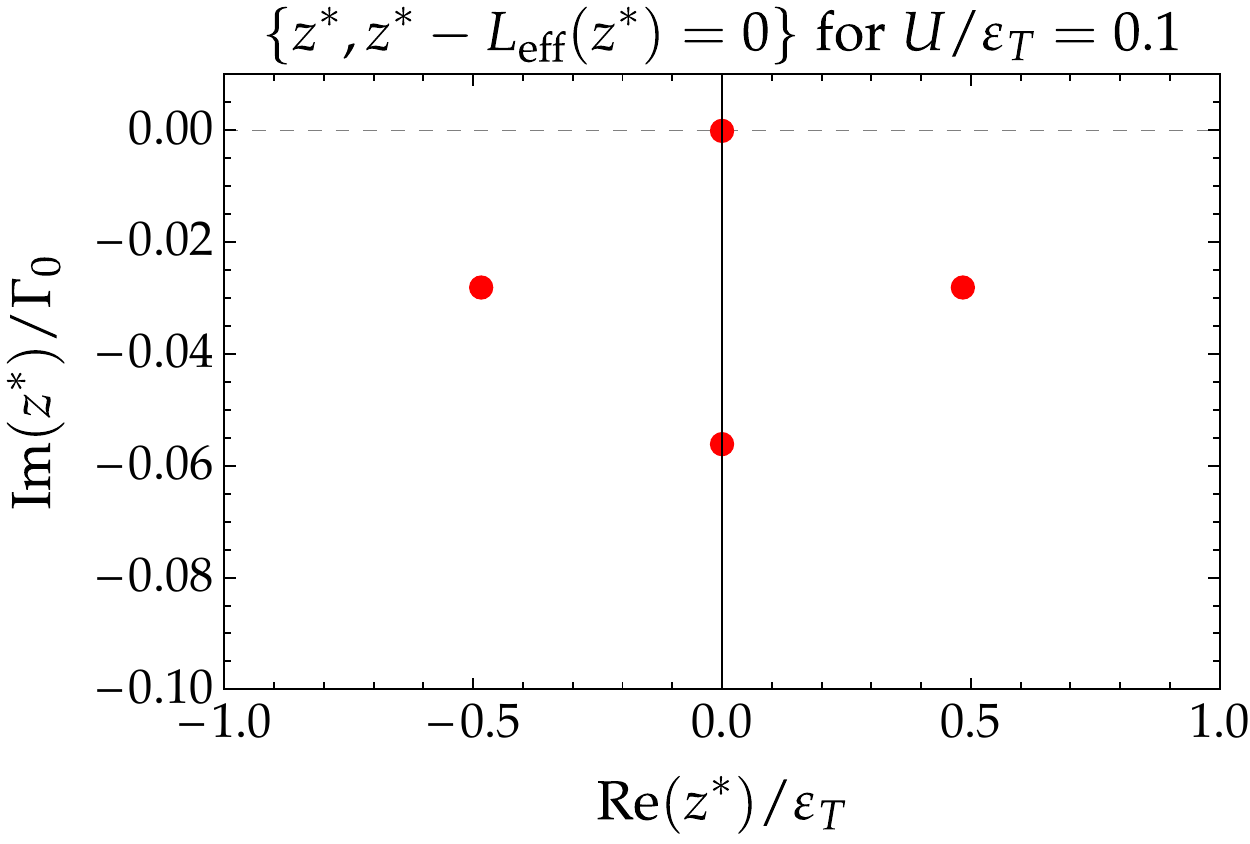}
  \small{(c)}
  \includegraphics[width=0.435\textwidth]{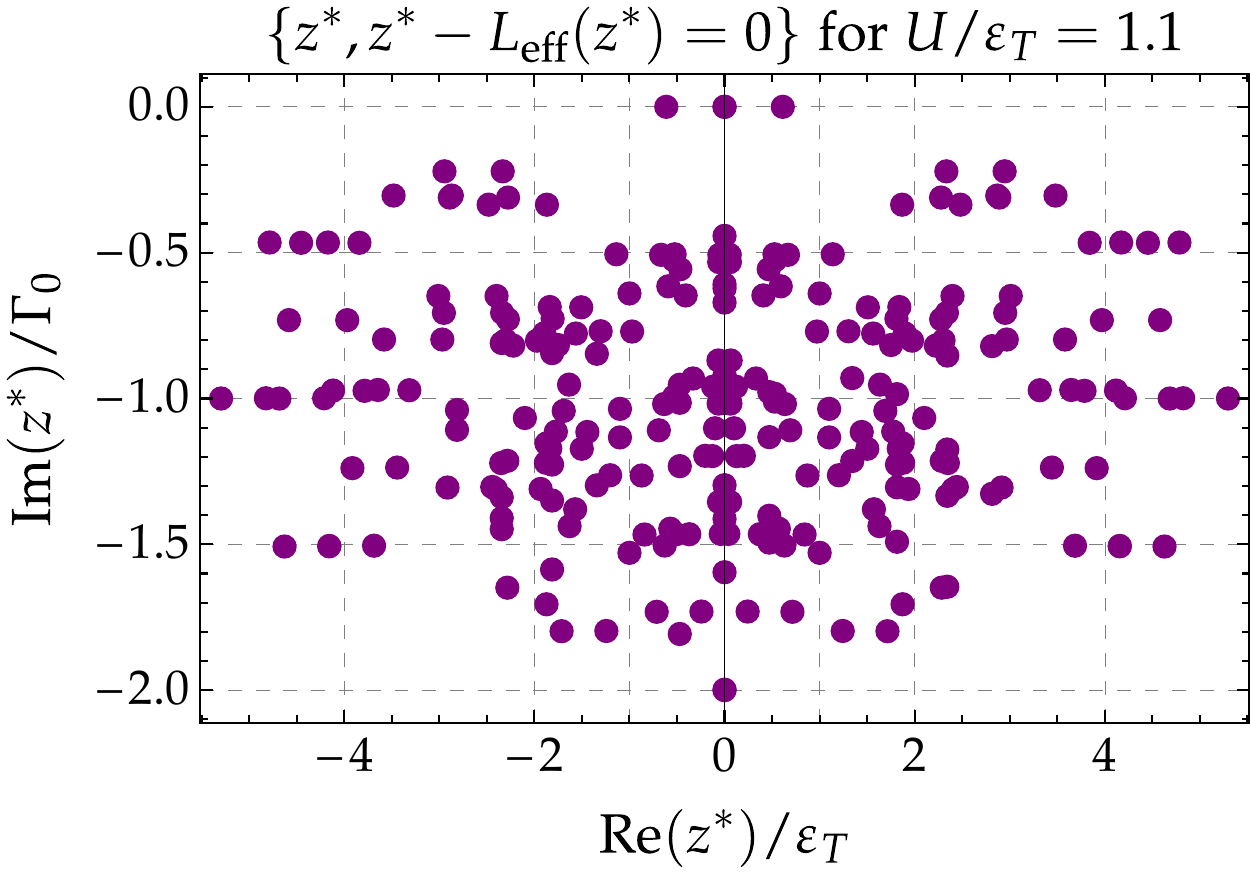}
  \small{(d)}
  \includegraphics[width=0.48\textwidth]{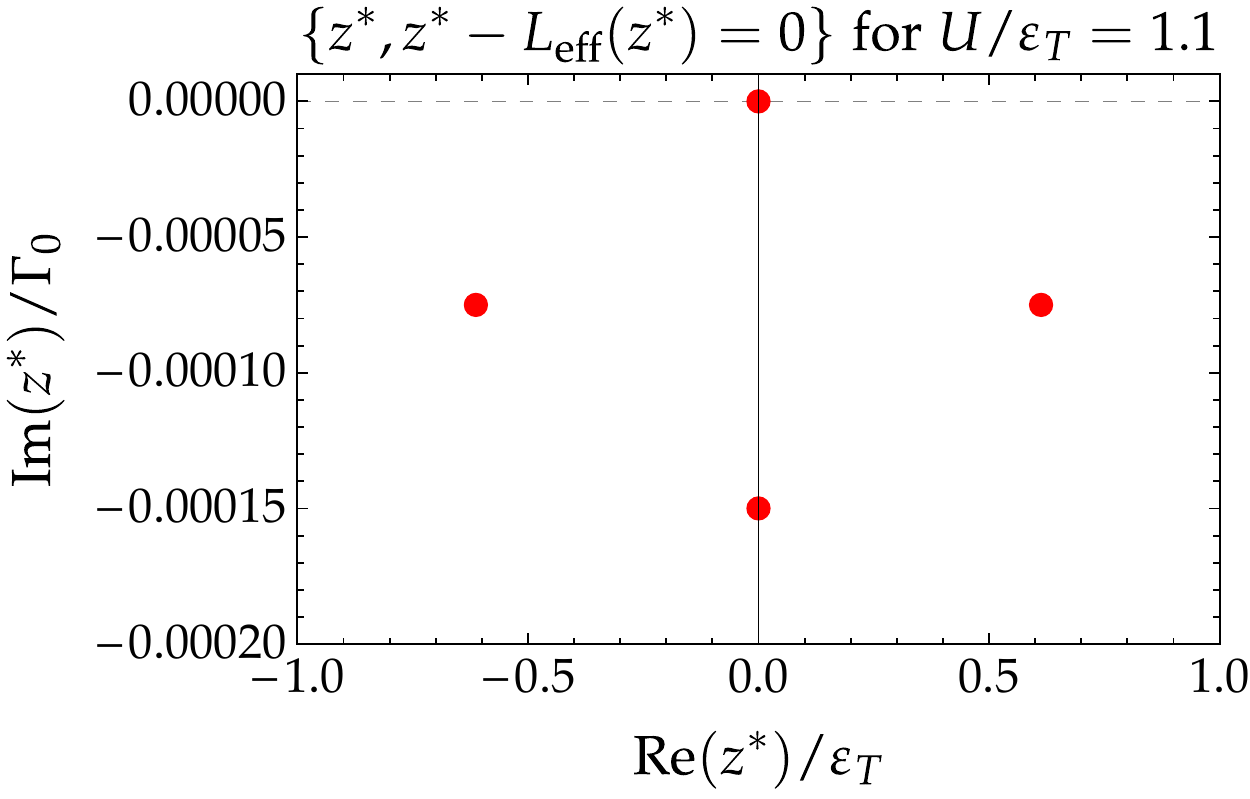}
  \caption{Roots $z^*$ of $z-L_{\text{eff}}(z)=0$. (a),(b):
  $U/\varepsilon_T=0.1$ and (c),(d): $U/\varepsilon_T=1.1$.
  We find that the imaginary part $\text{Im}(z^*)$ of the majority of
  roots $z^*$ is of
order $\mathcal{O}(\Gamma_0)$. In addition we find four
roots with an imaginary part orders of magnitude smaller than
$\Gamma_0$. In (b) and (d) we zoom in on these four roots. We find that the
imaginary part of three poles is two orders of magnitude smaller than
$\Gamma_0$ for $U/\varepsilon_T=0.1$ and five orders of magnitude
smaller for $U/\varepsilon_T=1.1$. The root with imaginary part $\Gamma\equiv0$ is associated
with the stationary state.}
  \label{fig:poles}
\end{figure}
We find that the imaginary part of all but four roots is of order
$\mathcal{O}(\Gamma_0)$. This means that almost all transient features
decay on the
expected time scale. In figure \ref{fig:poles} (d) we zoom in on the
four extraordinary roots. We see that their imaginary part is
$\text{Im}(z)\simeq \Gamma_0/100000$. This indicates a clear separation
of scales between the decay rate of these four roots and each 
other root. Such a separation of scales in the solutions to
(\ref{eq:roots}) is sometimes referred to as dissipative phase transition. The four roots that feature a small imaginary part belong to $\vert 2,g\rangle
\langle 2,g\vert$ with root $z=0-i0^{+}$, $\vert 2,e\rangle \langle
2,e\vert$ with root $z=0-i\Gamma_1$, $\vert 2,e\rangle \langle
2,g \vert$ with root $z=\varepsilon-i\Gamma$, and $\vert
2,g\rangle \langle 2,e\vert$ with root
$z=-\varepsilon-i\Gamma$.
\subsection{Inverse transformation to realtime}
The effective reduced density matrix matrix of the impurity in Laplace
space $\tilde{\rho}_{\text{ns}}(E)$ is given by
\begin{align}
  \tilde{\rho}_{\text{ns}}(E)=\frac{i}{E-L_{\text{eff}}(E)}\rho_{\text{ns}}(t_0)\,,
  \label{eq:effrho}
\end{align}
where each pole of the resolvent 
\begin{align}
  \frac{1}{E-\lambda_j(E)} \vert v_j ) ( v_j \vert\,,
  \label{}
\end{align}
has an imaginary part $\Gamma_j\leq 0$. We can therefore replace the
inverse Laplace transform by a Fourier transform and close the
integration contour in the lower half-plane such that
\begin{align}
  \tilde{\rho}_{\text{ns}} (t) &=\frac{1}{2\pi}\sum_j \int_{-\infty}^{+\infty} dE\,
  \frac{ie^{-i E (t-t_0)}}{E-\lambda_j(E)}\vert v_j )( v_j \vert\,
  \rho_{\text{ns}} (t_0)\\\nonumber
    &=\theta (t-t_0)\sum_j \exp(i\lambda_j t-\Gamma_j t)\vert v_j )( v_j
    \vert \,\rho_{\text{ns}} (t_0)\,.
  \label{}
\end{align}
Each pole of (\ref{eq:effrho}) corresponds to a transient feature with frequency
$\lambda_j$ and decay rate $\Gamma_j$.
\subsection{Coupling of the current operators to the pole
$\lambda_\varepsilon$}
\begin{figure}[]
  \centering
  \small{(a)}
  \includegraphics[width=0.46\textwidth]{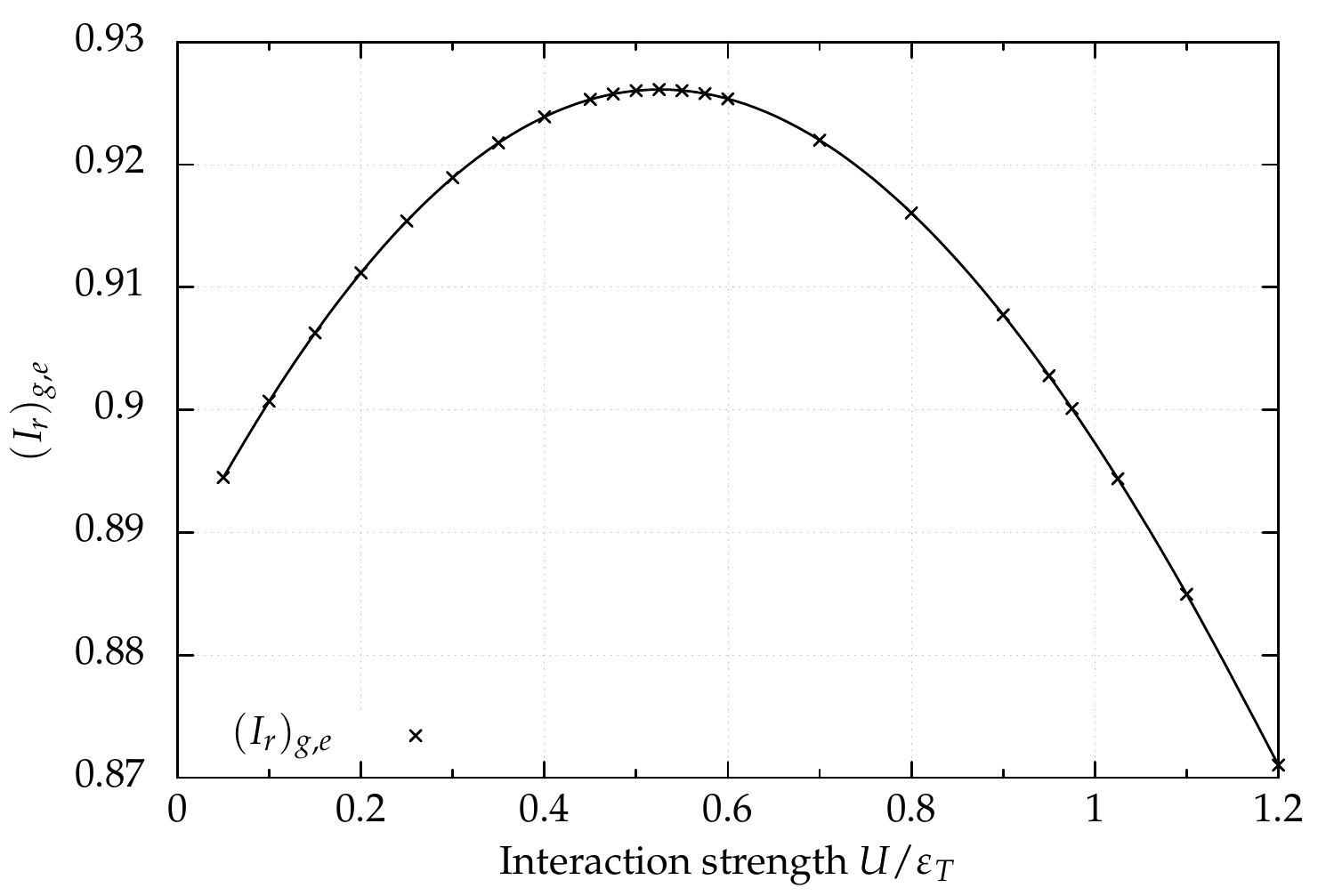}
  \small{(b)}
  \includegraphics[width=0.46\textwidth]{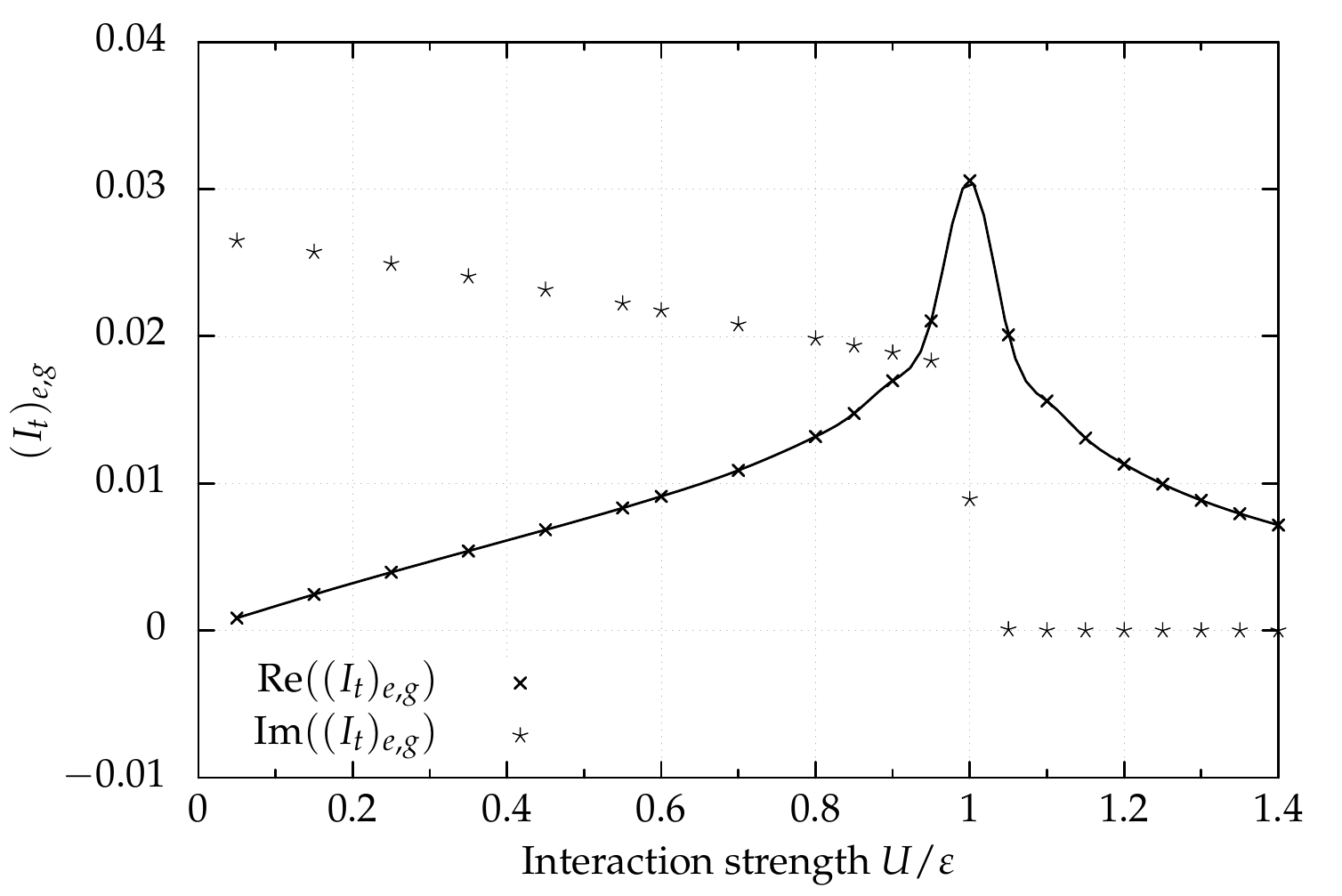}
  \caption{(a) Matrix element $(I_{r})_{g,e}$ of the ring current operator coupling
  to the transient feature of the reduced density matrix
  $\tilde{\rho}_{\text{ns}}(E)$ with oscillation frequency $\varepsilon$. (b)
  Real and imaginary part of the matrix element $(I_{t})_{e,g}$ of the transmitted current operator coupling
  to the transient feature of the reduced density matrix
  $\tilde{\rho}_{\text{ns}}(E)$ with oscillation frequency $\varepsilon$.}
  \label{fig:icoup}
\end{figure}

\paragraph{Ring current}
The operator $I_{\text{r}}$ measuring the local current in the ring reads
\begin{align}
I_{\text{r}}&= I_{u}-I_{l}\,,\\
I_{u}&=-i\left[n_2,H\right]\propto i
\left(d_1^{\dagger}d_2-d^{\dagger}_2 d_1\right)\,,\\
I_{l}&=-i\left[n_3,H\right]\propto i \left(d_1^{\dagger}d_3-d^{\dagger}_3
d_1\right)\,.
\end{align}
with
\begin{align}
  \langle I_{\text{r}} \rangle(t) = \text{Tr}_S\left[I_{\text{r}} \rho_{\text{ns}}(t)\right]\,.
  \label{}
\end{align}
After a transformation of the current operator and the reduced density
matrix $\rho_{\text{ns}}$ to the basis of the eigenstates of Hamiltonian, this becomes
\begin{align}
  \langle I_{\text{r}} \rangle (t) = \sum_{n=1}^{d(\mathcal{H})} \langle n \vert
  \underbrace{\left(U I_{\text{r}}
    U^{\dagger}\right)}_{=\tilde{I}_r} \rho_{\text{ns},n}(t) \vert n \rangle\,,
  \label{}
\end{align}
where $\rho_{\text{ns},n}(t)$ denotes the time-dependent reduced density matrix
expressed in the basis given by the eigenstates of the Hamiltonian. 
We are mainly interested in the matrix element of $\tilde{I}_r$ that couples to
the matrix element 
$\vert v_\varepsilon ) = \vert 2,e \rangle \langle 2,g \vert$ of the reduced density matrix
\begin{align}
  \langle I_{\text{r}} \rangle_{\text{osc.}}(t)=2 \langle 2,g \vert
  \left((\tilde{I}_r)_{g,e} \vert 2,g \rangle \langle 2,e \vert
  \right) \left[ \exp(i\lambda_\varepsilon t - \Gamma t)
  \vert 2,e \rangle \langle 2,g \vert \right] \vert 2,g \rangle\,.
  \label{}
\end{align}
In figure \ref{fig:icoup} (a) we plot the absolute value of the coupling of
the current operator to the off-diagonal matrix element $\vert e \rangle
\langle g \vert$ of the reduced density matrix. We find that after an
initial increase with interaction strength, the coupling decreases with
interaction strength. In the entire range of values for the interaction
strength that we have studied, the matrix element $(\tilde{I}_{r})_{g,e}$ exceed
every other matrix element of the current operator $\tilde{I}_{r}$.
\paragraph{Transmitted current}
We determine the extent to which the operator $I_{\text{t}}$, measuring the transmitted
current, couples to the matrix elements $\vert 2,e \rangle \langle
2,g\vert$ and $\vert 2,g \rangle \langle 2,e \vert$
of the reduced density matrix
$\tilde{\rho}_S(E)$ directly from the perturbation theory. The
expectation value of the transmitted current in Laplace space is given by
\begin{align}
  \langle I_{\text{t}} \rangle (E) = \text{tr}_{\text{ns}} \Sigma_{I_{\text{t}}}(E)
  \frac{1}{E-L_{\text{eff}}(E)}\rho_{\text{ns}}(t_0)\,,
\end{align}
where 
\begin{align}
  \Sigma_{I_{\text{t}}}(E)=\int_{-D}^{D}d\omega_1\,\sum_{p,p'=\pm} \sum_{1,1'}
  (I_{\text{t}})_1^p\frac{1}{\omega_1+E+\eta_1 \mu_1-L_{S}}G^{p'}_{1'}\gamma^{p
  p'}_{1 1'}\,.
\end{align}
The modified vertex superoperator is defined as
\begin{align}
  (I_{\text{t}})^p_1 A = \left\lbrace\begin{array}{ll}
    I_{\text{t}} A & p=+ \\
    \sigma^p A I_{\text{t}} & p=-
  \end{array}\right.\,,
\end{align}
where the operator $I_{\text{t}}$ acts in the Hilbert space $\mathcal{H}_{\text{r}}$ of the ring as
\begin{align}
  I_{\text{t}} = d^{\dagger}_1 -d_1\,.
\end{align}
The part of the transmitted current that acquires the small decay rate
$\Gamma$ is given by
\begin{align}
  (I_{\text{t}})_{e,g}=\sum_{n}^{d(\mathcal{H})}\langle n \vert \Sigma_{I_{\text{t}}}(\lambda_\varepsilon) \vert
  v_\varepsilon)\vert n \rangle\,,
  \label{}
\end{align}
where the vectors $\vert n \rangle$ form a basis of the Hilbert space of
the ring. In figure \ref{fig:icoup} (b) we plot the real and imaginary
part of $(I_{\text{t}})_{e,g}$ as a function of $U/\varepsilon_T$. We find that
the real part of $(I_{\text{t}})_{e,g}$ is small but features a reasonance at
$U=\varepsilon_T$. We thus find that only a small part of the
transmitted current $I_{\text{t}}$ decays with the decay rate $\Gamma$ while the
majority relaxes with the decay rate $\Gamma_0$.
\section{Schrieffer-Wolff transformation and perturbation theory}
\subsection{Schrieffer-Wolff transformation of the impurity
system}
In the limit $U/\varepsilon_T\rightarrow \infty$ the low-energy sector of the
spectrum of the uncoupled ring impurity features only 
the two charge density wave eigenstates $\vert 2,g \rangle$ and $\vert
2,e \rangle$. From figure 2 it becomes obvious that for
$U/J\gg 1$ the energy gap $\varepsilon$ between the two CDW eigenstates becomes small
compared to the energy separation between the CDW states and the
remainder of the spectrum. It is then intuitive to construct an
effective low-energy Hamiltonian in the subspace of the Hilbert space
$\mathcal{H}_{\text{r}}$, which is
spanned by the two CDW eigenstates. In the limit $U/\varepsilon_T \rightarrow
\infty$ and $U/J\rightarrow
\infty$ the CDW eigenstates take the form of simple product
states $\vert 2,g \rangle = \vert 1\rangle \vert 0\rangle \vert
0\rangle \vert 1 \rangle\equiv \vert 1001 \rangle$ and $\vert 2,e \rangle =
\vert 0\rangle \vert 1\rangle\vert 1\rangle\vert0 \rangle\equiv\vert
0110\rangle$. We define the operator that projects onto this low-energy subspace
as
\begin{align}
  P_0=\vert 1001 \rangle \langle 1001 \vert + \vert 0110 \rangle \langle
  0110 \vert\,.
  \label{}
\end{align}
The Hamiltonian of the full system can be separated into a 
contribution that is diagonal in this new basis 
\begin{align}
  H_0 &= \varepsilon_T n_2 + \sum_{\langle i,j \rangle} U\left(n_i n_j
  -\frac{n_i + n_j}{2}\right)\,,
\end{align}
and a contribution that connects the subspace spanned by the CDW
eigenstates with the rest of the Hilbert space, which reads
\begin{align}
  \hat{V} &= -J \sum_{\langle i,j \rangle} \left(d^{\dagger}_i d_j
  +\text{h.c.}\right)-J_{\text{c}}\left(d_1^{\dagger}c_{\text{L},0}+d_4^{\dagger}
  c_{\text{R},0}+\text{h.c.}\right)\,.
  \label{}
\end{align}
The second contribution can be regarded as a small perturbation. We then perform a
Schrieffer-Wolff transformation to project onto the subspace spanned by
the CDW states and to virtually include transitions to states orthogonal
to the CDW eigenstates. We follow the work by Bravyi, DiVincenzo and Loss [Ann. Phys. \textbf{326}, 2793 (2011)]  to expand the general expression for a transformation of the Hamiltonian
\begin{align}
  H_{\text{eff}}=P_0 \exp(S)(H_0 + \hat{V}) \exp(-S)P_0\,,
  \label{}
\end{align}
where $\exp(S)$ is a unitary operator, into a power series up to fourth order
in the perturbation
$\hat{V}=V_{\text{od}}+V_{\text{d}}$ reading
\begin{align}
  H_{\text{eff}}^{(4)}=H_0 P_0 &+ P_0 \hat{V} P_0+\frac{1}{2}P_0
  \left[\mathcal{L}(V_{\text{od}}),V_{\text{od}}\right] P_0 -
  \frac{1}{2}P_0
  \left([V_{\text{od}},[\mathcal{L}(V_{\text{d}}),[\mathcal{L}(V_{\text{d}}),\mathcal{L}(V_{\text{od}})]]]\right)
  P_0\\\nonumber
  &+\frac{1}{6}P_0\left(
  [V_{\text{od}},\mathcal{L}[\mathcal{L}(V_{\text{od}}),[\mathcal{L}(V_{\text{od}}),
  V_{\text{od}}]]]\right)P_0+\frac{1}{24}P_0\left([\mathcal{L}(V_{\text{od}}),
    [\mathcal{L}(V_{\text{od}}),[\mathcal{L}(V_{\text{od}}),V_{\text{od}}]]]\right)P_0 ,
  \label{}
\end{align}
where $V_{\text{od}}$ denotes the part of the perturbation $\hat{V}$ that
facilitates transitions between the low-energy Hilbert and the
complementary Hilbert space and $V_{\text{d}}$ denotes the part of the
perturbation that only connects states exclusively inside either subspace. We use the shorthand 
\begin{align}
  \mathcal{L}(X)=\frac{\langle i \vert X \vert j \rangle}{E_i-E_j}\vert
  i \rangle\langle j \vert \,,
  \label{}
\end{align}
to denote the inverse energy difference between two states $\vert i
\rangle$ and $\vert j \rangle$ that are connected through the operator $X$.
After evaluation of the commutators we arrive at the expression for the
effective Hamiltonian $H_{\text{eff}}^{(4)}$. It reads
\begin{align}
  H_{\text{eff}}^{(4)}=
H_0 P_0 &+\frac{1}{2}P_0
\left[\mathcal{L}(V_{\text{od}})V_{\text{od}}-V_{\text{od}}\mathcal{L}(V_{\text{od}})\right]P_0\\\nonumber
&-\frac{1}{2}
P_0\left[V_{\text{od}}(\mathcal{L}V_{\text{d}})^2\mathcal{L}(V_{\text{od}})-\mathcal{L}(\mathcal{L}(\mathcal{L}(V_{\text{od}})V_{\text{d}})V_{\text{d}})V_{\text{od}}\right]P_0\\\nonumber
&+\frac{1}{24}P_0\left[(\mathcal{L}(V_{\text{od}}))^3V_{\text{od}}-3\mathcal{L}(V_{\text{od}})^2V_{\text{od}}\mathcal{L}(V_{\text{od}})+3\mathcal{L}(V_{\text{od}})V_{\text{od}}\mathcal{L}(V_{\text{od}}
)^2-V_{\text{od}}\mathcal{L}(V_{\text{od}})^3\right] P_0\\\nonumber
&+\frac{1}{6}P_0\left[V_{\text{od}}\left(\mathcal{L}\mathcal{L}(V_{\text{od}})^2 
V_{\text{od}}-2\mathcal{L}\mathcal{L}(V_{\text{od}})V_{\text{od}}\mathcal{L}(V_{\text{od}})+\mathcal{L}V_{\text{od}}\mathcal{L}(V_{\text{od}})^2\right)\right.\\\nonumber
&\quad\quad -\left.
\mathcal{L}\left(\mathcal{L}(V_{\text{od}})^2V_{\text{od}}^2+2\mathcal{L}(V_{\text{od}})V_{\text{od}}\mathcal{L}(V_{\text{od}})V_{\text{od}}-V_{\text{od}}\mathcal{L}(V_{\text{od}})^2 V_{\text{od}}\right)\right]P_0\,,
\label{eq:sw}
\end{align}
where the first line includes all contributions up to second order in
the perturbation $\hat{V}$ and lines two through five contain the
contributions up to fourth order.
\paragraph{Second order correction}
Evaluating the diagonal contribution $H_0$ for the two CDW eigenstates
yields the effective Hamiltonian in zeroth order as
\begin{align}
  H_{\text{eff}}^{(0)}=H_0 P_0 = \left(\begin{array}{cc}
   \varepsilon_T & 0\\
   0 & 0 
  \end{array}\right)\,,
\end{align}
where from now on we treat the two CDW states like pseudo-spins defined as
\begin{align}
  \vert 0110 \rangle =& \left(\begin{array}{c}
   1 \\
   0
  \end{array}\right)\equiv\vert \uparrow \rangle\\
  \vert 1001 \rangle =& \left(\begin{array}{c}
    0\\
    1
  \end{array}\right)\equiv \vert \downarrow \rangle\,.
\end{align}
The leading order corrections to the effective Hamiltonian are of second
order in the perturbation $\hat{V}$. The corrections encompass 
two consecutive tunneling processes, either tunneling within the ring impurity
leading to corrections
$\propto J^2$, or tunneling between the ring and the leads and back
yielding corrections $\propto J^2_{\text{c}}$. In the follwing we show the
calculation of each correction term featured in
\begin{align}
  H^{(2)}_{\text{eff}} = H_0 P_0 + \frac{1}{2}P_0&
\left[\mathcal{L}(V_{\text{od}})V_{\text{od}}-V_{\text{od}}\mathcal{L}(V_{\text{od}})\right]P_0\,.
  \label{}
\end{align}
The first leading order correction term gives
\begin{align}
  -V_{\text{od}}\mathcal{L}(V_{\text{od}})P_0 =& J
  V_{\text{od}}\mathcal{L} \left(d^{\dagger}_2 d_1 + d^{\dagger}_4 d_2 +
  d^{\dagger}_3 d_4 + d^{\dagger}_1 d_3+d^{\dagger}_1 d_2 +
  d^{\dagger}_2 d_4 + d^{\dagger}_4 d_3+d^{\dagger}_3 d_1
  \right)P_0\\\nonumber
  &+J_{\text{c}} V_{\text{od}}\mathcal{L}\left(d^{\dagger}_1 c_{\text{L},0} +
  d^{\dagger}_4 c_{\text{R},0} + c^{\dagger}_{\text{L},0} d_1 + c^{\dagger}_{\text{R},0}
  d_4\right)P_0\\\nonumber
  =&J V_{\text{od}}\left(\frac{1}{U+\varepsilon_T}d^{\dagger}_2 d_1 +
  \frac{1}{U}d^{\dagger}_3 d_4 
  \frac{1}{U+\varepsilon_T}d^{\dagger}_2 d_4 +\frac{1}{U} d^{\dagger}_3 d_1 \right)\vert
  1001 \rangle\\\nonumber
  &+J V_{\text{od}}\left(\frac{1}{U-\varepsilon_T}d^{\dagger}_4 d_2 +
  \frac{1}{U}d^{\dagger}_1 d_3+\frac{1}{U-\varepsilon_T}d^{\dagger}_1 d_2 +
  \frac{1}{U}d^{\dagger}_4 d_3\right)\vert
  0110 \rangle\\\nonumber
  &+J_{\text{c}} V_{\text{od}}\left[\left(\frac{1}{U}d^{\dagger}_1 c_{\text{L},0} +
  \frac{1}{U}d^{\dagger}_4 c_{\text{R},0}\right)\vert 0110 \rangle
  +\left(\frac{1}{U} c^{\dagger}_{\text{L},0} d_1 + \frac{1}{U}c^{\dagger}_{\text{R},0}
  d_4\right)\vert 1001 \rangle\right]\\\nonumber
=& -J^2 \left(\frac{1}{U+\varepsilon_T}d^{\dagger}_3 d_4 d^{\dagger}_2 d_1 +
  \frac{1}{U}d^{\dagger}_2 d_1 d^{\dagger}_3 d_4 
  +\frac{1}{U+\varepsilon_T}d^{\dagger}_3 d_1 d^{\dagger}_2 d_4
  +\frac{1}{U}d^{\dagger}_2 d_4 d^{\dagger}_3 d_1 \right)\vert
  1001 \rangle\\\nonumber
  &-J^2 \left(\frac{1}{U+\varepsilon_T}d^{\dagger}_1 d_2 d^{\dagger}_2 d_1 +
  \frac{1}{U}d^{\dagger}_4 d_3 d^{\dagger}_3 d_4 
  +\frac{1}{U+\varepsilon_T}d^{\dagger}_4 d_2 d^{\dagger}_2 d_4
  +\frac{1}{U}d^{\dagger}_1 d_3 d^{\dagger}_3 d_1 \right)\vert
  1001 \rangle\\\nonumber
  &-J^2 \left(\frac{1}{U-\varepsilon_T}d^{\dagger}_1 d_3 d^{\dagger}_4 d_2 +
  \frac{1}{U}d^{\dagger}_4 d_2 d^{\dagger}_1
  d_3+\frac{1}{U-\varepsilon_T}d^{\dagger}_4 d_3 d^{\dagger}_1 d_2 +
  \frac{1}{U}d^{\dagger}_1 d_2 d^{\dagger}_4 d_3\right)\vert
  0110 \rangle\\\nonumber
  &-J^2\left(\frac{1}{U-\varepsilon_T}d^{\dagger}_2 d_4 d^{\dagger}_4 d_2 +
  \frac{1}{U}d^{\dagger}_3 d_1 d^{\dagger}_1
  d_3+\frac{1}{U-\varepsilon_T}d^{\dagger}_2 d_1 d^{\dagger}_1 d_2 +
  \frac{1}{U}d^{\dagger}_3 d_4 d^{\dagger}_4 d_3\right)\vert
  0110 \rangle\\\nonumber
  &-J_{\text{c}}^2\left[\frac{1}{U}\left(c^{\dagger}_{\text{L},0} d_1 d^{\dagger}_1 c_{\text{L},0} +
  c^{\dagger}_{\text{R},0} d_4 d^{\dagger}_4 c_{\text{R},0}\right)\vert 0110
  \rangle\right]\\\nonumber
  &-J_{\text{c}}^2\left[
  \frac{1}{U}\left(d^{\dagger}_1 c_{\text{L},0} c^{\dagger}_{\text{L},0} d_1 +
  d^{\dagger}_4 c_{\text{R},0} c^{\dagger}_{\text{R},0}
  d_4\right)\vert 1001 \rangle\right]\\\nonumber
=& -J^2 \left(\frac{1}{U+\varepsilon_T}d^{\dagger}_2 d^{\dagger}_3 d_4 d_1 +
  \frac{1}{U}d^{\dagger}_2 d^{\dagger}_3 d_4 d_1
  -\frac{1}{U+\varepsilon_T}d^{\dagger}_2 d^{\dagger}_3 d_4 d_1
  -\frac{1}{U}d^{\dagger}_2 d^{\dagger}_3 d_4 d_1 \right)\vert
  1001 \rangle\\\nonumber
  &-J^2 \left(\frac{2}{U+\varepsilon_T} +
  \frac{2}{U}\right) 
  \vert
  1001 \rangle\\\nonumber
  &-J^2 \left(\frac{-1}{U-\varepsilon_T}d^{\dagger}_1 d^{\dagger}_4 d_3 d_2 -
  \frac{1}{U}d^{\dagger}_1 d^{\dagger}_4
  d_3 d_2+\frac{1}{U-\varepsilon_T}d^{\dagger}_1 d^{\dagger}_4 d_3 d_2 +
  \frac{1}{U}d^{\dagger}_1 d^{\dagger}_4 d_3 d_2\right)\vert
  0110 \rangle\\\nonumber
  &-J^2\left(\frac{2}{U-\varepsilon_T}
  +\frac{2}{U}
  \right)\vert
  0110 \rangle\\\nonumber
  &-J_{\text{c}}^2\left(\frac{(1-n_{\text{L},0})+(1-n_{\text{R},0})}{U}\vert 1001 \rangle +
  \frac{n_{\text{L},0}+n_{\text{R},0}}{U}\vert 0110
  \rangle\right)\,,
\label{}
\end{align}
where $n_{\text{L},0}$ and $n_{\text{R},0}$ is the electron density on the site of the
lead 
closest to the impurity for the left and the right lead respectively.
The second term of the leading order correction yields 
\begin{align}
  \mathcal{L}(V_{\text{od}})V_{\text{od}}P_0
  =& J^2 \left(\frac{-1}{U}d^{\dagger}_3 d_4 d^{\dagger}_2 d_1 +
  \frac{-1}{U+\varepsilon_T}d^{\dagger}_2 d_1 d^{\dagger}_3 d_4 
  +\frac{-1}{U}d^{\dagger}_3 d_1 d^{\dagger}_2 d_4
  +\frac{-1}{U+\varepsilon_T}d^{\dagger}_2 d_4 d^{\dagger}_3 d_1 \right)\vert
  1001 \rangle\\\nonumber
  &+ J^2 \left(\frac{-1}{U+\varepsilon_T}d^{\dagger}_1 d_2 d^{\dagger}_2 d_1 +
  \frac{-1}{U}d^{\dagger}_4 d_3 d^{\dagger}_3 d_4 
  +\frac{-1}{U+\varepsilon_T}d^{\dagger}_4 d_2 d^{\dagger}_2 d_4
  +\frac{-1}{U}d^{\dagger}_1 d_3 d^{\dagger}_3 d_1 \right)\vert
  1001 \rangle\\\nonumber
  &+J^2 \left(\frac{-1}{U}d^{\dagger}_1 d_3 d^{\dagger}_4 d_2 +
  \frac{-1}{U-\varepsilon_T}d^{\dagger}_4 d_2 d^{\dagger}_1
  d_3+\frac{-1}{U}d^{\dagger}_4 d_3 d^{\dagger}_1 d_2 +
  \frac{-1}{U-\varepsilon_T}d^{\dagger}_1 d_2 d^{\dagger}_4 d_3\right)\vert
  0110 \rangle\\\nonumber
  &+J^2\left(\frac{-1}{U-\varepsilon_T}d^{\dagger}_2 d_4 d^{\dagger}_4 d_2 +
  \frac{-1}{U}d^{\dagger}_3 d_1 d^{\dagger}_1
  d_3+\frac{-1}{U-\varepsilon_T}d^{\dagger}_2 d_1 d^{\dagger}_1 d_2 +
  \frac{-1}{U}d^{\dagger}_3 d_4 d^{\dagger}_4 d_3\right)\vert
  0110 \rangle\\\nonumber
  &+J_{\text{c}}^2\left[\frac{-1}{U}\left(c^{\dagger}_{\text{L},0} d_1 d^{\dagger}_1 c_{\text{L},0} +
  c^{\dagger}_{\text{R},0} d_4 d^{\dagger}_4 c_{\text{R},0}\right)\vert 0110
  \rangle\right]\\\nonumber
  &+J_{\text{c}}^2\left[
  \frac{-1}{U}\left(d^{\dagger}_1 c_{\text{L},0} c^{\dagger}_{\text{L},0} d_1 +
  d^{\dagger}_4 c_{\text{R},0} c^{\dagger}_{\text{R},0}
  d_4\right)\vert 1001 \rangle\right]\\\nonumber
  =&J^2 \left(\frac{-1}{U}d^{\dagger}_2 d^{\dagger}_3 d_4 d_1 +
  \frac{-1}{U+\varepsilon_T}d^{\dagger}_2 d^{\dagger}_3 d_4 d_1
  -\frac{-1}{U}d^{\dagger}_2 d^{\dagger}_3 d_4 d_1
  -\frac{-1}{U+\varepsilon_T}d^{\dagger}_2 d^{\dagger}_3 d_4 d_1 \right)\vert
  1001 \rangle\\\nonumber
  &+J^2 \left(\frac{-2}{U+\varepsilon_T} +
  \frac{-2}{U}\right) 
  \vert
  1001 \rangle\\\nonumber
  &+J^2 \left(\frac{1}{U}d^{\dagger}_1 d^{\dagger}_4 d_3 d_2 -
  \frac{-1}{U-\varepsilon_T}d^{\dagger}_1 d^{\dagger}_4
  d_3 d_2+\frac{-1}{U}d^{\dagger}_1 d^{\dagger}_4 d_3 d_2 +
  \frac{-1}{U-\varepsilon_T}d^{\dagger}_1 d^{\dagger}_4 d_3 d_2\right)\vert
  0110 \rangle\\\nonumber
  &+J^2\left(\frac{-2}{U-\varepsilon_T}
  +\frac{-2}{U}
  \right)\vert
  0110 \rangle\\\nonumber
  &+J_{\text{c}}^2\left(-\frac{(1-n_{\text{L},0})+(1-n_{\text{R},0})}{U}\vert 1001 \rangle -
  \frac{n_{\text{L},0}+n_{\text{R},0}}{U}\vert 0110
  \rangle\right)\,.
  \label{}
\end{align}
Assuming that the mean electron density in both leads combined is
$n_{\text{L},x}+n_{\text{R},x}=1$, the two correction terms are identical.
The effective Hamiltonian in leading order then 
reads
\begin{align}
  H^{(2)}_{\text{eff}} =& 
  \left[\varepsilon_T-J^2\left(\frac{2}{U-\varepsilon_T}+\frac{2}{U}\right)-\frac{J^2_{\text{c}}}{U}\right]\left(\frac{1}{2}\mathbf{1}+S^z\right)
     +\left[-J^2
     \left(\frac{2}{U+\varepsilon_T}+\frac{2}{U}\right)-\frac{J^2_{\text{c}}}{U}\right]\left(\frac{1}{2}\mathbf{1}-S^z\right)\\\nonumber
     =&
     \left[\varepsilon_T-J^2\left(\frac{2}{U-\varepsilon_T}-\frac{2}{U+\varepsilon_T}\right)\right]S^z
     \equiv \varepsilon^{(2)} S^z\,.
  \label{}
\end{align}
We see that the effective energy gap $\varepsilon^{(2)}$ between the ground state $\vert \downarrow
\rangle$ and the excited state $\vert \uparrow \rangle$ is reduced
as compared to the bare energy gap $\varepsilon_T$ by the
perturbative corrections.
In leading order we furthermore find no off-diagonal terms and as it turns out
not in any higher order $(J^2)^n$ of perturbations $\hat{V}$ which feature
only in-ring hopping terms $\sim J d_i^{\dagger} d_j$.
Since the hopping between ring and leads alone cannot facilitate a
pseudo-spin flip, one finds that they are not possible in leading order.
\paragraph{Mirror symmetry}
\begin{figure}[]
  \centering
  \includegraphics[width=0.7\textwidth]{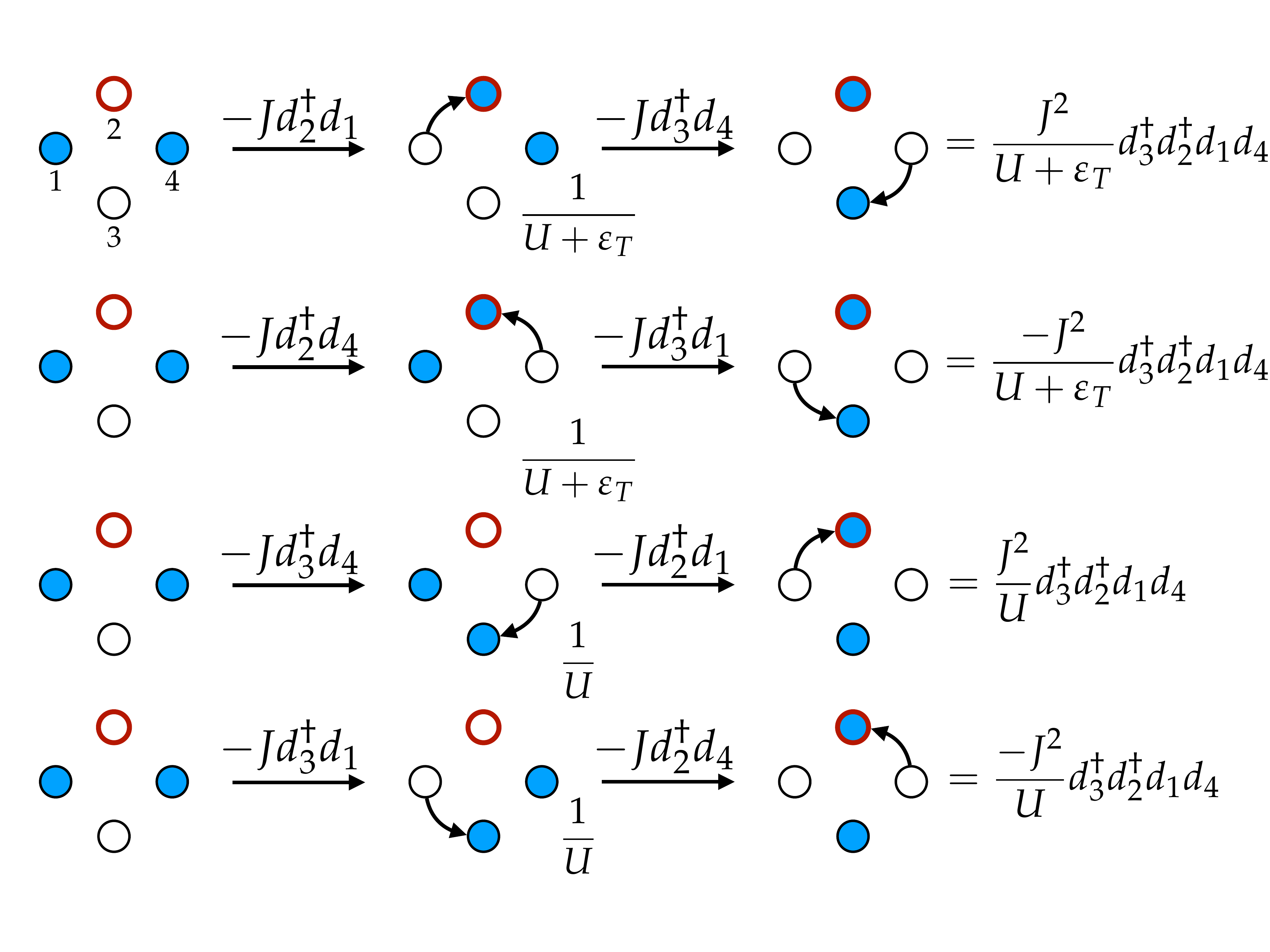}
  \caption{Schematic representation of the correction terms connecting
  $\vert 1001 \rangle$ and $\vert 0110 \rangle$ in second order in the
  perturbation $\hat{V}$. There are two sets of two processes with the same
amplitude that corresponds to processes which are mirror images of one
another in the axis through sites $2$ and $3$. These processes are of
opposite sign and we thus find pairwise cancellation of the off-diagonal
correction terms.}
  \label{fig:2order}
\end{figure}
We attribute the lack of the off-diagonal, pseudo-spin flip terms to a
symmetry of the nanostructure associated with the mirror symmetry in the
axis through lattice sites $2$ and $3$ or equivalently an exchange of
lattice sites
$1\leftrightarrow 4$. The operator $M$ corresponding to this
symmetry reads 
\begin{align}
  M &=
  \mathbf{1}+\left(d^{\dagger}_1-d^{\dagger}_4\right)\left(d_4-d_1\right)
  \\\nonumber
  &=\frac{1}{2}\left[-\left(d^{\dagger}_4 d^{\dagger}_1 d_1 d_4 +
    d^{\dagger}_1 d^{\dagger}_4 d_4 d_1\right) + \left(d_4 d_1 d^{\dagger}_1
  d^{\dagger}_4 + d_1 d_4 d^{\dagger}_4 d^{\dagger}_1\right)\right] +
  d^{\dagger}_1 d_4 + d^{\dagger}_4 d_1 \,.
  \label{}
\end{align}
The symmetry operator satisfies
\begin{align}
  M \vert 1001 \rangle &= - \vert 1001 \rangle = - \vert \downarrow
  \rangle\\
  M \vert 0110 \rangle &= + \vert 0110 \rangle = + \vert \uparrow
  \rangle\,,
  \label{}
\end{align}
as well as
\begin{align}
  M^2 = \mathbf{1}\,,
  \label{}
\end{align}
and 
\begin{align}
  \left[H_{\text{r}}, M \right]=0 \,.
  \label{}
\end{align}
The CDW eigenstates of the Hamiltonian for $U/J \rightarrow \infty$ are
also eigenstates of the symmetry operator $M$ with
eigenvalues $m=\pm 1$.
Since $M$ commutes with the Hamiltonian $H_{\text{r}}$ of the
uncoupled impurity for arbitrary $U/J$, the in-ring hopping terms of
$H_{\text{r}}$ cannot couple 
the different eigenstates of $M$. For the Hamiltonian $H_{\text{c}}$, which couples the ring to the leads,
we instead find $\left[ H_{\text{c}},M\right] \neq 0$. As a consequence we have
$\left[(H_{\text{c}}+H_{\text{r}})^2,M\right]\neq 0$, indicating that off-diagonal, pseudo-spin flip terms can
occur in higher orders of the perturbation.
We illustrate the connection between the mirror symmetry and the absence of
off-diagonal terms in leading order in figure \ref{fig:2order}. Each
process connecting $\vert \downarrow \rangle$ and $\vert \uparrow \rangle$ has a
mirror image with opposite sign leading to pairwise cancellation of all
terms.
\paragraph{Fourth order correction}
In order to obtain finite off-diagonal terms in the effective
Hamiltonian one needs to include the fourth
order corrections. Here we show an example calculation of one
such correction term. All other fourth order correction terms follow accordingly. 
\begin{align}
  V_{\text{od}}(\mathcal{L}V_{\text{d}})^2\mathcal{L}(V_{\text{od}})\vert
  1001 \rangle =&
  V_{\text{od}}(\mathcal{L}V_{\text{d}})^2\mathcal{L}\left[-J\left(d^{\dagger}_2
  d_1 + d^{\dagger}_2 d_4 + d^{\dagger}_3 d_4 + d^{\dagger}_3 d_1
  \right)\right.\\\nonumber
  &+\left.-J_{\text{c}} \left(c^{\dagger}_{\text{L},0} d_1 + c^{\dagger}_{\text{R},0}
  d_4\right)\right]\vert 1001 \rangle\\\nonumber
  =& V_{\text{od}} (\mathcal{L} V_{\text{d}})^2
  \left[\frac{-J}{U+\varepsilon_T}\left(d^{\dagger}_2 d_1 +
  d^{\dagger}_2 d_4\right)+\frac{-J}{U}\left( d^{\dagger}_3 d_4 +
  d^{\dagger}_3 d_1\right)\right.\\\nonumber
  &\left.+\frac{-J}{U}\left( c^{\dagger}_{\text{L},0} d_1 +
  c^{\dagger}_{\text{R},0} d_4 \right)\right]\vert 1001 \rangle\\\nonumber
  =&V_{\text{od}}\mathcal{L}V_{\text{d}}\left[\frac{J_{\text{c}}
  J}{(U+\varepsilon)^2}\left(\underline{d^{\dagger}_1 c_{\text{L},0}
  d^{\dagger}_2 d_1} +
  c^{\dagger}_{\text{R},0} d_4 d^{\dagger}_2 d_1+c^{\dagger}_{\text{L},0} d_1
  d^{\dagger}_2 d_4 + d^{\dagger}_4 c_{\text{R},0} d^{\dagger}_2 d_4\right)\right.\\\nonumber
  &\left. + \frac{J_{\text{c}} J}{U^2}\left(c^{\dagger}_{\text{L},0} d_1 d^{\dagger}_3
  d_4 + d^{\dagger}_4 c_{\text{R},0} d^{\dagger}_3 d_4+d^{\dagger}_1 c_{\text{L},0}
  d^{\dagger}_3 d_1 + c^{\dagger}_{\text{R},0} d_4 d^{\dagger}_3
  d_1\right)\right.\\\nonumber
  & + \frac{J J_{\text{c}}}{(U+\varepsilon_T)U}\left(d^{\dagger}_2 d_4
  c^{\dagger}_{\text{L},0} d_1 + d^{\dagger}_2 d_1 c^{\dagger}_{\text{R},0} d_4
  \right)\\\nonumber
  &\left. + \frac{J J_{\text{c}}}{U^2}\left(d^{\dagger}_3 d_4 c^{\dagger}_{\text{L},0}
  d_1 + d^{\dagger}_3 d_1 c^{\dagger}_{\text{R},0} d_4 \right)\right] \vert
  1001
  \rangle
  \label{eq:first4order}
\end{align}
The contributions that feature the underlined term in equation
(\ref{eq:first4order}) combine into
\begin{align}
  P_0 V_{\text{od}} \mathcal{L} V_{\text{d}}\frac{J_{\text{c}}
  J}{(U+\varepsilon_T)^2}d^{\dagger}_1 c^{\dagger}_{\text{L},0} d^{\dagger}_2
  d_1 \vert 1001 \rangle =&
  P_0\frac{J^2 J_{\text{c}}^2}{(U+\varepsilon_T)^3}\times\\\nonumber
  &\times\left(c^{\dagger}_{\text{L},0} d_1
  d^{\dagger}_3 d_4 d^{\dagger}_1 c_{\text{L},0} d^{\dagger}_2
  d_1 + c^{\dagger}_{\text{R},0} d_4 d^{\dagger}_3 d_1 d^{\dagger}_1 c_{\text{L},0} d^{\dagger}_2
  d_1\right.\\\nonumber
  &+ \left.d^{\dagger}_3 d_4 c^{\dagger}_{\text{L},0} d_1 d^{\dagger}_1 c_{\text{L},0} d^{\dagger}_2
  d_1 + d^{\dagger}_3 d_1 c^{\dagger}_{\text{R},0} d_4 d^{\dagger}_1 c_{\text{L},0} d^{\dagger}_2
  d_1 \right.\\\nonumber
  &+ \left. d^{\dagger}_1 d_2 c^{\dagger}_{\text{L},0} d_1 d^{\dagger}_1 c_{\text{L},0} d^{\dagger}_2
  d_1 + d^{\dagger}_4 d_2 c^{\dagger}_{\text{R},0} d_4 d^{\dagger}_1 c_{\text{L},0} d^{\dagger}_2
  d_1 \right)\vert 1001 \rangle\\\nonumber
  =&\frac{J^2 J^2_{\text{c}}}{(U+\varepsilon_T)^3}\left[\left(2 n_{\text{L},0}
  - 2 c^{\dagger}_{\text{R},0} c_{\text{L},0}\right) \vert 0110 \rangle\right.
  \\\nonumber
  &+\left.\left(n_{\text{L},0} + c^{\dagger}_{\text{R},0} c_{\text{L},0}\right) \vert 1001
  \rangle \right]\,.
  \label{}
\end{align}
We find that the pseudo-spins on the impurity
couple to a second spin-like degree of freedom in the leads which can be
associated with the symmetric
and antisymmetric modes in the leads. We define the annihilation
operator for an electron with pseudo-spin $\sigma$ in the leads as 
\begin{align}
  c_{\uparrow}=&\frac{1}{\sqrt{2}}\left(c_{\text{L},0}+c_{\text{R},0}\right)\\
  c_{\downarrow}=&\frac{1}{\sqrt{2}}\left(c_{\text{L},0}-c_{\text{R},0}\right)\,.
  \label{}
\end{align}
We use this pseudo-spin notation for the lead degrees of freedom and
collect the different correction terms up to fourth order in the perturbation.
The corrections read
\begin{align}
  V_{\text{od}}(\mathcal{L} V_{\text{d}})^2
  \mathcal{L}(V_{\text{od}})\vert 1001 \rangle =& 2J^2 J^2_{\text{c}}
  \left(\frac{3}{U^3}+\frac{1}{U^2
  (U+\varepsilon_T)}-\frac{1}{U
  (U+\varepsilon_T)^2}-\frac{3}{(U+\varepsilon_T)^3}\right)
  d^{\dagger}_{\uparrow} d_{\downarrow} c^{\dagger}_{\downarrow}
  c_{\uparrow}\\\nonumber
  &+2 J^2 J^2_{\text{c}}
  \left[\left(\frac{1}{U^3}+\frac{1}{(U+\varepsilon_T)^3}\right)d^{\dagger}_{\downarrow}
  d_{\downarrow} c^{\dagger}_{\uparrow} c_{\uparrow}\right. \\\nonumber
  &\left.+\left(\frac{4}{U^3}+\frac{1}{U^2
  (U+\varepsilon_T)}+\frac{2}{U(U+\varepsilon_T)^2}+\frac{1}{(U+\varepsilon_T)^3}\right)d^{\dagger}_{\downarrow}
  d_{\downarrow} c_{\downarrow} c^{\dagger}_{\downarrow}\right]\,,
  \label{}
\end{align}
\begin{align}
  -\mathcal{L}(\mathcal{L}(\mathcal{L}(V_{\text{od}})V_{\text{d}})V_{\text{d}})
  V_{\text{od}} \vert 1001 \rangle =& 2 J^2 J^2_{\text{c}}
  \left(\frac{3}{(U-\varepsilon_T)^3}+\frac{1}{U(U-\varepsilon_T)^2}-\frac{1}{U^2
  (U-\varepsilon_T)}-\frac{3}{U^3}\right) d^{\dagger}_{\uparrow}
  d_{\downarrow} c^{\dagger}_{\downarrow} c_{\downarrow}\\\nonumber
  &+2 J^2 J^2_{\text{c}}
  \left[\left(\frac{1}{U^3}+\frac{1}{(U+\varepsilon_T)^3}\right)d^{\dagger}_{\downarrow}
  d_{\downarrow} c^{\dagger}_{\uparrow} c_{\uparrow}\right. \\\nonumber
  &\left.+\left(\frac{4}{U^3}+\frac{1}{U^2
  (U+\varepsilon_T)}+\frac{2}{U(U+\varepsilon_T)^2}+\frac{1}{(U+\varepsilon_T)^3}\right)d^{\dagger}_{\downarrow}
  d_{\downarrow} c_{\downarrow} c^{\dagger}_{\downarrow}\right]\,,
\label{}
\end{align}
\begin{align}
  V_{\text{od}}(\mathcal{L} V_{\text{d}})^2
  \mathcal{L}(V_{\text{od}})\vert 0110 \rangle =& 2 J^2 J^2_{\text{c}}
  \left(\frac{3}{U^3}+\frac{1}{U^2(U-\varepsilon_T)}-\frac{1}{U(U-\varepsilon_T)^2}-\frac{3}{(U-\varepsilon_T)^3}\right)d^{\dagger}_{\downarrow}
  d_{\uparrow} c_{\downarrow} c^{\dagger}_{\uparrow}\\\nonumber
  &+2 J^2 J^2_{\text{c}}
  \left[\left(\frac{1}{U^3}+\frac{1}{(U-\varepsilon_T)^3}\right)
  d^{\dagger}_{\uparrow} d_{\uparrow} c_{\uparrow}
  c^{\dagger}_{\uparrow}\right.\\\nonumber
  &\left. +
  \left(\frac{4}{U^3}+\frac{1}{U^2(U-\varepsilon_T)}+\frac{2}{U(U-\varepsilon_T)^2}+\frac{1}{(U-\varepsilon_T)^3}\right)
  d^{\dagger}_{\uparrow} d_{\uparrow} c^{\dagger}_{\downarrow}
  c_{\downarrow}\right]\,,
  \label{}
\end{align}
\begin{align}
  -\mathcal{L}(\mathcal{L}(\mathcal{L}(V_{\text{od}})V_{\text{d}})V_{\text{d}})V_{\text{od}} \vert 0110 \rangle =& 2 J^2 J^2_{\text{c}}
  \left(\frac{3}{(U+\varepsilon_T)^3}+\frac{1}{U(U+\varepsilon_T)^2}-\frac{1}{U^2(U+\varepsilon_T)^2}-\frac{3}{U^3}\right)\times
  \\\nonumber
  &\times d^{\dagger}_{\downarrow}
  d_{\uparrow} c_{\downarrow} c^{\dagger}_{\uparrow}\\\nonumber
  &+2 J^2 J^2_{\text{c}}
  \left[\left(\frac{1}{U^3}+\frac{1}{(U-\varepsilon_T)^3}\right)
  d^{\dagger}_{\uparrow} d_{\uparrow} c_{\uparrow}
  c^{\dagger}_{\uparrow}\right.\\\nonumber
  &\left. +
  \left(\frac{4}{U^3}+\frac{1}{U^2(U-\varepsilon_T)}+\frac{2}{U(U-\varepsilon_T)^2}+\frac{1}{(U-\varepsilon_T)^3}\right)
  d^{\dagger}_{\uparrow} d_{\uparrow} c^{\dagger}_{\downarrow}
  c_{\downarrow}\right]\,.
  \label{}
\end{align}
First, we identify the off-diagonal terms which couple $\vert 1001 \rangle =
\vert \downarrow \rangle$ and $\vert 0110 \rangle = \vert \uparrow
\rangle$. Now, unlike the leading order corrections, the off-diagonal terms are finite
and cause a simultaneous pseudo-spin-flip on the impurity and in the leads. We can
thus express the off-diagonal terms as
\begin{align}
  -J^2 J^2_{\text{c}}
  &\sum_{\sigma}\left[\frac{1}{U^2(U+\varepsilon_T)}-\frac{1}{U^2(U-\varepsilon_T)}-\left(\frac{1}{U(U+\varepsilon_T)^2}-\frac{1}{U(U-\varepsilon_T)^2}\right)\right.\\\nonumber
  &\left.-\left(\frac{3}{(U+\varepsilon_T)^3}-\frac{3}{(U-\varepsilon_T)^3}\right)\right]
  d^{\dagger}_{\sigma} d_{\bar{\sigma}} c^{\dagger}_{\bar{\sigma}}
  c_{\sigma} \\\nonumber
  =&-2J^2 J_{\text{c}}^2 \sum_{p=\pm}\left( \frac{p}{U^2
(U+p\varepsilon)}-\frac{p}{U
(U+p\varepsilon)^2}-\frac{3p}{(U+p\varepsilon)^3}\right)\left(S^x S^x_{\text{res}} +
S^y S^y_{\text{res}}\right)\,.
  \label{}
\end{align}
We find that the off-diagonal terms describe a spin-spin interaction in the
x and y direction with an amplitude $J_{\perp}(U,\varepsilon_T)$. The
diagonal correction terms are 
\begin{align}
  -2 J^2 J^2_{\text{c}}
  \left(\frac{1}{U^3}+\frac{1}{(U+\varepsilon_T)^3}\right)&\left(\frac{1}{2}\mathbf{1}-S^z\right)\left(\frac{1}{2}\mathbf{1}+S^z_{\text{res}}\right)\\\nonumber
   -2 J^2 J^2_{\text{c}} 
   \left(\frac{4}{U^3}+\frac{1}{U^2(U+\varepsilon_T)}+\frac{2}{U(U+\varepsilon_T)^2}+\frac{1}{(U+\varepsilon_T)^3}\right)
  &\left(\frac{1}{2}\mathbf{1}-S^z\right)\left(\frac{1}{2}\mathbf{1}+S^z_{\text{res}}\right)\\\nonumber
  -2 J^2 J^2_{\text{c}}
  \left(\frac{1}{U^3}+\frac{1}{(U-\varepsilon_T)^3}\right)&\left(\frac{1}{2}\mathbf{1}+S^z\right)\left(\frac{1}{2}\mathbf{1}-S^z_{\text{res}}\right)\\\nonumber
  -2 J^2 J^2_{\text{c}}
  \left(\frac{4}{U^3}+\frac{1}{U^2(U-\varepsilon_T)}+\frac{2}{U(U-\varepsilon_T)^2}+\frac{1}{(U-\varepsilon_T)^3}\right)&\left(\frac{1}{2}\mathbf{1}+S^z\right)\left(\frac{1}{2}\mathbf{1}-S^z_{\text{res}}\right)\,.
\end{align}
The diagonal correction terms contain three different couplings. A spin-spin interaction in the
z-direction with amplitude $J_z(U,\varepsilon_T)$, a correction to the effective magnetic field on
the impurity and a small effective magnetic field
$h^{*}(U,\varepsilon_t)$ on the sites of the
leads  
next to the impurity. The corrections expressed in the pseudo-spin notation
are
\begin{align}
  &+2 J^2 J^2_{\text{c}} \sum_{p=\pm}
  \left(\frac{5}{U^3}+\frac{1}{U^2(U+p\varepsilon_T)}+\frac{2}{U(U+p\varepsilon_T)^2}+\frac{2}{(U+p\varepsilon_T)^3}\right)S^z
  S^z_{\text{res}}\\\nonumber
  &- J^2 J^2_{\text{c}} \sum_{p=\pm}p
  \left(\frac{5}{U^3}+\frac{1}{U^2(U+p\varepsilon_T)}+\frac{2}{U(U+p\varepsilon_T)^2}+\frac{2}{(U+p\varepsilon_T)^3}\right)\left(S^z_{\text{res}}-S^z\right)\,.
\label{eq:effhlead}
\end{align}
In the last step we collect the correction terms arising from lines three through
five of (\ref{eq:sw}). These read
\begin{align}
  \frac{1}{24}P_0 &\left[(\mathcal{L}(V_{\text{od}}))^3V_{\text{od}}-3\mathcal{L}(V_{\text{od}})^2V_{\text{od}}\mathcal{L}(V_{\text{od}})+3\mathcal{L}(V_{\text{od}})V_{\text{od}}\mathcal{L}(V_{\text{od}}
)^2-V_{\text{od}}\mathcal{L}(V_{\text{od}})^3\right] P_0\\\nonumber
=&-\left[\sum_{p=\pm}\frac{4p}{3}J^4
\left(\frac{1}{(U+p\varepsilon_T)^3}+\frac{1}{U^2
(U+p\varepsilon_T)}+\frac{1}{U(U+p\varepsilon_T)^2}\right)\right.\\\nonumber
&\left. +\frac{2p}{3}J^2 J^2_{\text{c}} \left(\frac{1}{U^2
(U+p\varepsilon_T)}+\frac{1}{U
(U+p\varepsilon_T)^2}\right)\right] S^z\,,
  \label{}
\end{align}
\begin{align}
 &\frac{1}{6}\left\lbrace P_0\left[V_{\text{od}}\left(\mathcal{L}\mathcal{L}(V_{\text{od}})^2 
 V_{\text{od}}-2\mathcal{L}\mathcal{L}(V_{\text{od}})V_{\text{od}}\mathcal{L}(V_{\text{od}})+\mathcal{L}V_{\text{od}}\mathcal{L}(V_{\text{od}})^2\right)\right]P_0\right.\\\nonumber
 &\left.+P_0\left[-\mathcal{L}\left(\mathcal{L}(V_{\text{od}})^2
V_{\text{od}}^2+2\left(\mathcal{L}(V_{\text{od}})V_{\text{od}}\right)^2-V_{\text{od}}\mathcal{L}(V_{\text{od}})^2
V_{\text{od}}\right)\right]P_0\right\rbrace\\\nonumber
&=
\left[\sum_{p=\pm}p\frac{J^4}{3}\left(\frac{18}{(U+p\varepsilon_T)^3}+\frac{14}{U^2
(U+p\varepsilon_T)}+\frac{14}{U(U+p\varepsilon_T)^2}\right)\right.\\\nonumber
&\left. +p\frac{J^2
J^2_{\text{c}}}{6}\left(\frac{2}{(U+p\varepsilon_T)^3}+\frac{14}{U^2(U+p\varepsilon_T)}+\frac{14}{U(U+p\varepsilon_T)^2}\right)\right]S^z\,.
  \label{}
\end{align}
Adding the correction terms up to fourth order in the perturbation
$\hat{V}$ to the Hamiltonian describing
the lead degrees of freedom we arrive at the effective low-energy Hamiltonian
$H^{(4)}_{\text{eff}}$ which reads
\begin{align}
  H_{\text{eff}}^{(4)}=&h(U,\varepsilon_T) S^z 
  +\tilde{h}(U,\varepsilon_T) S^z_{\text{res}}+J_z(U,\varepsilon_T)S^z S^z_{\text{res}}+
J_{\perp}(U,\varepsilon_T)(S^x S^x_{\text{res}} + S^y S^y_{\text{res}})\\\nonumber	
&+\sum_{k,\sigma}\varepsilon_k
c^{\dagger}_{k,\sigma}c_{k,\sigma}+\theta(-t)\frac{V}{2}\sum_{k,\sigma,\sigma'} 
c^{\dagger}_{k,\sigma}\tau^x_{\sigma
\sigma'}c_{k,\sigma}\,,
\end{align}
where $h(U,\varepsilon_T)$ constitutes an effective magnetic
field on the impurity and incorporates all terms coupling to $S^z \otimes
\mathbf{1}_r$, $\tilde{h}(U,\varepsilon_T)$ denotes the terms proprotional to
$\mathbf{1}_d\otimes S^z_{\text{res}}$, and $J_z(U,\varepsilon_T)$ and
$J_{\perp}(U,\varepsilon_T)$ feature all terms coupling the impurity
spin and the lead spins in the
z-direction or x-y-direction respectively.
This effective model is reminiscent of the anisotropic
single-channel Kondo model  
with anisotropic coupling between lead spins and impurity spin in the z-direction and 
the x-y-plane as well as a magnetic field $h(\varepsilon_T)\simeq
\mathcal{O}(\varepsilon_T)$ on the
impurity.
In our effective model the spin degrees of freedom do not correspond to physical
spins. Instead, we identify pseudo-spins  $\vert \downarrow \rangle \equiv \vert
g \rangle$ and $\vert \uparrow \rangle \equiv \vert e \rangle$ on the
impurity. For vanishing bias voltage, $V=0$, we can identify $\vert \downarrow \rangle$ with the antisymmetric and
$\vert \uparrow \rangle$ with symmetric modes in the leads. The
operators creating these modes read
\begin{align}
  c^{\dagger}_{\uparrow,k}=&\frac{1}{\sqrt{2}}\left(c^{\dagger}_{\text{L},k}+c^{\dagger}_{\text{R},k}\right)\,,\\\nonumber
  c^{\dagger}_{\downarrow,k}=&\frac{1}{\sqrt{2}}\left(c^{\dagger}_{\text{L},k}-c^{\dagger}_{\text{R},k}\right)\,.
  \label{}
\end{align}
For finite
bias voltage, $V\neq 0$, the term
\begin{align}
  \frac{V}{2}\sum_{k,\sigma,\sigma'} 
c^{\dagger}_{k,\sigma}\tau^x_{\sigma,\sigma'}c_{k,\sigma'}\,,
  \label{}
\end{align}
leads to a hybridization of these two modes, which means they are no longer eigenstates
of the lead Hamiltonian. The conserved quantum number becomes the lead
index $\alpha=L,R$ instead. The linear dependance between lead index 
$\alpha$ and pseudo-spin index $\sigma$ in the leads proves to be responsible for
differing properties, e.g. decay rates, of our effective model as compared to the
anisotropic Kondo model.
\subsection{Schrieffer-Wolff transformation of the current operators}
To determine how the operator representing the ring current $I_{\text{r}}$ couples to
to the matrix elements of the reduced impurity density matrix
$\rho_{\text{ns}}$ we perform a second Schrieffer-Wolff
transformation up to leading order in $J^2/U$. From this we obtain an effective ring
current operator $I_{\text{eff}}$ acting in the subspace $P_0$. 
The operators measuring the local currents in the ring were previously defined as
\begin{align}
    I_{\text{r}}&= I_{\text{u}}-I_{\text{l}}\,,\\
    I_{\text{u}}&=-i\left[n_2,H\right]=ieJ\left(d_1^{\dagger}d_2-d^{\dagger}_2
  d_1\right)\,,\\
  I_{\text{l}}&=-i\left[n_3,H\right]=ieJ\left(d_1^{\dagger}d_3-d^{\dagger}_3
 d_1\right)\,.
\end{align}
The Schrieffer-Wolff of the current operators has been performed in the same fashion as the
transformation for the effective Hamiltonian.
The resulting effective ring current operator in leading order reads
\begin{align}
    I_{\text{r,eff}}= P_0 \exp(S)\left(I_{\text{u}}-I_{\text{l}}\right) \exp(-S) P_0&\simeq 
P_0
\left(1+\mathcal{L}(V_{\text{od}})\right)\left(I_{\text{u}}-I_{\text{l}}\right)\left(1-\mathcal{L}(V_{\text{od}})\right)P_0\\\nonumber
&=J^2\left(\frac{2}{U}+\frac{1}{U-\varepsilon_T}+\frac{1}{U+\varepsilon_T}\right)S^y\,.
 \end{align}
One consequently finds that
$(I_{\text{r,eff}})_{\uparrow,\downarrow}=(I_{\text{eff}})^{\dagger}_{\downarrow,\uparrow}\propto
 J^2/U$ and
 $(I_{\text{r,eff}})_{\downarrow,\downarrow}=(I_{\text{eff}})_{\uparrow,\uparrow}\equiv
 0$.
The effective current operator couples exclusively to the off-diagonal
matrix elements $\rho_{\uparrow, \downarrow}$ and $\rho_{\downarrow,\uparrow}$
of the reduced impurity density
matrix $\rho_{\text{ns}}$ in the context the effective low-energy decription. 
The transient decay rate of the ring current is therefore determined
by the decay rate of these two particular matrix elements.

In contrast, the current, which is transmitted through the ring, is equivalent to 
\begin{align}
    I_{\text{t}}=I_{\text{u}}+I_{\text{l}}\,,
    \label{eq:Itr}
\end{align}
as the sum of the two local currents amounts to the total current flowing from one lead through the ring to the second lead.
Using the effective operators for the local currents $I_{\text{u}}$ and $I_{\text{l}}$ after a Schrieffer-Wolff transformation,
we find for the transmitted current in leading order the effective operator
\begin{align}
    I_{\text{t,eff}} = J^2 \left(\frac{2}{U}-\frac{1}{U-\varepsilon_T}-\frac{1}{U+\varepsilon_T}\right) S^y\,.
    \label{eq:ItrSW}
\end{align}
It is immediately obvious that the effective operator for the transmitted current $I_{\text{t,eff}} \rightarrow 0$ for $U/\varepsilon_T \rightarrow \infty$.
As a consequence, the coupling of the transmitted current to the off-diagonal elements of the effective reduced density matrix,
which are the ones exhibiting the small decay rate, is strongly suppressed for strong interaction.
This is consistent with our numerical data from tdDMRG, where we also do not 
observe a slow decay of the transmitted current, but a decay on the time scale given by the hybridization $\Gamma_0$.
\subsection{Limitations on the viability of the effective low-energy model}
As a consistency check of the effective model $H_{\text{eff}}^{(4)}$ up to order $U^{-4}$
we
perform a series expansion of the perturbative corrections in $U^{-n}$ around
$U/J \rightarrow \infty$. The results of this series expansion for the
amplitudes of the spin-spin
interaction terms read 
\begin{align}
J_{z}&=2J^2 J_{\text{c}}^2\sum_{p=\pm}\frac{5}{U^3}+ \frac{1}{U^2
(U+p\varepsilon_T)}+\frac{2}{U 
(U+p\varepsilon_T)^2}+\frac{2}{(U+p\varepsilon_T)^3}\simeq
\frac{10}{U^3}J^2 J_{\text{c}}^2+\mathcal{O}\left(U^{-4}\right)\,,\\
J_{\perp}&=-2J^2 J_{\text{c}}^2 \sum_{p=\pm} \frac{p}{U^2
(U+p\varepsilon_T)}-\frac{p}{U
(U+p\varepsilon_T)^2}-\frac{3p}{(U+p\varepsilon_T)^3}\simeq
0+\mathcal{O}\left(U^{-5}\right)\,.
\end{align}
When expanding the expression for $J_{\perp}$ up to fourth order in the
inverse interaction strength $U^{-1}$ we encounter an inconsistency of
our 
Schrieffer-Wolff transformation as $J_{\perp}$ vanishes up to this
order. We can therefore not assume with certainty that the
spin-spin interaction $J_{\perp}$
in the effective model is finite.
\subsection{Perturbation theory for the effective model 
in the limit $T\rightarrow 0$}
In our effective model the hybridization between the impurity and the
leads satisfies
\begin{align}
\sqrt{\Gamma_0}\propto\text{max}(J_z,J_{\perp})\propto\mathcal{O}(J^2
J_{\text{c}}^2/U^3)\,,
\end{align}
which is small in the limit $U\gg \text{max}(J,\varepsilon)$ even in the
case of a chosen bare coupling $J_{\text{c}}=\mathcal{O}(J)$. In the limit $U/\varepsilon \gg 1$ we can
thus perform a perturbation theory calculation for the effective low-energy model
whilst employing the same values for the bare model parameters $J$, $J_{\text{c}}$ as in our
initial DMRG
calculations. This way we can compare the results from both methods for the decay rate of
the off-diagonal matrix elements $\rho_{ \uparrow,\downarrow }$
and $\rho_{\downarrow , \uparrow }$
and thus the decay rate of the transient ring current in the strong interaction limit. In contrast to the earlier perturbation
theory calculation, we no longer study charge fluctuations on the
impurity but pseudo-spin fluctuations instead. This requires a few 
modifications to the procedure outlined in the previous section on the
perturbation theory. The coupling Liovilliain $L_V$ now features two
field superoperators for the leads and the impurity, instead of just one. It
reads
\begin{align}
  L_V=G^{p_1 p_2}_{12}:J^{p_1}_1 J^{p_2}_2:\,,
  \label{}
\end{align}
where
\begin{align}
  G^{p_1 p_2}_{12}A=\delta_{p_1 p_2}\left\lbrace\begin{array}{ll}
    d_1 d_2 A & p_1 = +\\
    -A d_1 d_2 & p_1 = -
  \end{array}\right. \,.
  \label{}
\end{align}
The field superoperators for the leads remain unchanged 
\begin{align}
  J^p_1 A = \left\lbrace \begin{array}{ll}
    c_1 A &p=+ \\
    A c_1 &p=- 
  \end{array}\right.\,.
  \label{}
\end{align}
As a consequence, the first order corrections to the Liouvillian now contain
two reservoir contractions $\gamma^{pp'}_{11'}$ which in turn requires
integration over two reservoir frequencies $\omega_1$ and $\omega_2$. The
perturbative correction reads
\begin{align}
  \Sigma^{(1)}(E)=\sum_{p_1 p_2 p_3 p_4}\sum_{1 2 3 4}G^{p_1
  p_2}_{12}\frac{1}{E_{12}+\bar{\omega}_{12}-L_{\text{ns}}}G^{p_3
  p_4}_{34} \gamma^{p_1 p_4}_{14} \gamma^{p_2 p_3}_{23}\,.
  \label{}
\end{align}
It is again possible to separate the corrections into a symmetric and an
antisymmetric part. In the zero temperature limit the two contributions read 
\begin{align}
  \Sigma_s(E)&=\frac{1}{4}\sum_{\nu_1,\eta_1}\sum_{\nu_2,\eta_2}\bar{G}_{12}\int_{-\infty}^{\infty}d\omega_1
  d\omega_2\,\rho(\omega_1)\rho(\omega_2)
  \frac{1+\text{sign}\left(\omega_1\right)\text{sign}\left(\omega_2\right)}{E+\omega_1+\omega_2+\eta_1\mu_1+\eta_2\mu_2-L_{\text{ns}}}\bar{G}_{\bar{2}\bar{1}}\,,\\
  \Sigma_a(E)&=-\frac{1}{2}\sum_{\nu_1,\eta_1}\sum_{\nu_2,\eta_2}\bar{G}_{12}\int_{-\infty}^{\infty}d\omega_1
  d\omega_2\,\rho(\omega_1)\rho(\omega_2)
  \frac{\text{sign}\left(\omega_2\right)}{E+\omega_1+\omega_2+\eta_1\mu_1+\eta_2\mu_2-L_{\text{ns}}}\tilde{G}_{\bar{2}\bar{1}}\,.
  \label{}
\end{align}
First we discuss the integrals over the reservoir frequencies $\omega_1$
and $\omega_2$. For this we introduce the density of states
\begin{align}
  \rho(\omega) =2 \rho_0 \theta\left(\omega - \vert D \vert \right)\,.
\end{align}
The integral in the symmetric part of the self-energy correction
evaluates to
\begin{align}
  \int_{-\infty}^{\infty}\int_{-\infty}^{\infty}  d\omega_1
  d\omega_2\, \frac{\rho(\omega_1) \rho(\omega_2)}{4}
  \frac{1+\text{sign}(\omega_1)\text{sign}(\omega_2)}{z+\omega_1+\omega_2}=&\rho_0^2\left(\int_{-D}^{0}\int_{-D}^{0}
  d\omega_1 d\omega_2 \frac{1}{z+\omega_1+\omega_2}\right.\\\nonumber
  &\left. +\int_{0}^{D}\int_{0}^{D}d\omega_1 d\omega_2
  \frac{1}{z+\omega_1+\omega_2}\right)\\\nonumber
  =&\rho_0^2 \left[2 z\log z + (z-2D)\log(z-2D)\right.\\\nonumber
  &+2(z-D)\log(z-D)-2(z+D)\log(z+D)\\\nonumber
&\left.+(z+2D)\log(z+2D)\right]\\\nonumber
  \simeq& \rho_0^2 \left[2z\log z+ (z-2D)\left(i\pi+\log D +\log 2 -
  \frac{z}{2D}\right)\right.\\\nonumber
  &+(z+2D)\left(\log D + \log 2 + \frac{z}{2D}\right)\\\nonumber
  &-2(z+D)\left(\log D +\frac{z}{D}\right)\\\nonumber
&\left.+2(-z+D)\left(i\pi+\log D -\frac{z}{D}\right)\right]\\\nonumber
  \simeq&\rho_0^2 \left[ 2z\log
  \left(\frac{2z}{D}\right)-2z-i\pi z\right]\\\nonumber
  =&\rho^2_0 \left[ 2z \log \left(\left\vert \frac{2z}{D}\right\vert -1\right)
  -i\pi\vert z \vert \right]\,.
  \label{}
\end{align}
Similarly, the integral in the antisymmetric self-energy correction
gives
\begin{align}
  -\int_{-\infty}^{\infty}\int_{-\infty}^{\infty} d\omega_1 d\omega_2
  \,\frac{ \rho(\omega_1) \rho(\omega_2)}{2}
  \frac{\text{sign}(\omega_2)}{z+\omega_1+\omega_2}=&-\int_{-D}^{D}
  2 d\omega_1
  \left(\int_{0}^{D}\frac{\rho_0^2
  d\omega_2}{z+\omega_1+\omega_2}-\int_{-D}^{0}\frac{\rho_0^2 d\omega_2}{z+\omega_1+\omega_2}\right)\\\nonumber
  =&-2\rho_0^2\left\lbrace \int_{-D}^{D} d\omega_1
  \left[-\log(\omega_1
  +z)+\log(\omega_1+z-D)\right]\right.\\\nonumber
  &\left.+\int_{-D}^{D}d\omega_1\left[-\log(\omega_1+z)+\log(\omega_1+z+D)\right]\right\rbrace\\\nonumber
  =&-2\rho_0^2\left[2(z-D)\log(z-D)-2(z+D)\log(z+D)\right.\\\nonumber
  &\left.-(z-2D)\log(z-2D)+(z+2D)\log(z+2D)\right]\\\nonumber
  \simeq& -2\rho_0^2\left[2(z-D)\left(i\pi+\log D
    -\frac{z}{D}\right)\right.\\\nonumber
    &-2(z+D)\left(\log D+\frac{z}{D}\right)\\\nonumber
  &+(z+2D)\left(i\pi+\log D + \log 2 -\frac{z}{2D}\right)\\\nonumber
&\left. -(z-2D)\left(i\pi+\log D + \log 2 -
\frac{z}{2D}\right)\right]\\\nonumber
\simeq&-2\rho_0^2\left(i\pi z+4D \log 2\right)\,.
  \label{}
\end{align}
Next, we discuss the superoperators acting in the Liouville space of the
impurity. We follow the notation introduced in Schoeller and Reininghaus [Phys. Rev. B \textbf{80}, 045117 (2009)]. First we define
the Liouville superoperators that act as the spin operators
$\underline{S}=(S^x,S^y,S^z)$ on the impurity. These Liouville superoperators are
\begin{align}
  \underline{L}^{\pm} = (L^{\pm x},L^{\pm y},L^{\pm z})\,,
  \label{}
\end{align}
where the sign $p=\pm$ indicates the order of the operators as
\begin{align}
  \underline{L}^+ A = \underline{S} A\quad ,\quad \underline{L}^- A = -A
  \underline{S}\,.
  \label{}
\end{align}
A matrix representation of these superoperators in the basis $\vert \uparrow
\uparrow)$, $\vert \downarrow \downarrow )$, $\vert \uparrow \downarrow
)$, $\vert \downarrow \uparrow)$ reads
\begin{align}
  L^{+x}=\begin{pmatrix}
  0 & 0 & 0 &\frac{1}{2} \\
  0 & 0 &\frac{1}{2} & 0 \\
  0 & \frac{1}{2} & 0 & 0 \\
  \frac{1}{2} & 0 & 0 & 0
  \end{pmatrix},\quad
  L^{+y}=\begin{pmatrix}
    0 & 0 & 0 & -\frac{i}{2} \\
    0 & 0 & \frac{i}{2} & 0 \\
    0 & -\frac{i}{2} & 0 & 0 \\
    \frac{i}{2} & 0 & 0 & 0
  \end{pmatrix},\quad
  L^{+z}=\begin{pmatrix}
    \frac{1}{2} & 0 & 0 & 0 \\
    0 &-\frac{1}{2} & 0 & 0 \\
    0 & 0 & \frac{1}{2} & 0 \\
    0 & 0 & 0 & -\frac{1}{2}
  \end{pmatrix}\,,
  \label{}
\end{align}
\begin{align}
  L^{-x}=-\begin{pmatrix}
  0 & 0 & \frac{1}{2} & 0\\
  0 & 0 & 0 & \frac{1}{2} \\
  \frac{1}{2} & 0 & 0 & 0 \\
  0 & \frac{1}{2} & 0 & 0
  \end{pmatrix},\quad
  L^{-y}=\begin{pmatrix}
    0 & 0 & 0 & -\frac{i}{2} \\
    0 & 0 & \frac{i}{2} & 0 \\
    0 & -\frac{i}{2} & 0 & 0 \\
    \frac{i}{2} & 0 & 0 & 0
  \end{pmatrix},\quad
  L^{-z}=\begin{pmatrix}
    \frac{1}{2} & 0 & 0 & 0 \\
    0 &-\frac{1}{2} & 0 & 0 \\
    0 & 0 & \frac{1}{2} & 0 \\
    0 & 0 & 0 & -\frac{1}{2}
  \end{pmatrix}\,.
  \label{}
\end{align}
From these superoperators we can construct a basis of superoperators
sufficient to describe the spin-spin interaction processes between
impurity and lead pseudo-spins. We further introduce the
'scalar' superoperators
\begin{align}
  L^a &= \frac{3}{4}\mathbf{1} + \underline{L}^+
  \cdot\underline{L}^- \,,\\\nonumber
  L^c &= \frac{1}{2}\mathbf{1} + 2 L^{+z} L^{-z} \,, \\\nonumber
  L^h &= L^{+z} + L^{-z}\,,
  \label{}
\end{align}
as well as the vector superoperators
\begin{align}
  \underline{L}^1 &=\frac{1}{2}\left(\underline{L}^+ - \underline{L}^- -
  2i \underline{L}^+ \times \underline{L}^-\right) \,, \\\nonumber
  \underline{L}^2 &= -\frac{1}{2}\left(\underline{L}^+ +
  \underline{L}^-\right)\,,\\\nonumber
  \underline{L}^3 &= \frac{1}{2}\left(\underline{L}^+ - \underline{L}^-
  +2 i \underline{L}^+ \times \underline{L}^-\right)\,.
  \label{}
\end{align}
Due to the anisotropy of the interactions we need to introduce a third
set of superoperators, which reads
\begin{align}
  L^4_{\pm} &= L^{2x}\pm i L^{2y} \pm \left[\left( L^{+x}\pm i
    L^{+y}\right) L^{-z} +
  L^{+z}\left( L^{-x} \pm i L^{-y} \right)\right]\,, \\\nonumber
  L^5_{\pm} &= L^{2x}\pm i L^{2y} \mp \left[\left( L^{+x}\pm i
    L^{+y}\right) L^{-z} +
  L^{+z} \left(L^{-x} \pm i L^{-y} \right)\right]\,, \\\nonumber
  L^6_{\pm} &= L^c \pm
  \frac{1}{2}\left[\left(L^{3x}+iL^{3y}\right)\left(
    L^{1x}+iL^{1y}\right) + \left(L^{3x}-iL^{3y}\right)\left(
    L^{1x}-iL^{1y}\right)
  \right]\,.
\end{align}
In terms of these basis superoperators the bare Liouvillian is given as
\begin{align}
  L_{\text{ns}} = h(U,\varepsilon_T) L^h\,,
\end{align}
where $h(U,\varepsilon_T)$ is the effective magnetic field on
the impurity and $L^h$ represents the action of $[S^z,\bullet]$ on the
impurity.
The effective Liouvillian for the impurity in first order perturbation
theory reads
\begin{align}
  L_{\text{eff}}(E)=L_{\text{ns}}+\Sigma^{(1)}(E)\,.
  \label{}
\end{align}
The first order, energy-dependent self-energy corrections are 
\begin{align}
  \Sigma^{(1)}(E)=&
  \rho_0^2\sum_{\substack{\nu_1,\eta_1 \\
  \nu_2,\eta_2}}\bar{G}_{12}\left[2 (E+\mu_{12}-L_{\text{ns}})\left(\log\left\vert
    \frac{2(E+\mu_{12}-L_{\text{ns}})}{D}\right\vert 
  -1\right)\right]\bar{G}_{\bar{2}\bar{1}}\\\nonumber
  &+\rho_0^2 \sum_{\substack{\nu_1.\eta_1\\ \nu_2,\eta_2}} \bar{G}_{12}
  \left[-i\pi \left\vert
  E+\mu_{12}-L_{\text{ns}}\right\vert\right]\bar{G}_{\bar{2}\bar{1}}\\\nonumber
  &+\rho_0^2\sum_{\substack{\nu_1,\eta_1\\
  \nu_2,\eta_2}}\bar{G}_{12}\left[-8 D \log 2
  -2\pi i(E+\mu_{12}-L_{\text{ns}})\right]\tilde{G}_{\bar{2}\bar{1}}\,,
\end{align}
where $\mu_{12}=\eta_1 \mu_1 + \eta_2 \mu_2$ and
\begin{align}
  \bar{G}_{12} = \left\lbrace \begin{array}{cc}
    +J_{\perp} \left(\tau_{\sigma_1 \sigma_2}^x L^{2x} + \tau_{\sigma_1
    \sigma_2}^y L^{2y} \right) + J_z \tau_{\sigma_1 \sigma_2} L^{2z} &
    \eta_1 = -\eta_2 = + \\
    -J_{\perp} \left(\tau_{\sigma_2 \sigma_1}^x L^{2x} + \tau_{\sigma_2
    \sigma_1}^y L^{2y}\right) - J_z \tau_{\sigma_2 \sigma_1} L^{2z} &
    \eta_1 = -\eta_2 = -
  \end{array}\right. \,,
\end{align}
as well as
\begin{align}
  \tilde{G}_{12} = \left\lbrace \begin{array}{cc}
    +J_{\perp} \left[\tau_{\sigma_1 \sigma_2}^x \left(L^{1x} +
      L^{3x}\right)+ \tau_{\sigma_1
      \sigma_2}^y \left(L^{1y}+L^{3y}\right) \right] + J_z
      \tau_{\sigma_1 \sigma_2} \left(L^{1z}+L^{3z}\right) &
    \eta_1 = -\eta_2 = + \\
    -J_{\perp} \left[\tau_{\sigma_2 \sigma_1}^x
      \left(L^{1x}+L^{3x}\right) + \tau_{\sigma_2
      \sigma_1}^y \left(L^{1y}+L^{3y}\right)\right] - J_z \tau_{\sigma_2
      \sigma_1} \left(L^{1z}+L^{3z}\right) &
    \eta_1 = -\eta_2 = -
  \end{array}\right. \,.
\end{align}
It is then straightforward to calculate the self energy corrections that
are proportional to simple products
$\bar{G}_{12}\bar{G}_{\bar{2}\bar{1}}$ and $\bar{G}_{12}
\tilde{G}_{\bar{2}\bar{1}}$.
The two different products of superoperators evaluate to
\begin{align}
  \bar{G}_{12} \bar{G}_{\bar{2}\bar{1}} =&\sum_{j} \bar{G}_{12} \vert
  v_j )( v_j \vert \bar{G}_{\bar{2}\bar{1}}\\\nonumber
  =& \frac{J^2_{\perp}}{2}L^c
  +\frac{J^2_{\perp}}{2}L^c +
  J^2_{\perp}(L^a-L^c)+\frac{J^2_z}{2}(L^c+L^h)+J^2_{\perp}(L^a-L^c)+\frac{J^2_z}{2}(L^c-L^h)\,,
  \label{}
\end{align}
and
\begin{align}
  \bar{G}_{12} \tilde{G}_{\bar{2}\bar{1}} = \sum_j \bar{G}_{12} \vert
  v_j )( v_j \vert \tilde{G}_{\bar{2}\bar{1}}= \frac{J^2_{\perp}}{2}L^h - \frac{J^2_{\perp}}{2}L^h + J^2_{\perp}
  L^{3 z} - J^2_{\perp} L^{3 z}\,.
  \label{}
\end{align}
To calculate terms involving $\bar{G}f(\mu_{12})\bar{G}$ and $\bar{G}
f(\mu_{12}) \tilde{G}$ a rotation to a different
basis is necessary.
\subsection{Modifications to the perturbation theory for spin
  fluctuations due to the
  linear dependence between pseudo-spin and lead index}
From eq. (\ref{eq:effhlead}) we see that the magnetic field
$\tilde{h}(U,\varepsilon_T)$,
experienced by the pseudo-spins on the lead sites closest to the
impurity, is small up to order $\mathcal{O}(U^{-4})$, ie 
$h^{*}\ll h,V$ and we can thus neglect it. It is then more useful to express the Hamiltonian
in the basis of the lead index $\alpha\in\lbrace
L,R\rbrace$ eigenstates, which corresponds to a rotation 
$\tau^x_{\sigma_1 \sigma_2} \rightarrow \tau^z_{\alpha_1 \alpha_2}$,
$\tau^y_{\sigma_1 \sigma_2} \rightarrow -\tau^y_{\alpha_1 \alpha_2}$,
$\tau^z_{\sigma_1 \sigma_2}
\rightarrow \tau^x_{\alpha_1 \alpha_2}$ in the leads.
In terms of the rotated operators, $c_{\alpha,k}=1/\sqrt{2}(c_{\downarrow,k}\pm
c_{\uparrow,k})$, the effective Hamiltonian reads
\begin{align}
  H_{\text{eff}}^{(4)}=&hS^z+\sum_{k,\alpha}\varepsilon_k
c^{\dagger}_{k,\alpha}c_{k,\alpha}+\frac{V}{2}\sum_{k,\alpha,\alpha'} 
c^{\dagger}_{k,\alpha}\tau^z_{\alpha \alpha'}c_{k,\alpha'}
\\\nonumber
&+\frac{J_z}{2}\sum_{\substack{k,k'\\ \alpha,\alpha'}}
S^z c^{\dagger}_{k,\alpha}\tau^x_{\alpha \alpha'}c_{k',\alpha'}
+\frac{J_{\perp}}{2}\sum_{\substack{k,k'\\ \alpha,\alpha'}} c^{\dagger}_{k,\alpha}c_{k',\alpha'}(
S^x \tau^z_ { \alpha \alpha'}+S^y (-\tau^y_{\alpha\alpha'}))\,.
\end{align}
In this rotated basis the part of the Hamiltonian acting exclusively on the
leads is diagonal so
the reservoir contractions reduce to simple fermionic
distribution functions. After rotation the vertex
superoperators read
\begin{align}
  \bar{G}_{12}=
    \left\lbrace \begin{array}{cc}
    +\left[J_{\perp}L^{2x}\tau^{z}_{\alpha_1 \alpha_2}+J_{\perp}L^{2y}(-\tau^{y}_{
    \alpha_1 \alpha_2})+J_{z}L^{2z}\tau^{x}_{\alpha_1 \alpha_2}\right] &
\eta_1=-\eta_2=+ \\
-\left[J_{\perp}L^{2x}\tau^{z}_{\alpha_2 \alpha_1}+J_{\perp}L^{2y}(-\tau^{y}_{\alpha_2 \alpha_1})+J_{z}L^{2z}\tau^{x}_{\alpha_2 
\alpha_1}\right] & \eta_1=-\eta_2=-
\end{array}\right.\,,
\end{align}
and
\begin{align}
\tilde{G}_{12}=\left\lbrace
\begin{array}{cc}
+\left[J_{\perp}(L^{1x}+L^{3x})\tau^{z}_{\alpha_1 \alpha_2}+J_{\perp}(L^{1y}+L^{3y}
)(-\tau^{y}_{\alpha_1 \alpha_2})+J_{z}(L^{1z}+L^{3z})\tau^{x}_{\alpha_1
\alpha_2}\right] & \eta_1=-\eta_2=+ \\
-\left[J_{\perp}(L^{1x}+L^{3x})\tau^{z}_{\alpha_2
\alpha_1}+J_{\perp}(L^{1y}+L^{3y})(-\tau^{y}_{\alpha_2
\alpha_1})+J_{z}(L^{1z}+L^{3z})\tau^{x}_{
\alpha_2 \alpha_1}\right] & \eta_1=-\eta_2=-
\end{array}\right.\,,
\end{align}
where we have dropped the factor $1/2$ resulting from the substitution
$S_{\sigma_1 \sigma_2} \rightarrow \tau_{\alpha_1 \alpha_2}$ for
convenience. We reintroduce the factor in the final result.
The first set of self-energy corrections that are affected by the
linear dependance between pseudo-spin index and lead index involve terms
proportional to
$\bar{G}_{12}\mu_{12} \bar{G}_{\bar{2} \bar{1}}$ and
$\bar{G}_{12}\mu_{12} \tilde{G}_{\bar{2}\bar{1}}$.
The first term reads
\begin{align}
  \bar{G}_{12}\mu_{12}\bar{G}_{\bar{2}\bar{1}}=\sum_{\alpha_1,\alpha_2}\sum_{\eta_1=-\eta_2}\sum_{l,k=x,y,z}
  \tau^l_{\alpha_1 \alpha_2}\tau^k_{\alpha_2 \alpha_1} L^{2
l}\left(\eta_1 V_{\alpha_1}+\eta_2 V_{\alpha_2}\right)L^{2 k}\,.
\end{align}
In the following we evaluate the cases $l=k$ and
$l\neq k$ separately. For each example calculation we
set $\eta_1=-\eta_2=+$ without loss of generality. For $l=k$
we find
\begin{align}
\bar{G}_{12}\mu_{12}\bar{G}_{\bar{2}\bar{1}}=& J_l^2 \sum_{\substack{\alpha_1,\alpha_2=1,2 \\ l=x,y,z}} \tau^l_{\alpha_1 \alpha_2} 
\tau^l_{\alpha_2 \alpha_1}\left(V_{\alpha_1}-V_{\alpha_2}\right) L^{2l}
L^{2l}\\\nonumber
=&J_l^2 \sum_{\substack{\alpha_1,\alpha_2=1,2 \\
l=x,y,z}}\tau^l_{\alpha_1 \alpha_2} \tau^l_{\alpha_2 \alpha_1}
\left[\text{sign} (V_{\alpha_1}) 
(1-\delta_{\alpha_1 \alpha_2})\right] L^{2 l} L^{2 l}\\\nonumber
=&J_l^2
V\sum_{l=x,y,z}\left[\text{sign}(V_2)\tau^l_{21}\tau^l_{12}+\text{sign}(V_{1})\tau^l_{12} \tau^l_{21}\right] L^{2l} L^{2l}=0\,,
\end{align}
and for $l\neq k$ we obtain
\begin{align}
\bar{G}_{12}\mu_{12}\bar{G}_{\bar{2}\bar{1}}=&J_l J_k
\sum_{\substack{l\neq k \\ l,k=x,y,z}}\sum_{\alpha_1,\alpha_2=1,2}\tau^l_{\alpha_1 \alpha_2} 
\tau^k_{\alpha_2 \alpha_1} \left[\text{sign} (V_{\alpha_1})
(1-\delta_{\alpha_1 \alpha_2})\right] L^{2 l} L^{2 k}\\\nonumber
=&J_l J_k
V\sum_{\substack{l\neq
k\\ l,k=x,y,z}}\left[\text{sign}(V_2)\tau^l_{21}\tau^k_{12}+\text{sign}(V_{1})\tau^l_{12}
\tau^k_{21}\right] L^{2l} L^{2k}\\\nonumber
=&\left(J_z J_{\perp}\right)V L^{2x}\,.
\end{align}
This term, proportional to the bias voltage $V$, does not appear in the
perturbation theory of
the regular anisotropic Kondo model. The correction term still satisfies 
$\text{Tr}_{S}(L^{2x})=0$ such that $\text{Tr}_S(L_{\text{eff}})=0$, a
necessary requirement for the validity of the perturbation theory.
Similarly for $\bar{G}_{12} \mu_{12} \tilde{G}_{\bar{2}\bar{1}}$ we find
\begin{align}
\bar{G}_{12}\mu_{12}\tilde{G}_{\bar{2}\bar{1}}=&J_l J_k
\sum_{\substack{l\neq k \\ l,k=x,y,z}}\sum_{\alpha_1,\alpha_2=1,2}\tau^l_{\alpha_1 \alpha_2} 
\tau^k_{\alpha_2 \alpha_1} \left[\text{sign} (V_{\alpha_1})
(1-\delta_{\alpha_1 \alpha_2})\right] L^{2 l} \left(L^{1 k}+L^{3
k}\right)\\\nonumber
=&J_l J_k
V\sum_{\substack{l\neq
k \\ l,k=x,y,z}}\left[\text{sign}(V_2)\tau^l_{21}\tau^k_{12}+\text{sign}(V_{1})\tau^l_{12}
\tau^k_{21}\right] L^{2l} \left(L^{1 k}+L^{3 k}\right)\\\nonumber
=&\left(J_z J_{\perp}\right)V L^{3x}\,,
\end{align}
which satisfies $\text{Tr}_S(L^{3x})=0$ as well.
Next we discuss the
correction terms proportional to $\bar{G}_{12}\vert
E+\mu_{12}-L_{\text{ns}}\vert\bar{G}_{\bar{2}\bar{1}}$. We know
that $\Gamma_0\ll h$ which means that the 
perturbative corrections to the roots of the
unperturbed Liovillian $L_{\text{ns}}$ are small. We can thus safely assume
$\lambda_{\pm}^{*}=L_{\text{eff}}(\lambda_{\pm}^{*})\simeq \pm
h$. As such we
evaluate the correction terms proportional to $\bar{G}_{12}\vert
E+\mu_{12}-L_{\text{ns}}\vert\bar{G}_{\bar{2}\bar{1}}$ for $E\simeq \pm h$ and obtain 
\begin{align}
  \bar{G}_{12}\vert h+\mu_{12}-L_{\text{ns}}\vert\bar{G}_{\bar{2}\bar{1}}=&J_l J_k
\sum_{\substack{\alpha_1,\alpha_2=1,2 \\ l,k=x,y,z}}\tau^l_{\alpha_1
\alpha_2}
\tau^k_{\alpha_2 \alpha_1}L^{2 l} \left\vert h-L_{\text{ns}}+\eta_1
V_{\alpha_1}+\eta_2 V_{\alpha_2}\right\vert L^{2 k}\\\nonumber
=& \frac{J^2_{\perp}}{4}\left(\vert h+V\vert + \vert
h-V\vert \right) L^6_+ + \frac{J^2_{\perp}}{2}\vert
h \vert L^6_-\\\nonumber
& +\frac{J^2_{\perp}}{4}\left(2\vert \delta h \vert + \vert
\delta h +V \vert + \vert
\delta h-V\vert\right)\left(L^a-L^c\right)\\\nonumber
&+\frac{J^2_{\perp}}{4}\left( 2\vert 2h\vert
+ \vert 2h+V \vert + \vert 2h-V \vert \right)
\left(L^a-L^c\right)\\\nonumber
& + \frac{J_{\perp} J_z}{4}\left(\vert \delta h+V\vert - \vert
\delta h-V\vert\right) \left(L^4_-+L^5_+\right) \\\nonumber
& +\frac{J_{\perp} J_z}{4}\left(\vert 2h+V\vert - \vert
2h-V\vert \right) \left(L^{4}_{+}+ L^{5}_{-}\right)\\\nonumber
& +\frac{J^2_z}{4}\left(\vert\delta h+ V\vert + \vert
\delta h-V\vert \right)
\left(L^c+L^h\right)\\\nonumber
&+\frac{J^2_z}{4}\left(\vert 2h+V\vert
+\vert 2h-V\vert \right) \left(L^c-L^h\right)\,,
\end{align}
where $\delta h=h-h_0$ and we verify that
$\text{Tr}_S(L^4_{\pm})=\text{Tr}_S(L^5_{\pm})=\text{Tr}_S(L^6_{\pm})=0$.
Lastly we discuss the correction term that involves the logarithm of the
Liouvillian $L_{\text{ns}}$. We abbreviate $z=E+\mu_{12}-L_{\text{ns}}$ and approximate $E=\pm h$. The correction term
then evaluates to 
\begin{align}
  \bar{G}_{12} z\log\left\vert
  \frac{2z}{D}\right\vert
  \bar{G}_{\bar{2}\bar{1}}=&\frac{J^2_{\perp}}{4}h\log\left\vert\frac{2h}{D}\right\vert
  L^6_-\\\nonumber
  &+\frac{J^2_{\perp}}{4}\left(
  (h+V)\log\left\vert \frac{2(h+V)}{D}\right\vert
  +(h-V)\log\left\vert
  \frac{2(h-V)}{D}\right\vert\right) L^6_+\\\nonumber
  &+ \frac{J^2_{\perp}}{4}\left(\delta h\log\left\vert
  \frac{2\delta h}{D}\right\vert+(2h)\log\left\vert
  \frac{4h}{D}\right\vert\right) \left(L^a-L^c\right)\\\nonumber
  &+\frac{J^2_z}{4}\left( (\delta h+V)\log\left\vert
  \frac{2(\delta h+V)}{D}\right\vert
  +(\delta h-V)\log\left\vert\frac{2(\delta h-V)}{D}\right\vert\right)\left(L^c+L^h\right)\\\nonumber
  &+\frac{J^2_z}{4}\left( (2h+V)\log\left\vert
  \frac{2(2h+V)}{D}\right\vert
  +(2h-V)\log\left\vert\frac{2(2h-V)}{D}\right\vert\right)\left(L^c-L^h\right)\\\nonumber
&+\frac{J_z J_{\perp}}{4}\left(
  (\delta h+V)\log\left\vert\frac{2(\delta h+V)}{D}\right\vert-(\delta h-V)\log\left\vert
  \frac{2(\delta h-V)}{D}\right\vert\right)\left(L^{4}_{-}+L^{5}_{+}\right)\\\nonumber
&+\frac{J_z J_{\perp}}{4}\left(
  (2h+V)\log\left\vert\frac{2(2h+V)}{D}\right\vert-(2h-V)\log\left\vert
  \frac{2(2h-V)}{D}\right\vert\right)\left(L^{4}_{+}+L^{5}_{-}\right)\\\nonumber
  &+\frac{J^2_{\perp}}{4}\left( (\delta h+V)\log\left\vert
  \frac{2(\delta h+V)}{D}\right\vert+(\delta h-V)\log\left\vert\frac{2(\delta h-V)}{D}\right\vert
  \right)\left(L^a-L^c\right) \\\nonumber 
  &+\frac{J^2_{\perp}}{4}\left( (2h+V)\log\left\vert
  \frac{2(2h+V)}{D}\right\vert+(2h-V)\log\left\vert\frac{2(2h-V)}{D}\right\vert
  \right)\left(L^a-L^c\right)\,.
  \label{}
\end{align}
With all the self-energy terms evaluated we can determine the
eigenvalues of the effective Liouvillian $L_{\text{eff}}(E)$. To obtain
an analytical result for the roots $\pm h$ we perform the
diagonalization of $L_{\text{eff}}$ perturbatively as well. In first
order
\begin{align}
  h^{(1)}=(\uparrow \downarrow \vert
  L_{\text{eff}}(h) \vert \uparrow \downarrow )= - ( \downarrow
  \uparrow \vert L_{\text{eff}}(-h) \vert \downarrow \uparrow
  )\,,
  \label{}
\end{align}
we find
\begin{align}
  h=h_0 +\frac{\rho_0^2}{4}&\left[-2\left(J^2_{\perp}
  h + J_z^2 \delta h\right)
  + \frac{J^2_{\perp}}{2} h \log
  \left\vert\frac{2h}{D}\right\vert\right.\\\nonumber
  &+ \frac{J^2_{\perp}}{2}
  \left( (h+V)\log \left\vert
  \frac{2(h+V)}{D}\right\vert+(h-V)\log\left\vert
  \frac{2(h-V)}{D}\right\vert \right)\\\nonumber
  &+J^2_{z}
  \left( (\delta h+V)\log \left\vert
  \frac{2(\delta h+V)}{D}\right\vert+(\delta h-V)\log\left\vert
  \frac{2(\delta h-V)}{D}\right\vert \right)\\\nonumber
  &\left. -i\frac{\pi}{4} J_{\perp}^2\left(\vert h+V \vert +
  \vert h-V \vert + 2 \vert h \vert \right)
  -i\frac{\pi}{2} J^2_z\left(\vert \delta h+V\vert +
\vert \delta h-V \vert \right)\right]\,,
\end{align}
where $\delta h=h-h_0\simeq 0$ and $h_0$ denotes the root of the bare Liouvillian
$L_{\text{ns}}$. The imaginary part of the
root $h$, which corresponds to its transient decay rate, reads
\begin{align}
  \text{Im}(h)\simeq -i\rho_0^2\frac{\pi}{16} J_{\perp}^2\left(\vert h+V \vert +
  \vert h-V \vert + 2 \vert h \vert \right)
  -i\rho_0^2\frac{\pi}{8} J^2_z\left(\vert \delta h+V\vert +
  \vert \delta h-V \vert \right)\,.
 \label{eq:decayanalytical}
\end{align}
We see that for $V\rightarrow 0$ the imaginary part of $h$ is essentially given
by the terms proportional to $J_{\perp}^2$. For $V=0$ we thus find a power law
decrease of the decay rate with $U^{-\alpha}$ and $\alpha\geq 8$. For
finite bias voltage and $U\rightarrow \infty$ the terms proportional to
$J_z^2$ become dominant and we observe a power law decrease of the decay
rate with $U^{-\beta}$
and $\beta=6$.
\begin{figure}[ ]
  \centering
  \small{(a)}
  \includegraphics[width=0.8\textwidth]{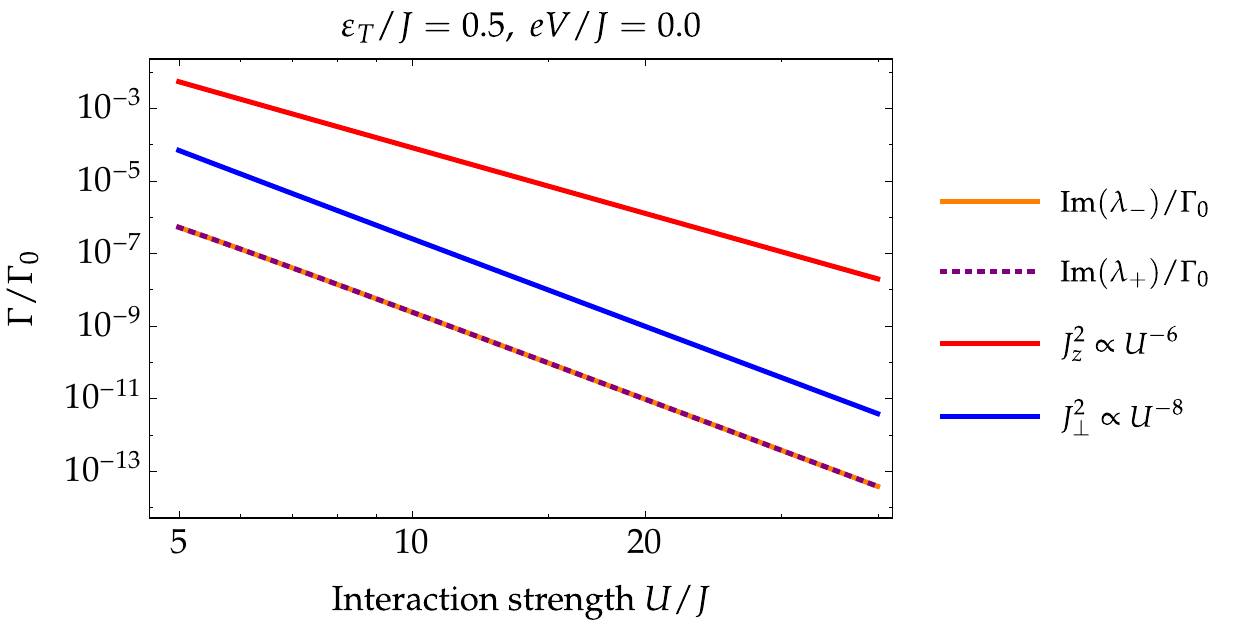}\\
  \small{(b)}
  \includegraphics[width=0.8\textwidth]{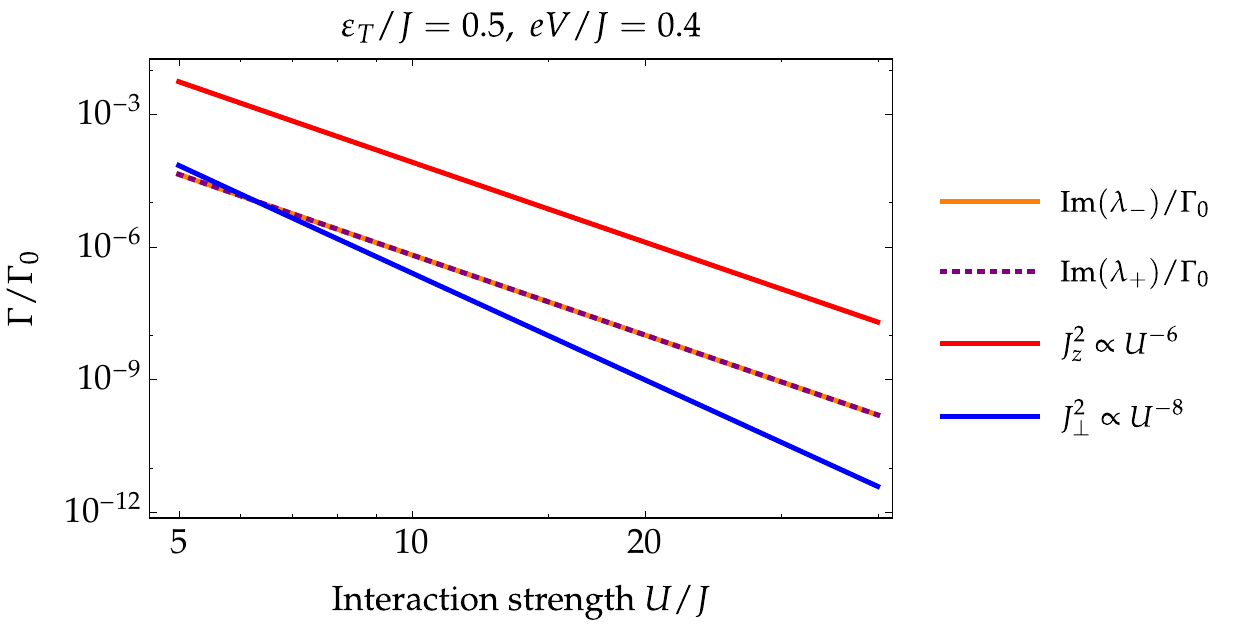}
  \caption{Decay rate $\Gamma/\Gamma_0$ of the roots $\lambda_+$
  and $\lambda_-$ for (a): $\varepsilon_T/J=0.5$, $J_{\text{c}}/J=0.5$,
$eV/J=0$ and (b): $\varepsilon_T/J=0.5$, $J_{\text{c}}/J=0.5$, $eV/J=0.4$.}
\label{fig:decayratesg2}
\end{figure}
In figure \ref{fig:decayratesg2} we plot our numerical results for the
decay rates $\text{Im}(\lambda^*_{\pm})$ with $\lambda^*_{\pm}=\pm
h$. We find that for $V=0$ the
decay rates obey a power law, $\text{Im}(\lambda^*_{\pm})(U)\propto
U^{-8}$,
the same as the spin-flip interaction $J_{\perp}^2(U)$. For finite bias
voltage we observe a different power law, $\text{Im}(\lambda^*_{\pm})(U)
\propto U^{-6}$, a behavior shared by $J_z^2(U)$. Our numerical findings
support our perturbative result for the decay rates
(\ref{eq:decayanalytical}). The perturbation theory for the effective
model finds that $\Gamma \rightarrow 0$ for $U\rightarrow \infty$ and
supports our findings from DMRG calculations and perturbation theory in
the limit of small coupling which see very long life times $\tau \gg
\Gamma_0^{-1}$ of the ring current oscillations.

\end{document}